\documentclass{svmult}

\usepackage{mathptmx}       % selects Times Roman as basic font
\usepackage{helvet}         % selects Helvetica as sans-serif font
\usepackage{courier}        % selects Courier as typewriter font
\usepackage{type1cm}        % activate if the above 3 fonts are
\usepackage{makeidx}         % allows index generation
\usepackage{graphicx}        % standard LaTeX graphics tool
\usepackage{multicol}        % used for the two-column index
\usepackage[bottom]{footmisc}% places footnotes at page bottom

\usepackage{amsmath}
\usepackage{amssymb}
\usepackage{bm}
\usepackage{comment}
\usepackage{enumerate}

\makeindex             % used for the subject index
\begin{document}

\title*{What can we learn from transfer, and how is best to do it?}
\titlerunning{What can we learn from transfer, and how is best to do it?}
\author{Wilton N. Catford}
\institute{Wilton Catford \at
Department of Physics, University of Surrey, Guildford GU2 7XH, United Kingdom, \email{w.catford@surrey.ac.uk}
}
\maketitle

\abstract{
An overview of the experimental aspects of nucleon transfer reactions with radioactive beams is presented, aimed principally at a researcher who is beginning their work in this area. Whilst the physics motivation and the means of theoretical interpretation are briefly described, the emphasis is on the experimental techniques and the quantities that can be measured. General features of the reactions which affect experimental design are highlighted and explained. A range of experimental choices for performing the experiments is described, and the reasons for making the different choices are rationalised and discussed. It is often useful to detect gamma-rays from electromagnetic transitions in the final nucleus, both to improve the precision and resolution of the excitation energy measurements and to assist in the identification of the observed levels. Several aspects related to gamma-ray detection and angular correlations are therefore included. The emphasis is on single-nucleon transfer reactions, and mostly on (d,p) transfer to produce nuclei that are more neutron-rich, but there are also brief discussions of other types of transfer reactions induced by both light ions and heavy ions.
}

%%%%%%%%%%%%%%%%%%%%%%%%%%%%%%

%-------------------------
%-------------------------------------------------------------------------------------------------
\section{Motivation to study single-nucleon transfer using radioactive beams}
\label{sec:motivation}
A single-nucleon transfer reaction is a powerful experimental tool to populate a certain category of interesting states in nuclei in a selective manner. These states have a structure that is given by the original nucleus as a core, with the transferred nucleon in an orbit around it. Nucleon transfer is thus an excellent way to probe the energies of shell model orbitals and to study the changes in the energies of these orbitals as we venture away from the stable nuclei. Despite a large number of detailed issues that complicate this simple picture, it remains the case that nucleon transfer reactions preferentially populate these "single particle" states in the final nucleus and also that these states are of especial interest, theoretically. Therefore, transfer reactions promise to be one of the most important sources of nuclear structure information about exotic nuclei, as more beams become available at radioactive beam facilities.

\begin{figure}[b]
\sidecaption
\includegraphics[width=.7\textwidth]{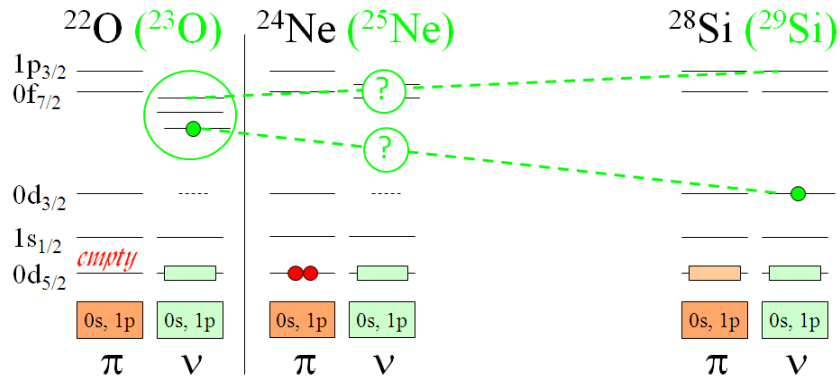}
%If the width of the Figure is less than 7.8 cm use the \texttt{sidecapion} command to flush the caption on the left side of the page. If the Figure is positioned at the top of the page, align the sidecaption with the top of the Figure -- to achieve this you simply need to use the optional argument \texttt{[t]} with the \texttt{sidecaption} command}
\caption{The effective energies of the valence neutron orbitals are modified according to the number of protons present in the $0d_{5/2}$ orbital. The effect is to replace the $N=20$ neutron shell gap by a gap at $N=16$ when the nucleus becomes more exotic. }
\label{fig:1}       % Give a unique label
\end{figure}

The factors that complicate the interpretation of the experiments arise primarily from the theoretical interpretation of the data. Experimentally, the selectivity of the transfer reactions is usually clear, and the states of interest - those having a large overlap with the simple core-plus-particle picture - are emphatically favoured. Often, these states will be embedded within a background of other nuclear levels. This selectivity on structural grounds is itself useful, and often allows immediate associations to be inferred between experimentally observed states and the predictions from, for example, shell model calculations. The states that are suppressed will have more complex wave functions that mix a number of configurations and are intrinsically more difficult to describe theoretically. In the first instance, it is in many ways best to focus upon the more simple states that are selected by transfer reactions, and to use these to refine the theory. Complications begin to arise when we seek to quantify the degree to which the wave function of a particular state overlaps with the simple core-plus-particle wave function. At that level, many debates occur, regarding the quantitative interpretation of data. With suitably stated assumptions, however, quantitative analyses of experiments can be performed and confronted with theory. Thus, on a qualitative and on a quantitative level, transfer reactions provide an indispensable tool for uncovering the structures of exotic nuclei.

\subsection{Migration of shell gaps and magic numbers, far from stability}
\label{subsec:migration}

Figure \ref{fig:1} shows a simplified representation of the proton and neutron shell model orbital energies and occupancies for some light nuclei. In the nuclear shell model, each nucleon is assumed to occupy an energy level (or orbital) that can be obtained by solving the Schr\"odinger equation for a mean field potential. This potential represents the average binding effect of all of the other nucleons. In the simplest model, the nuclear structure is obtained by filling orbitals from the lowest energies, obeying the Pauli exclusion principle. In a more sophisticated model, the interactions between valence nucleons in different orbitals (or in the same orbital) are taken into account. This allows significant mixing between different simple configurations that all have the same spin and parity and about the same (unmixed) energy. Some degree of mixing will even occur over a wide range of configuration energies. In principle the valence nucleon interaction energies, which can be represented as matrix elements in some suitable basis, can be calculated from the solutions for the mean field and an expression for the nucleon-nucleon interaction (with all of its dependence on spatial variables, spin and orbital angular momentum). In practice, the best shell model calculations in terms of agreement with experimental data are those in which the calculated matrix elements are subsequently varied by fitting them to a selection of experimental data, thus establishing an {\it effective interaction} that is valid in a particular model space that was used for the fitting procedure. Once we accept that valence nucleons will have an interaction potential, and hence some energy associated with the interaction, it naturally becomes possible that the valence interactions can actually change to some degree the effective energies of the orbitals themselves. Slightly more technically, the interaction potentials can be analysed in terms of a multipole expansion. It is the monopole term in the expansion that has the effect of changing the effective energies of orbitals. The energy of a single valence nucleon in a particlar orbital is determined by the energy of the orbital plus the sum of the monopole components of its interaction with other active valence nucleons. A closed shell has no net effect, so it is the interactions with partially filled orbitals that needs to be considered. Whilst both the interactions between like nucleons ({\it p-p}) or ({\it n-n}) and interactions between different nucleons ({\it p-n}) are all important, the strongest effects occur when it is a proton-neutron interaction between active valence nucleons. After that, the strongest effects are between orbitals of the same number of radial nodes, and then if the angular momentum is the same this makes the effect is even stronger. This arises from the degree of spatial overlap of the wave functions. For example, the interaction between protons in an open $0d_{5/2}$ orbital and neutrons in an open $0d_{3/2}$ orbital is particularly strong.

In Figure \ref{fig:1}, the structure of the $N=14$ isotones is shown, with the additional odd neutron for $N=15$ being shown in an otherwise vacant $0d_{3/2}$orbital. The $0d_{5/2}$ neutron orbital is filled, at $N=14$. On the right hand side, we see the stable nucleus $^{28}$Si, wherein the 14 protons also fill the proton $0d_{5/2}$ orbital. The $3/2^+$ state in $^{28}$Si is therefore at a relatively low energy, because the $sd$-shell orbitals are relatively closely spaced, all lying below the $N=20$ shell gap. As successive pairs of protons are removed from $0d_{5/2}$, the diagram indicates that the energy of the $0d_{3/2}$ orbital increases. This is actually in accord with detailed calculations and can be understood in terms of the monopole interaction \cite{otsuka1,otsuka2} and a version of this diagram can be found in ref. \cite{otsuka1}. By the time we reach the neutron rich $^{22}$O, the $0d_{3/2}$ orbital has risen to such an extent that the shell gap is now below that orbital, at $N=16$. The orbital that has moved up in energy has $j=\ell - 1/2$ and the reason for its change is that there are fewer protons in $0d_{5/2}$ (where $j=\ell + 1/2$) with which a valence $d_{3/2}$ neutron can interact. This proton-neutron interaction between $\ell + 1/2$ and $\ell - 1/2$ nucleons is attractive \cite{otsuka1}, and hence this reduction in $0d_{5/2}$ protons causes the raising in energy of the neutron $0d_{3/2}$ orbital. This is, in fact, essentially the explanation for $^{24}$O being the heaviest bound oxygen isotope (with the neutrons just filling the $1s_{1/2}$ orbital). Neutrons in the $0f_{7/2}$ and $1p_{3/2}$ orbitals, with $j=\ell + 1/2$, experience a repulsive interaction with the $0d_{5/2}$ protons and hence they are lowered in energy as these protons are removed. This further confounds the previous $N=20$ gap seen for nuclei near stability, and also tends to displace the $N=28$ gap to a higher number ($N=34$). In the present work, several of the example nuclei studied using transfer ($^{25,27}$Ne, $^{21}$O) are directly of interest because of this particular migration of orbital energies. What we measure experimentally are the energies of actual states in these nuclei, and not the energies of the shell model orbitals {\it per se}, but there is a strong connection between the energies of the states and the orbitals in the cases that are studied here.

\subsection{Coexistence of single particle structure and other structures}
\label{subsec:coexistence}

\begin{figure}[h]
\sidecaption
\includegraphics[width=.55\textwidth]{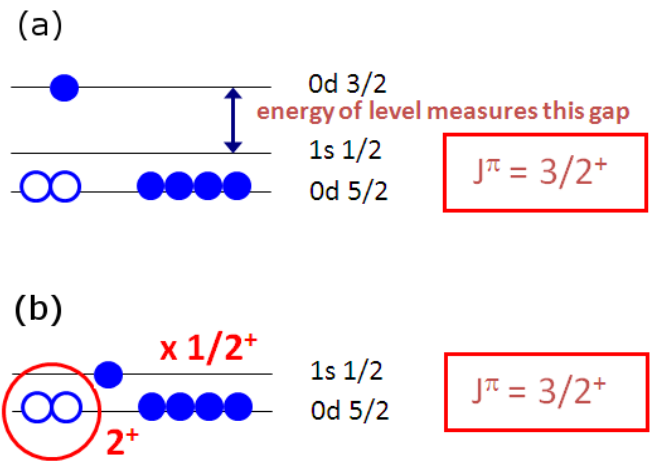}
%If the width of the Figure is less than 7.8 cm use the \texttt{sidecapion} command to flush the caption on the left side of the page. If the Figure is positioned at the top of the page, align the sidecaption with the top of the Figure -- to achieve this you simply need to use the optional argument \texttt{[t]} with the \texttt{sidecaption} command}
\caption{Neutron structure for states in $^{21}$O: (a) a low-lying $3/2^+$ state can be made by transferring a neutron into the vacant $0d_{3/2}$ orbital , or (b) by having a neutron in $1s_{1/2}$ coupled to a $2^+$ $^{20}$O core, where the two holes in $0d_{5/2}$ are coupled to spin 2.}
\label{fig:2}       % Give a unique label
\end{figure}

Of course, it is not the case that all of the excited states in the final nucleus will have a structure that is simply explained by a neutron orbiting the original core nucleus. Such states are an important but (usually) small subset of the states in the final nucleus, and are selectively populated by transfer reactions. To measure the energy of the  $0d_{3/2}$ neutron orbital, say, in $^{21}$O we could imagine an experiment to add a neutron to $^{20}$O and then deduce the energy of the $3/2^+$ excited state, and hence the $0d_{3/2}$ orbital energy relative to the $0d_{5/2}$ orbital of the ground state. This is shown conceptually in Figure \ref{fig:2}(a). In the lowest energy configuration, the two holes in the neutron $0d_{5/2}$ orbital are coupled to spin zero. This association of the energy of the state directly with that of the orbital is overly simplified because the state will not have a pure configuration. In Figure \ref{fig:2}(b), another relatively low energy configuration is shown, which also has spin and parity $3/2^+$. Here, the holes in $0d_{5/2}$ are coupled to spin 2, as they are in the $2^+$ state of $^{20}$O. A neutron in $1s_{1/2}$ can then couple with this to produce two states in $^{21}$O, one of which is $3/2^+$. The residual interactions between valence nucleons will mix these two configurations and the nucleus $^{21}$O will have the single particle amplitude split between the two states. Indeed, in the real nucleus, there will be even more components contributing to the wave functions with various smaller amplitudes.

\subsection{Description of single particle structure using spectroscopic factors}
\label{subsec:SFs}

\begin{figure}[h]
\sidecaption
\includegraphics[width=.7\textwidth]{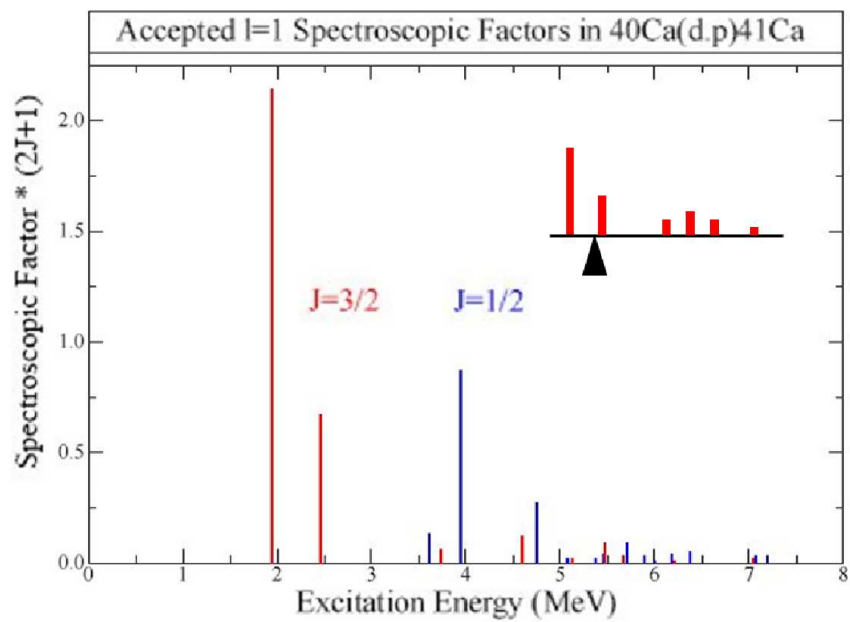}
%If the width of the Figure is less than 7.8 cm use the \texttt{sidecapion} command to flush the caption on the left side of the page. If the Figure is positioned at the top of the page, align the sidecaption with the top of the Figure -- to achieve this you simply need to use the optional argument \texttt{[t]} with the \texttt{sidecaption} command}
\caption{Spectroscopic factor versus excitation energy for $3/2^-$ (red) and $1/2^-$ (blue) states in $^{41}$Ca. The strength for a given spin is split between different states in the final nucleus. This Figure is due to John Schiffer \protect
\cite{JS_ORNL}. Here, we have added the inset to indicate schematically how the weighted average of the $3/2^-$ excitation energies can be calculated, which gives a measure of the energy of the $1p_{3/2}$ single particle orbital. }
\label{fig:3}       % Give a unique label
\end{figure}

Due to the mixing of different states with the same spin and parity, the single particle state produced by a nucleon orbiting the core of the target, in an otherwise vacant orbital,  will be mixed with other nuclear states of different structures. Usually, these will be of more complex structures, or core excited structures. The contribution that this single particle amplitude makes, to the different states, will result in these states all being populated in a nucleon transfer reaction. The strength of the population of each state in the reaction will depend on the intensity of the single particle component. This intensity is essentially the quantity that is called the {\it spectroscopic factor}. Experimentally, it is measured by taking the cross section that is calculated for a pure single particle state and comparing it to the cross section that is measured. More specifically, this comparison is performed using differential cross sections, which are a function of the scattering angle. If the picture described here is correct, then the experimental cross section should have the same shape as the theory, and simply be multiplied by a number less than one - that is, the spectroscopic factor. To describe the sharing of intensity between states, we say that the single particle strength will be spread across a range of states in the final nucleus. This is represented in Figure \ref{fig:3}, where the single particle strength (represented as the spectroscopic factor) is plotted as a function of excitation energy. The weighted average of the excitation energies, for all states containing strength from a particular $\ell j$ orbital, will give the energy of that orbital. Note that, for experiments with radioactive beams, the limited intensity of the beam is likely to preclude the possibility of identifying and measuring all of the spectroscopic strength, which was traditionally the aim of transfer experiments. A different approach will often be dictated by these circumstances, wherein only the strongest states are located experimentally. Then, placing more reliance on theory than was formerly done, an association can be made between the strong states experimentally and the states predicted by the theory to be the strongest. We then need to see whether the experimental data, in terms of the energies and spectroscopic factors for the strongest states, can give us enough clues about how to adapt the theoretical calculations to give an improved set of predictions. If applied consistently across a range of nuclei, using the same theory, this approach can reasonably be expected to yield good results.

\subsection{Disclaimer: what this article is, and is not, about}
\label{subsec:disclaimer}

This article is intended to describe briefly the general motivations for studies using (mostly) single nucleon transfer, and to provide in some detail the background, insights and perspectives relevant to designing and performing the experiments. For more details about the nuclear structure motivations in terms of nuclear structure and monopole shift the reader is referred to several excellent reviews \cite{sorlin,sorlin2,otsuka}.

This article most definitely does not seek to summarise or describe the theories that are used to interpret the data from nucleon transfer reactions, although some general features of the theoretical predictions are discussed and a justification is given for the model of choice for the examples of analysis that are described here. Detailed descriptions of the relevant reaction theory can be found in several well-known articles and books, such as those by Glendenning \cite{Glendenning-Cerny, Glendenning} and Satchler \cite{SatchlerIntro, SatchlerBig}. An excellent and up-to-date introduction and overview with particular reference to weakly bound and unbound states is given in this volume by G\'omez Camacho and Moro \cite{GCM}.

With regard to the experimental results, although the main objectives of most of these measurements is to obtain the differential cross sections for individual final states, just a small number of illustrative results are shown here. In all of the discussions, the references are given for the original work, and it is to those publications that reference may be made in order to study the extent and quality achieved for the various differential cross section measurements. It is through the measurement and interpretation of these differential cross sections that the assignments of angular momentum and determinations of spectroscopic single-particle strength are made, for the nuclear states.

%-------------------------------------------------------------------------------------------------
\section{Choice of the reaction and the bombarding energy}
\label{sec:2:reactions}

In this section, some features of transfer reactions as traditionally performed using stable targets and a low-mass beam (for example, the (d,p) reaction) are reviewed. Some of the differences in the case of inverse kinematics are introduced.

\subsection{Kinematics and measurements using normal kinematics}
\label{subsec:kinematics}

\begin{figure}[h]
\sidecaption
\includegraphics[width=.7\textwidth]{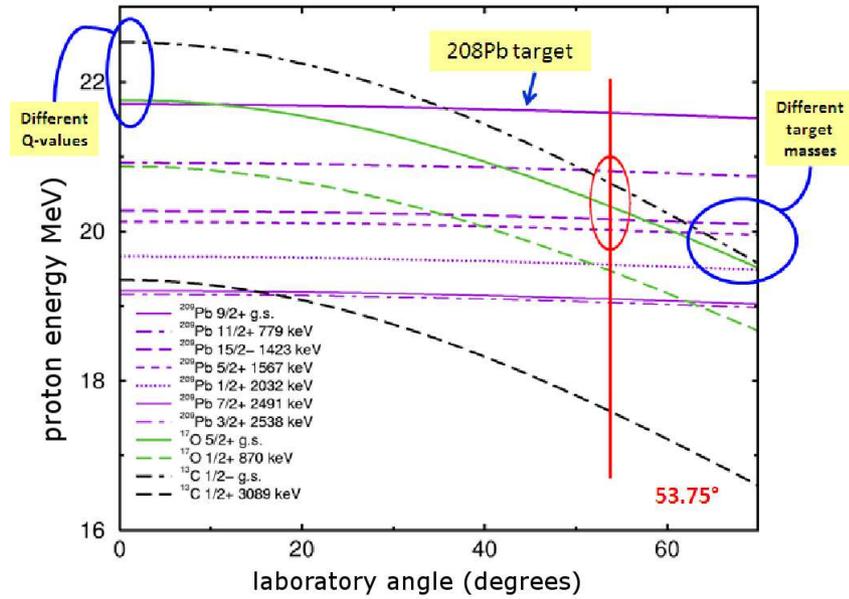}
%If the width of the Figure is less than 7.8 cm use the \texttt{sidecapion} command to flush the caption on the left side of the page. If the Figure is positioned at the top of the page, align the sidecaption with the top of the Figure -- to achieve this you simply need to use the optional argument \texttt{[t]} with the \texttt{sidecaption} command}
\caption{Kinematics plots showing proton energy as a function of laboratory angle for the reaction (d,p) initiated by 20 MeV deuterons. Different curves represent the population of different excited states formed by reactions on $^{208}$Pb, $^{12}$C and $^{16}$O (see text). The angle $53.75^\circ$ is relevant to Figure \protect\ref{fig:5}. }
\label{fig:4}       % Give a unique label
\end{figure}

\begin{figure}[h]
\sidecaption
\includegraphics[width=.7\textwidth]{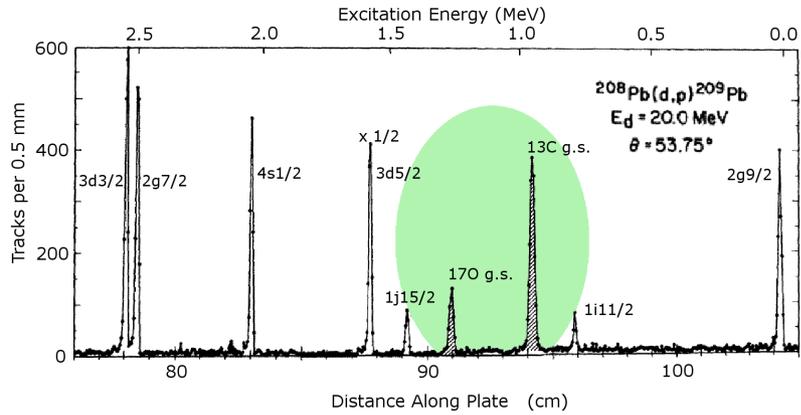}
%If the width of the Figure is less than 7.8 cm use the \texttt{sidecapion} command to flush the caption on the left side of the page. If the Figure is positioned at the top of the page, align the sidecaption with the top of the Figure -- to achieve this you simply need to use the optional argument \texttt{[t]} with the \texttt{sidecaption} command}
\caption{Magnetic spectrometer data for protons from $^{208}$Pb(d,p)$^{209}$Pb at a beam energy of $E_d=20.0$ MeV and a laboratory angle of $53.75^\circ$. Data are from ref. \protect\cite{Kovar}. Excitation energy increases from right to left and the unshaded peaks correspond to states in $^{209}$Pb. The shaded region is where reactions on the $^{12}$C and $^{16}$O in the target produce contaminant peaks. }
\label{fig:5}       % Give a unique label
\end{figure}

A good way to measure (d,p) reactions when using a beam of deuterons and a stable target is to use a high resolution magnetic spectrometer to record the protons from the reaction, because this can be done with a high precision and a low background. The proton peaks observed at a particular angle will have different energies for different excited states and hence will be dispersed across the focal plane of the spectrometer. The spacings of the proton energies will be almost the same as the spacings of the energy levels in the final nucleus. An example of the kinematical variation of proton energies with laboratory angle is shown in Figure \ref{fig:4}. The lines that are almost horizontal are calculations of the proton energies from the (d,p) reaction on a $^{208}$Pb target, with the uppermost being for protons populating the ground state of $^{209}$Pb. The energies have little variation with laboratory angle because the very heavy recoil $^{209}$Pb nucleus takes away very little kinetic energy. The uppermost line, with a much bigger slope, is for the (d,p) reaction on a much lighter target,  $^{12}$C. The lines of intermediate slope are for the (d,p) reaction on a target of $^{16}$O. The energy at zero degrees is different for the different targets because of the different reaction Q-values, whereas the slopes reflect the target mass. The carbon and oxygen calculations are shown, because these isotopes are typical target contaminants. In a study of $^{208}$Pb(d,p)$^{209}$Pb, the contaminant reactions will give proton energies that overlap the energy region of interest for the $^{209}$Pb states, but these can be identified by comparing data taken at different laboratory angles, since the contaminant peaks will shift in energy, relative to the $^{209}$Pb peaks. The example of the proton energies seen in a measurement made at $53.75^\circ$ is shown in Figures \ref{fig:4} and \ref{fig:5}.

The peaks corresponding to different final states in $^{209}$Pb, measured at $53.75^\circ$ for the (d,p) reaction \cite{Kovar}, are shown in Figure \ref{fig:5}. The different intensities reflect both the spectroscopic strengths and the dynamical effects of different angular momentum transfers. It is apparent that different states can easily be resolved and studied. The peaks in the shaded region of Figure \ref{fig:5} correspond to the reactions populating the ground states of $^{17}$O and $^{13}$C from the oxygen and carbon contamination in the target. At increasing laboratory angles, these peaks would be seen to move to the left in the spectrum, relative to the $^{209}$Pb peaks.

\subsection{Differential cross sections: dependence on beam energy and $\ell$ transfer}
\label{subsec:crosssection}
	
\begin{figure}[h]
\sidecaption
\includegraphics[width=.6\textwidth]{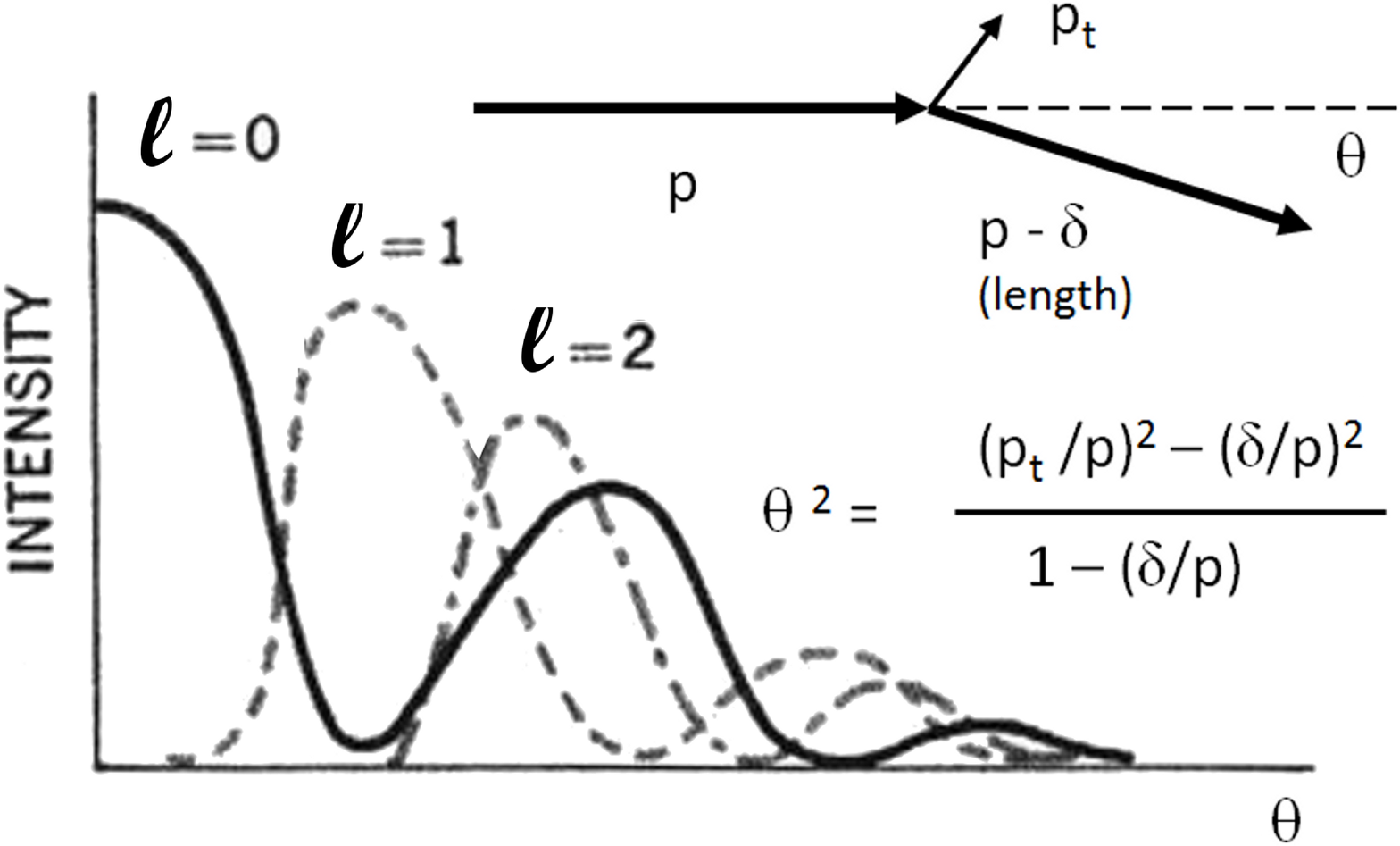}
%If the width of the Figure is less than 7.8 cm use the \texttt{sidecapion} command to flush the caption on the left side of the page. If the Figure is positioned at the top of the page, align the sidecaption with the top of the Figure -- to achieve this you simply need to use the optional argument \texttt{[t]} with the \texttt{sidecaption} command}
\caption{A consideration of the conservation of linear momentum in transfer implies a relationship of the laboratory scattering angle to the transferred momentum, and therefore to the transferred orbital angular momentum, $\ell$. This implies that the location of the primary maximum in the angular distribution will be approximately proportional to the transferred $\ell$ (see text). }
\label{fig:6}       % Give a unique label
\end{figure}

The principal piece of information (after excitation energy) that is measured directly, via transfer studies, is the orbital angular momentum that is transferred to the target nucleus. This comes from the shape of the differential cross section. Next, the magnitude of the cross section can tell us the magnitude of the single-particle component of the wave function, or the spectroscopic factor.

\begin{figure}[h]
\sidecaption
\includegraphics[width=1.0\textwidth]{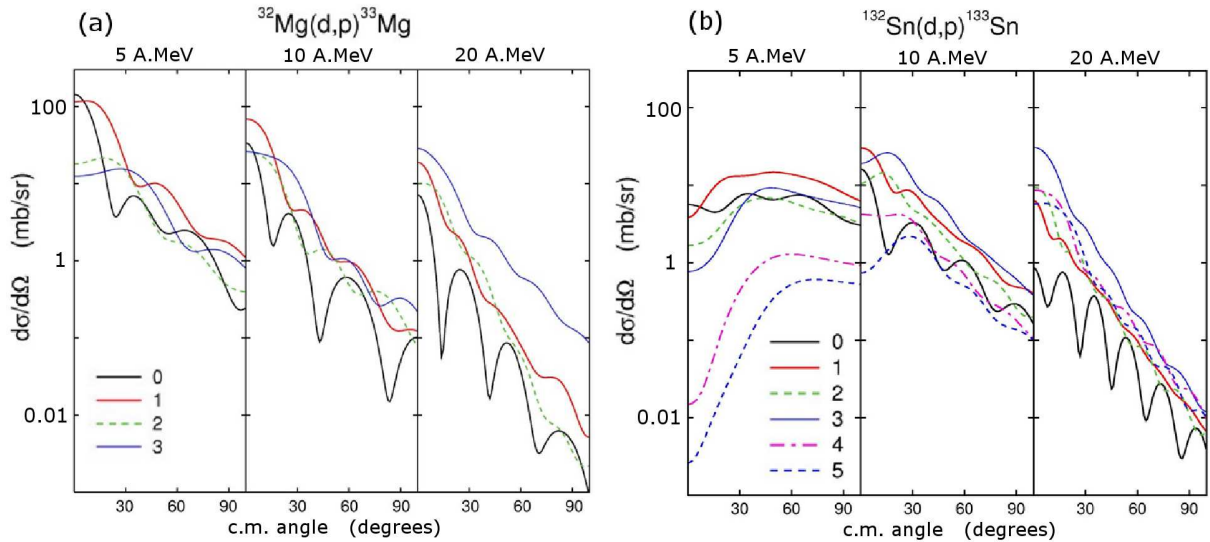}
%If the width of the Figure is less than 7.8 cm use the \texttt{sidecapion} command to flush the caption on the left side of the page. If the Figure is positioned at the top of the page, align the sidecaption with the top of the Figure -- to achieve this you simply need to use the optional argument \texttt{[t]} with the \texttt{sidecaption} command}
\caption{Differential cross sections for single nucleon transfer in (a) $^{32}$Mg(d,p)$^{33}$Mg, and (b) $^{132}$Sn(d,p)$^{133}$Sn. The three panels in each case are for three different bombarding energies, namely 5, 10 and 20 A.MeV. Each panel shows calculations for several different $\ell$-transfers. The ADWA model was used (see text, subsection \protect\ref{subsec:adwa}). These plots are in terms of the centre of mass reaction angle, $\theta _{{\rm c.m.}}$.}
\label{fig:7}       % Give a unique label
\end{figure}

The transferred angular momentum will indicate, for single-nucleon transfer, into which orbital the nucleon has been transferred. The transferred angular momentum is measured via the angular distribution of the reaction products. In this type of reaction, the differential cross section will tend to have some diffraction-like oscillatory behaviour, with the angle of the main maximum being related to the magnitude of the transferred angular momentum. We can see how the transferred angular momentum affects the angular distribution by considering a simple momentum diagram. Suppose as in the inset of Figure \ref{fig:6} that the incident projectile has momentum of magnitude $p$ and that the momentum transferred to the target nucleus has magnitude $p_t$. For a small scattering angle, $\theta$, the beam particle will have only a small reduction in the magnitude of its momentum, as seen by construction of the vector diagram for momentum conservation (cf. Figure \ref{fig:6}). From the application of the cosine rule to this triangle, the formula for $\theta ^2$ as shown in the Figure can be derived, where we make use of the expansion to second order for cosine: $2(1- \cos \theta ) \approx 2(1 - [1 - \theta ^2 /2! ]) = \theta^2$. From inspection of the diagram, the reduction $\delta$ in the length of the $p$ vector is small compared to the magnitude of the actual transferred momentum, $p_t$. Hence, we can drop the terms in $(\delta /p)$ in the expression for $\theta ^2$ and we obtain $\theta ^2 \approx (p_t /p)^2$. If the nucleon is transferred at the surface of the target nucleus, which has radius $R$, then the transferred angular momentum $\ell$ is given by $p_t \times R = \sqrt{\ell (\ell +1)} \hbar$. This immediately indicates that $\theta \approx {\rm constant} \times \ell$, and in a full quantum mechanical treatment we will not see a single angle but can expect a peak to occur in the differential cross section, at a laboratory angle that is approximately proportional to the transferred angular momentum, $\ell$. This is shown schematically in Figure \ref{fig:6}, which also includes the diffractive effects in a schematic fashion. In fact, for deuterons incident at a kinetic energy of $E$ (MeV) on a target of mass $A$ this simple picture gives $\theta {\rm (degrees)} \approx 217/(\sqrt{E} \times A^{1/3}) \times \sqrt{\ell (\ell+1)}$. For 20 MeV deuterons incident on a target of mass 32, the constant term evaluates to $15^\circ$, which of course can serve only as a guide, but is in reasonable agreement with the trend in the primary maxima observed in the middle panel of Figure \ref{fig:7}(a) which shows proper calculations for (d,p) on $^{32}$Mg at 10 A.MeV. The Figure is actually plotted in terms of $\theta _{{\rm c.m.}}$, but would look very similar when plotted in terms of $\theta _{{\rm lab}}$ for normal kinematics (which refers to the situation where the target is heavier than the projectile). The preceding discussion of the vector diagram is adapted from reference \cite{Cohen}.

Calculations are shown in Figure \ref{fig:7} for various $\ell$-transfers at several different bombarding energies, and for two different targets. The first point to note is that the shapes of the distributions for different $\ell$-transfer are distinctive, especially for 10 A.MeV (the middle panels). For the light nucleus $^{32}$Mg, the 5 A.MeV distributions are also characteristic of the transferred $\ell$. For the heavier $^{132}$Sn target, the distributions are less distinctive due to the forward angle parts (small $\theta _{{\rm c.m.}}$) being suppressed. This is due to the Coulomb repulsion between the projectile and the target, which means that the small angle scattering (especially) has a suppressed nuclear component. In addition to the above considerations, there is a general trend towards lower cross sections as the bombarding energy increases, of around half to one order of magnitude per 10 A.MeV. Taken together, this information suggests that 10 A.MeV is an ideal bombarding energy for this type of study, and this can be relaxed down to 5 MeV perhaps, for lighter nuclei. The remaining question is whether the existing theories are equally valid at all energies, and the ADWA model used here (see section \ref{subsec:adwa}) should have good validity at both 5 and 10 A.MeV, although probably not at energies much lower than this.

The aim of a typical nucleon transfer experiment is to measure the differential cross sections for different states in the final nucleus. From the shape of the cross section plot, the transferred angular momentum can then be deduced. The calculations shown in Figure \ref{fig:7} are for pure single-particle states. That is, it is assumed that the structure of the final state is given perfectly by the picture of the target core with the transferred nucleon in an associated shell model orbital. Hence, another important experimental result will be the scaling factor between the theoretical calculation and the data, which will give the experimental value for the spectroscopic factor.

\begin{figure}[h]
\sidecaption
\includegraphics[width=1.0\textwidth]{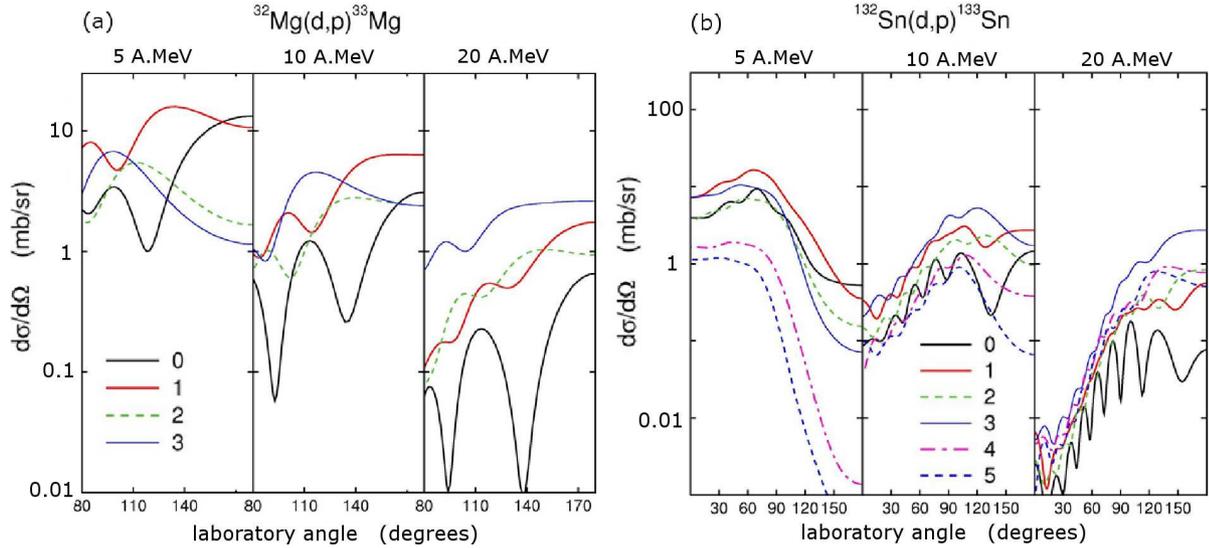}
%If the width of the Figure is less than 7.8 cm use the \texttt{sidecapion} command to flush the caption on the left side of the page. If the Figure is positioned at the top of the page, align the sidecaption with the top of the Figure -- to achieve this you simply need to use the optional argument \texttt{[t]} with the \texttt{sidecaption} command}
\caption{As for Figure \protect\ref{fig:7}, but in terms of the laboratory angle for the detected proton. In this reference frame, the extreme right of each panel corresponds to the point at the extreme left of the panels in Figure \protect\ref{fig:7}. }
\label{fig:8}       % Give a unique label
\end{figure}

If experiments are performed in normal kinematics, with a deuteron beam, then the cross sections will look much like Figure \ref{fig:7} whether we plot them in terms of the centre of mass angles or the laboratory angles. However, in reality the isotopes $^{32}$Mg and $^{132}$Sn used in these examples are radioactive and the experiments need to be performed in inverse kinematics: where the deuteron is the target and the heavier particle is the projectile. In Figure \ref{fig:8}, the same calculations as in Figure \ref{fig:7} are plotted, but using the laboratory angles and assuming an inverse kinematics experiment. The same relative velocities of beam and target, i.e. the same values of the beam energy in MeV per nucleon, are employed. It can be seen that the structure characteristic of $\ell$ is maintained, for the cases where it was previously evident. The transformation takes zero degrees in the centre of mass frame to $180^\circ$ in the laboratory frame. Now, the first peak observed relative to $180^\circ$ is further from $180^\circ$ as the $\ell$-transfer increases. From inspection, an experimental measurement should include at least the region from $90^\circ$ to $180^\circ$ in order to allow an assignment of the transferred $\ell$ according to the observed shape of the distribution. The situation with the heavier target is more problematic, especially at the lowest energy shown here.

The transformation from the centre of mass to the laboratory reference frame, and in particular the transformation of the solid angle, is discussed further in section \ref{subsec:com}.

\subsection{Choice of a theoretical reaction model: the ADWA description}
\label{subsec:adwa}

The perfect theoretical model to interpret experimental data for transfer reactions does not exist. The scattering theory is most often treated in an {\it optical model} approach, where the scattering potential is complex and has attractive and absorptive components. As in optical light being scattered from a cloudy crystal sphere, the loss of flux (by whatever process) is represented mathematically by the imaginary part of the potential. Most often, but not of necessity, the final state populated in a reaction such as (d,p) is represented as a core (being the original target nucleus) with the transferred nucleon in an eigenstate of the potential that arises due to the core. This implies a perfect single-particle structure for the final state, and the ratio between the experimental and theoretical cross sections is then the spectroscopic factor, as previously discussed. The simplest scattering theory, described in introductory quantum mechanics texts (for example ref. \cite{Schiff}), is in terms of a {\it plane wave Born approximation} (PWBA). An improved model \cite{Schiff} replaces the plane waves by the wave solutions that are distorted by the presence of the scattering potential, giving the {\it distorted wave Born approximation} (DWBA).
Even though transfer has been a widely used and valuable tool in nuclear spectroscopy for well over 50 years, there are still new and important developments occurring in quite fundamental aspects of the theory. One aspect of this concerns the spatial localisation of the transferred nucleon in the projectile and the final nucleus, or what is known as the {\em form factor}. Another important aspect, particularly for the (d,p) reaction, concerns the coupling to continuum states. Because the deuteron is weakly bound, it very easily disintegrates in the field of the target nucleus when used as a projectile. When the (d,p) reaction is applied to weakly bound exotic nuclei, the problem also occurs for the final nucleus. What is more, the coupling is not necessarily one-way: continuum states can couple back to the bound state, which can have important effects on the reaction cross section. One way to take this into account is via {\em coupled reaction channels} (CRC) calculations, in which all of the different contributing reaction pathways are explicitly included in the calculation. In order to include the continuum contribution, the theory usually considers hypothetical energy bins in the continuum and treats them as different states that can couple into the intermediate stages of the reaction. These are {\em coupled discretized continuum channels} (CDCC) calculations. The challenges of such calculations are many, including the computational power required and the choices of parameters for the various coupling strengths.

An ingenious analytical short-cut to include continuum contributions was developed by Johnson and Soper \cite{ADWA}. In the scattering process, the neutron and the proton inside the deuteron have complex histories, and in particular when continuum states for the neutron are included - that is, deuteron or final-nucleus breakup. The Johnson-Soper method relied on the observation that certain integrations over all spatial coordinates are dominated by the contributions wherein the neutron and proton are within a range determined by the neutron-proton interaction. Within a {\em distorted wave} formulation, certain energy differences are ignored, which means that the approximations of the model become less applicable at lower beam energies. However, at 5 to 10 A.MeV they should remain substantially valid. The coupling to the continuum, subject to these approximations, is included exactly and to all orders by means of the simplified integrations. This theoretical method has become known as the {\em adiabatic distorted wave approximation}, or ADWA. A convenient feature is that the calculations are largely identical (but with different input) to those required for the DWBA, and hence the pre-existing DWBA computer codes can be adapted to perform ADWA calculations. The DWBA remains another popular choice for the analysis of transfer reactions. Descriptions can be found, for example, in the articles and books by Glendenning \cite{Glendenning-Cerny, Glendenning} and Satchler \cite{SatchlerIntro, SatchlerBig}. The DWBA uses imaginary potentials to take into account the loss of reaction flux from the elastic channel, which allows for deuteron breakup but not for a proper two-way coupling with the continuum. The extensions via CDCC are computationally intensive and often incomplete in terms of the contributing physics. Therefore, the ADWA has important advantages in the case of (d,p) reactions and is adopted for all such analysis in the present work. The calculations are performed using a version of the code TWOFNR \cite{TWOFNR}. The ADWA method has recently been refined to take into account the zero-point motion of the neutron and proton inside the bound deuteron \cite{TimofeyukJS}.
	
\subsection{Comparisons: other transfer reactions and knockout reactions}
\label{subsec:knockout}
	
In the discussion in this article the emphasis is on single-nucleon transfer, and primarily (d,p) reactions, studied in inverse kinematics with radioactive beams. In terms of physics, the aim which is emphasised is the understanding of single particle structure and the evolution of shell orbitals and shell gaps as nuclei become more exotic. There are certainly other types of transfer reaction and other ways with which to probe single particle structure. Some of those topics are briefly described here. This article aims to identify the experimental challenges and techniques of transfer reaction studies, rather than to provide a review of all such studies in the literature; some more details of other work can be found, for example, in ref. \cite{KateNobel} or in other papers by many of the groups cited in section \ref{sec:4:lightion}.

Nucleon removal reactions include (p,d) and (d,t) which are discussed in sections \ref{subsec:spectrometer} and \ref{subsec:dt} respectively, and (d,$^3$He). An alternative to the first two is ($^3$He,$\alpha$) whilst an alternative to (d,p) is ($\alpha$,$^3$He). The choice of which reaction to use should not be random. The helium-induced reactions will generally show a different selectivity due to the different reaction Q-value, and could be chosen to highlight higher-$\ell$ transfers. In terms of the discussion in section \ref{subsec:crosssection}, a more negative Q-value will reduce the kinetic energy in the exit channel so that the exiting particle takes away less orbital angular momentum than it brings in. This will tend to favour the higher $\ell$-transfers. In practice, the helium-induced reactions are harder to study using radioactive beams. No simple and thin solid helium target exists, so it is necessary either to use a gas target (a windowed cell, or a differentially pumped jet) or an implanted helium-in-metal target or a cryogenic target. Each has its own challenges, but can be built and will find an increased application in the future.

Another important type of transfer reaction is when a cluster is transferred. It is always the case that when multiple particles are transferred then the process could be single-step (when the whole cluster is preformed and is transferred) or could have two or even more steps involved. Multiple-step processes are modelled theoretically using {\it coupled reaction channel} (CRC) extensions of the DWBA. In the case of heavy-ion transfer, they can also be modelled semiclassically, as mentioned in section \ref{subsec:hi-selectivity}. Traditionally, anything heavier than helium is called a heavy ion, and two heavy-ion induced transfer reactions of particular importance are ($^6$Li,d) and ($^7$Li,t), which transfer an $\alpha$-particle. Various heavy-ion transfer reactions, including $\alpha$-transfer, are discussed for example in ref. \cite{AnyasWeiss}.

The simplest form of cluster transfer is probably the (t,p) reaction in which the transfer of two neutrons, coupled to spin and relative orbital angular momentum zero, is the dominant mechanism. These can carry various amounts of angular momentum with them as a cluster, into the final nucleus. Experimentally, it is a challenging reaction: historically, the tritium nucleus was the projectile and would pose particular problems due to its radioactivity, and with the advent of radioactive beams the tritium has to be incorporated into a compact target and then be bombarded, which potentially poses even greater problems. Nevertheless, these problems have been solved in a study of shape coexistence in $^{32}$Mg via the (t,p) reaction in inverse kinematics \cite{Wimmer}. The beam was $5 \times 10^4$ pps of $^{30}$Mg at 1.8 A.MeV at ISOLDE, CERN. This experiment used the T-REX array which is described in section \ref{subsec:silicon}. The target was a foil of titanium metal ($500\mu$g/cm$^2$) into which $40\mu$g/cm$^2$ of $^3$H had been absorbed. There was a ratio of $\approx 1.5$ of hydrogen atoms to lattice Ti atoms, giving a radioactivity of the target of 10 GBq. For stable targets, the (t,p) reaction tends to have a large positive Q-value. For the more neutron-rich radioactive isotopes, such as $^{30}$Mg, the Q-value drops to be close to zero but the kinematics remain quite similar to (d,p), which is discussed in section \ref{subsec:inversekinematics}.

With the advent of radioactive beams at the extremes of measured nuclear existence, obtained via intermediate and high energy fragmentation reactions at laboratories such as MSU, GANIL, GSI and RIKEN, a new type of nucleon removal reaction was developed and exploited. This type of reaction is sometimes called a {\it knockout} reaction, but it is completely separate from true knockout reactions such as $(e,e^\prime p)$ and $(p,p^\prime p)$. The nucleus in the beam is incident on a light target nucleus that acts like a black disk and ideally cannot be internally excited without disintegrating - the usual choice is $^9$Be. The experimental requirement is that the projectile survives the reaction, with just the single nucleon removed, and this automatically selects very peripheral collisions. The black disk essentially erases some part of the tail of the wave function of the removed nucleon \cite{Hansen}. This method, which was originally developed to study the ground states of halo nuclei, provides another way in which to study the single particle structure of nuclei. Nucleon removal from the ground state of a projectile simultaneously studies the structure of the projectile state and the structure of the final nucleus. Individual states in the final nucleus can be identified using gamma-ray spectroscopy. The angular momentum transfer and the spectroscopic factor are deduced, respectively, from the width of the longitudinal momentum distribution of the beam fragment and from the magnitude of the cross section. A very successful method of analysing these reactions was developed using high-energy Glauber approximations that were previously used to describe high energy deuteron-induced reactions (the deuteron being the archetypal halo nucleus) and this theory is outlined in ref. \cite{JATLewes}, with a more extensive discussion of results in the review of ref. \cite{JATreview}. A currently very topical result from the extensive studies using knockout reactions is the apparent quenching of single-particle spectroscopic factors relative to the predictions of large-basis shell model calculations \cite{Gade}. The quenching appears to be correlated with the binding energy of the removed nucleon, which suggests some connection with higher-order correlations of nucleons, coupling to configurations outside of the shell model basis. Various different explanations have been advanced for this effect, for example those in refs. \cite{TimofeyukSource,Barbieri,barbieri2}. The observations appear to be consistent with previously observed quenching of spectrocopic strength in stable nuclei using $(e,e^\prime p)$. One way to investigate this for radioactive nuclei, and also to check the reaction dependence, is via $(p,p^\prime p)$ knockout reactions such as those performed in Japan \cite{Kobayashi} and GSI \cite{GSIp2p}. Another is to compare neutron and proton knockout with results from (d,t) and (d,$^3$He) studies, as has been performed for the neutron deficient $^{14}$O nucleus \cite{Freddy}.

%-------------------------------------------------------------------------------------------------
\section{Experimental features of transfer reactions in inverse kinematics}
\label{sec:3:inverse}

This section addresses some simple and rather general features of reactions such as (d,p) and (p,d) when studied in inverse kinematics. Instead of the centre of mass frame being almost at rest in the laboratory frame, as in normal kinematics experiments, the centre of mass frame moves with nearly the beam velocity. The kinematical variation of energy with angle therefore bears no resemblance to the situation for normal kinematics shown in Figure \ref{fig:4}. In a (d,p) or (p,d) reaction, the mass of the light (target) particle is substantially changed by the transfer, being halved in (d,p) or doubled in (p,d). This in itself turns out to be a major factor in determining the two-body kinematics of the reaction. In order to illustrate this, it is convenient to use velocity addition diagrams, where we add the velocities of particles as measured in the centre of mass frame to a vector representing the velocity of the centre of mass frame in the laboratory. The resultant vectors give the velocities of the final particles in the laboratory frame, and of course this is using the Galilean transformation and thus is strictly correct only for non-relativistic situations. This is no great problem if we are working at the energies of order 10 A.MeV that were suggested in section \ref{subsec:crosssection}. The discussion in the following section follows that in ref. \cite{Divonne}.

\begin{figure}[h]
\sidecaption
\includegraphics[width=.7\textwidth]{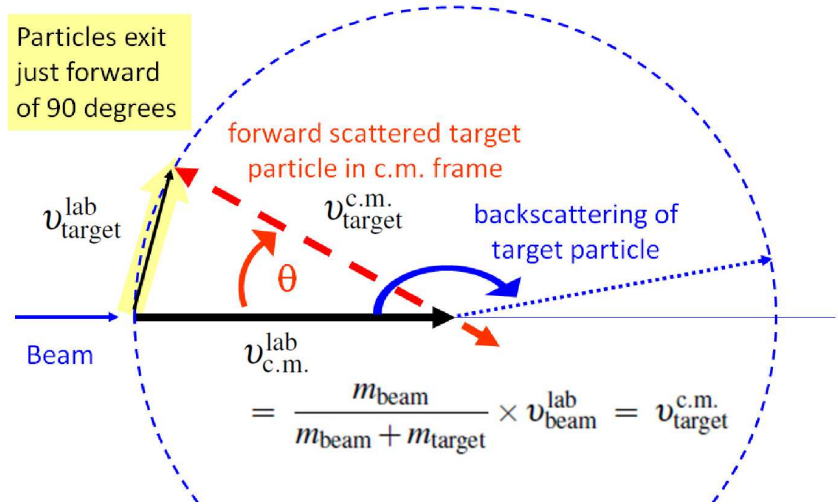}
%If the width of the Figure is less than 7.8 cm use the \texttt{sidecapion} command to flush the caption on the left side of the page. If the Figure is positioned at the top of the page, align the sidecaption with the top of the Figure -- to achieve this you simply need to use the optional argument \texttt{[t]} with the \texttt{sidecaption} command}
\caption{Classical velocity addition diagram for elastic scattering in inverse kinematics, showing that the light (target) particles emerge at angles just forward of $90^\circ$ for small centre of mass scattering angles. }
\label{fig:9}       % Give a unique label
\end{figure}

\subsection{Characteristic kinematics for stripping, pickup and elastic scattering}
\label{subsec:inversekinematics}

The vector diagram describing elastic scattering in inverse kinematics is shown in Figure \ref{fig:9}. The velocity of the centre of mass in the laboratory frame is given by a large fraction of the beam velocity, since the target is light. Measured in the centre of mass frame, taking into account conservation of momentum, we can also note that the velocities of the two particles after the collision must be in inverse proportion to their masses. Thus, the target-like particle has a velocity $\upsilon _{{\rm target}}^{{\rm c.m.}}$ that is much greater than that of the beam-like particle in this frame. This is shown in the Figure by the red dashed vectors. Furthermore, the target particle is initially at rest and hence the length of the target-like vector $\upsilon _{{\rm target}}^{{\rm c.m.}}$ is equal to the length of the centre of mass velocity as measured in the laboratory frame, $\upsilon _{{\rm c.m.}}^{{\rm lab}}$. The scattering angle as measured in the centre of mass frame is given by the angle enclosed between $\upsilon _{{\rm c.m.}}^{{\rm lab}}$ and $\upsilon _{{\rm target}}^{{\rm c.m.}}$, indicated by $\theta$ in Figure \ref{fig:9}. For a scattering angle of zero in the centre of mass frame, the light particle in the final state is stationary. For small scattering angles (where the cross section is highest, for elastic scattering) the light particles emerge just forward of $90^\circ$ and with a velocity (energy) that increases approximately linearly (quadratically) with centre of mass angle. Also, the centre of mass angle is simply twice the difference between the laboratory angle and $90^\circ$ in this classical approximation, since the velocity addition triangle is isosceles. The beam particle continues in the forward direction with little change in either energy or direction. In the case of backscattering in the centre of mass frame, the light particles travel rapidly in the direction of the incoming beam, and the beam particle also continues in that direction, being just slightly slowed down. An important result here, of experimental significance, is that elastically scattered target particles will be detected just forward of $90^\circ$ and their energies will increase rapidly with angle. In general, they will require a thick detector for them to be stopped and their energy measured precisely.

\begin{figure}[h]
\sidecaption
\includegraphics[width=1.0\textwidth]{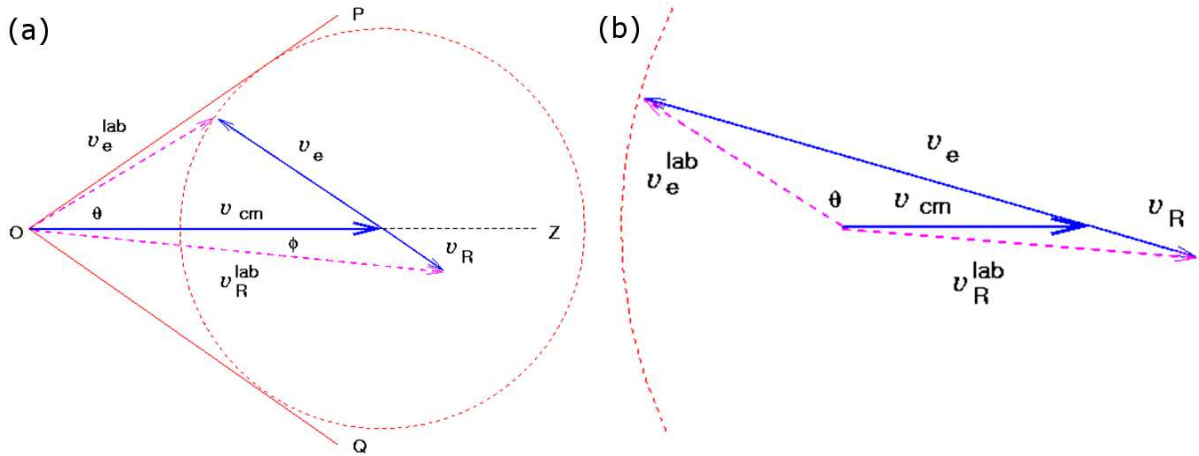}
%If the width of the Figure is less than 7.8 cm use the \texttt{sidecapion} command to flush the caption on the left side of the page. If the Figure is positioned at the top of the page, align the sidecaption with the top of the Figure -- to achieve this you simply need to use the optional argument \texttt{[t]} with the \texttt{sidecaption} command}
\caption{Velocity addition diagrams (a) for a typical {\em pickup} reaction such as (p,d) or (d,t), and (b) for a typical {\em stripping} reaction such as (d,p). Certain assumptions about the beam energy and the reaction Q-value are described in the text. }
\label{fig:10}       % Give a unique label
\end{figure}

The vector diagrams describing reactions in which there is {\em pickup} of a nucleon by the light particle, or {\em stripping} of a nucleon from the light particle are shown in Figure \ref{fig:10} (adapted from ref. \cite{Divonne}). The lengths of the vectors in these diagrams are given in terms of the masses involved, and the reaction Q-value, by formulae included in refs. \cite{Divonne, Kinematic}. As shown by those formulae, the diagrams shown here implicitly assume a small reaction Q-value, or at least that the Q-value in units of MeV is small compared to the energy of the beam as expressed in MeV per nucleon. Especially for reactions involving exotic neutron rich projectiles, the Q-values for neutron addition or removal will typically be small, and similarly for a reaction such as (d,$^3$He) on the proton-rich side of the nuclear chart.

In the case of a reaction such as (p,d), corresponding to Figure \ref{fig:10}(a), it is easy to obtain a rough estimate of the length of the light particle vector in the centre of mass, labelled $\upsilon _e$ in the Figure. Firstly, the heavy particle is going to continue with little change in velocity or direction, much as in the case of elastic scattering. Now, the centre of mass vector in elastic scattering was required to be the same length as the centre of mass velocity vector in the laboratory frame, denoted by $\upsilon _{{\rm cm}}$ in Figure \ref{fig:10}. In the case of (p,d), the mass of the light particle is doubled relative to the elastic scattering situation, but the momentum that this particle must carry in the centre of mass frame is about the same as in the elastic case, which follows from the remark about the velocity of the beam particle. Thus, this vector $\upsilon _e$ is about half the length of $\upsilon _{{\rm cm}}$. The precise value depends upon the reaction Q-value of course, but the basic form of the vector diagram is always the same, subject to the assumptions mentioned above. The result is that the light reaction products are forward focussed into a cone of angles or around $40^\circ$ relative to the beam direction. There will be two energy solutions for each angle, within this cone, wherein the lower energy corresponds to the smaller centre of mass reaction angle and hence (typically) the higher cross section. The low energy solution may be very low indeed, in energy.

In the case of a reaction such as (d,p), corresponding to Figure \ref{fig:10}(b), the mass of the light particle is halved in the reaction and hence its velocity vector in the centre of mass frame is approximately doubled in length, in the approximate picture. The small centre of mass angles (and typically the higher cross section) will correspond to light particles that emerge travelling opposite to the beam direction. They will have energies that may be quite low, and will increase in energy all the way to zero degrees in the laboratory frame, which corresponds to a centre of mass angle of $180^\circ$. For reactions that populate an excited state in the final nucleus, there will be less energy available in the final state than for the ground state, and hence the vectors in the centre of mass are shorter and the laboratory energies of the light particle will be lower than for the ground state, at all laboratory angles.

When planning an experiment in inverse kinematics, it can be useful to construct a velocity addition diagram such as those in Figure \ref{fig:10}. It allows an intuition about the reaction kinematics to be gained, easily. The form of the diagram depends only on the ratio of the length of $\upsilon _e$ to that of $\upsilon _{{\rm cm}}$. This ratio is given \cite{Divonne} by
$$ \frac{\upsilon _e}{\upsilon _{{\rm cm}}} = \left( q f ~\frac{M_R}{M_P} \right)^{1/2} \approx \sqrt{qf} {\rm ~~if~} M_R \approx M_P  $$
where the masses of the projectile and recoil are denoted by $M_P$ and $M_R$. The quantity $f$ is related to the change in mass of the target particle, $f=M_T / M_e$ where $M_T$ and $M_e$ are the masses of the target and light ejectile respectively. The quantity $q$ is of order unity but has a Q-value dependence and typically varies between 1 and 1.5. Specifically, $q=1+Q/E_{{\rm cm}}$ where $Q = Q_{{\rm g.s.}} - E_{{\rm x}}$ for an excited state and $E_{{\rm cm}}$ is the kinetic energy in the centre of mass frame. Given that the target is much lighter than the projectile, most of the kinetic energy in the centre of mass frame is carried by the target particle, so $E_{{\rm cm}} \approx M_T (E/A)_{{\rm beam}}$. Then $q \approx 1+ Q/2 (E/A)_{{\rm beam}}$ and $q$ is closer to unity for small Q-values or as the $E/A$ for the beam is increased. In the limit that $q=1$ then for a pickup reaction such as (p,d) or (d,t) the size of the cone around the beam direction that contains all of the events is given by $\theta _{{\rm max}} =  \sin ^{-1} \sqrt{f}$ where $f=1/2$ for (p,d)  and $f=2/3$ for (d,t). This gives, as a first approximation, a cone of about $50^\circ$ degrees half-angle in each case. Similarly, it is possible to estimate that in (d,p) the laboratory angle corresponding to $30^\circ$ in the centre of mass frame is about $110^\circ$, so a (d,p) experiment will typically need to measure at least the angular range from $110^\circ$ to near $180^\circ$ in the laboratory.

Another interesting feature of the velocity addition diagram is how it scales with the $E/A$ of the beam \cite{enam}. Whilst the relative lengths of the vectors are determined largely by the masses of the various particles, with some residual dependence on the Q-value and the beam energy, the length scale (as given in ref. \cite{Kinematic}) is $\sqrt{2q(M_R + M_e )}$ which, with the assumption again that $M_P \gg M_T$ is approximately proportional to $\sqrt{(E/A)_{{\rm beam}} / M_P}$. Now, with the assumption that $M_P \approx M_R$ (because the transfer hardly changes the mass), the lengths of the vectors such as $\upsilon _e$ and $\upsilon _{{\rm cm}}$ scales as $\sqrt{M_P}$. Thus, the whole diagram scales as the product of these lengths and the length scale itself, and the $\sqrt{M_P}$ factor cancels. The diagram therefore scales roughly as $\sqrt{(E/A)_{{\rm beam}}}$ and the energies will scale roughly as $(E/A)_{{\rm beam}}$. The approximation is better, the closer the Q-value is to zero, but the expression at least gives a guide to the behaviour that can be expected: the detected energy scales with the beam energy. For elastic scattering, the Q-value is zero, so the result is accurate: the rate of increase of the energy of the scattered particle with angle, for angles moving forward of $90^\circ$, scales with the beam energy.

\begin{figure}[h]
\sidecaption
\includegraphics[width=1.0\textwidth]{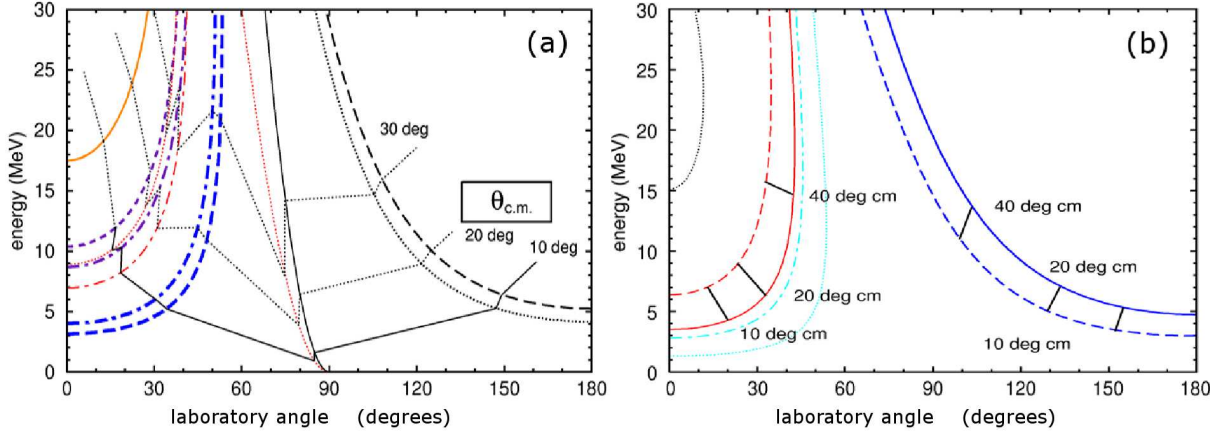}
%If the width of the Figure is less than 7.8 cm use the \texttt{sidecapion} command to flush the caption on the left side of the page. If the Figure is positioned at the top of the page, align the sidecaption with the top of the Figure -- to achieve this you simply need to use the optional argument \texttt{[t]} with the \texttt{sidecaption} command}
\caption{Two-body relativistic kinematics calculations for two very different beams in terms of mass and energy, including results for elastic scattering and several different single-nucleon transfer reactions: (a) $^{16}$C at 35 A.MeV, (b) $^{74}$Kr at 8.16 A.MeV. }
\label{fig:11}       % Give a unique label
\end{figure}

The results of proper (relativistically correct) kinematics calculations for two very different beams and energies are shown in Figure \ref{fig:11}. In Figure \ref{fig:11}(a), the results are for a beam of $^{16}$C at 35 A.MeV such as might be produced by a fragmentation beam facility. The central solid line near $90^\circ$ shows the energy of elastically scattered deuterons, rising steeply as the centre of mass angle increases and the laboratory angle slightly decreases. On the right hand side of Figure \ref{fig:11}(a) are the results for the protons from the (d,p) reaction populating the ground state in $^{17}$C (upper curve) and a hypothetical excited state at 4 MeV excitation energy. The pair of curves with the lowest energies at zero degrees are for the (d,t) reaction. The faint dotted line near $90^\circ$ shows the energies of elastically scattered protons, if there were to be any $^1$H in the target along with the $^2$H (a situation commonly encountered experimentally). The curve with the highest energy at zero degrees is for tritons from the (p,t) reaction populating the ground state of $^{14}$C. The remaining curves at the intermediate energies at zero degrees are for the reactions (d,$^3$He) and (p,d) initiated by the different isotopes in the target. Lines connecting all of the curves show the points corresponding to the indicated centre of mass angles. Note that the energies of the particles from (d,p) and (d,t) are less than or equal to 5 MeV over the most interesting range of relatively small centre of mass angles, where the differential cross section will be largest and most structured. Also, the maximum energies reached over the interesting range are all less than about 30 MeV. Figure \ref{fig:11}(b) is for $^{74}$Kr at 8.16 A.MeV. The curves on the right are for (d,p) to the ground state of $^{75}$Kr and a hypothetical state at 5 MeV excitation. At forward angles, the two lower curves are for (d,$^3$He) from this neutron-deficient nucleus. The next two curves are for (d,t). In each of these cases, the calculations are for the ground state and a 5 MeV state. The final kinematic curve in Figure \ref{fig:11}(b), intersecting at 15 MeV at $0^\circ$,  is for the (p,d) reaction to the ground state of $^{73}$Kr. Once again, the particles of principal interest are generally of about 5 MeV or less, and the energies of interest range up to about 30 MeV. This consistency of the relevant kinematic energy-angle domains has important implications for the design of particle detection systems aimed at studying transfer in inverse kinematics. It indicates that a static array could be optimised to such measurements and would be applicable to a wide range of reaction studies.

\subsection{Laboratory to centre of mass transformation}
\label{subsec:com}

It is common to transform results for the measurement of differential cross sections from the laboratory frame into the centre of mass frame, for comparison with the results of reaction theory calculations. The theory is of course naturally calculated in the centre of mass frame. In the days when the experiments were performed in normal kinematics, the shape of the cross section plot would be similar in both the laboratory and the centre of mass reference frame, because the target was typically much heavier than the incident deuteron. In the case of inverse kinematics, this is no longer the case, as shown by comparing Figures \ref{fig:7} and \ref{fig:8}. It is important to note that it is not simply a transformation from one angle to another that changes the differential cross sections between the two reference frames, but the solid angle is also transformed. The ratio of differential factors that describes this transformation is known as the Jacobian. Inspection of Figure \ref{fig:10}(b), which describes the (d,p) reaction, shows that for backward laboratory angles (as illustrated) the laboratory angle (for $\upsilon_e^{lab}$ measured relative to $\upsilon_{cm}$) varies much more rapidly than the centre of mass angle (enclosed between $\upsilon _{cm}$ and $\upsilon _e$). In the diagram there is a factor of about two, between the rates of change. This means that a small solid angle in the centre of mass frame is spread out over a rather large solid angle in the laboratory frame. Thus, while $d\sigma /d\Omega _{{\rm c.m.}}$ is largest at small $\theta _{{\rm c.m.}}$ or near $180^\circ$ in the laboratory frame, the effect of the Jacobian is that $d\sigma /d\Omega _{{\rm lab}}$ is reduced relative to less backward angles. That is, while the very backward laboratory angles are still important in (d,p) measurements, for determining the shape of the differential cross section, there are very few counts there.

The transformation from centre of mass to laboratory angles, as just mentioned, has the effect of spreading out the counts from (d,p) at small centre of mass angles, so that they are spread over a wider solid angle. This reduces the yield of counts observed near $180^\circ$ in the laboratory frame. A completely separate effect to also remember is the ``$\sin \theta$'' effect. This will further emphasise the importance of detectors close to $90^\circ$ compared to those near $180^\circ$, assuming that the charged particle detection is cylindrically complete, or approaching this. Then, since the solid angle in a range $d\theta $ at angle $\theta$ is given by $2\pi \sin \theta d\theta$, the cross section that measures the number of counts in an experiment is not $d\sigma / d\Omega$ but
$$\frac{d\sigma }{d\theta} = 2\pi \sin \theta \frac{d\sigma }{d\Omega }~.$$
Thus, if a coincidence experiment is considered, for example measuring gamma-rays in a (d,p$\gamma$) experiment, many of the gamma-rays will come in association with particles detected towards $90^\circ$.

The transformation between the centre of mass and laboratory reference frames, for the differential cross section, is given for example by Schiff in his classic {\em Quantum Mechanics} text \cite{Schiff}:
$$ \frac{d\sigma}{d\Omega}_{{\rm lab}} = \frac{(1+\gamma^2 +2\gamma \cos \theta _{{\rm c.m.}})^{3/2}}{\mid 1+\gamma \cos \theta _{{\rm c.m.}}\mid} \times
\frac{d\sigma}{d\Omega}_{{\rm c.m.}} $$
where $\gamma = \upsilon_{{\rm c.m.}}/\upsilon_e$. This complicated transformation, as noted above, changes the shape of the curves significantly. Therefore, a plot in the centre of mass frame of experimental data for the differential cross section will retain very little information about any experimental constraints or impacts of the laboratory angles. For example, the physical gaps between detectors, or the differing thicknesses of target through which the particles must exit: these often have important implications for the data but the information is lost in the transformation to the centre of mass frame. For this reason, some workers choose to plot experimentally measured cross sections in the laboratory frame, for inverse kinematics experiments, following the ethos of presenting the data in a form as close as possible to what is actually measured experimentally.
	
\subsection{Strategies to combat limitations in excitation energy resolution}
\label{subsec:resolution}
	
In trying to do experiments using radioactive beams, there are two properties of the beams that tend to influence the experimental design more than any others. Firstly, the beams are radioactive. That means that care must be taken regarding the eventual dumping of the beam and also, quite often, to deal with the angular scattering of the beam in the target \cite{TaLL3}. Secondly, the beams are generally weak, maybe up to a million times weaker than a typical stable beam that one might have used for an equivalent normal kinematics experiment with a stable target. This means that, in practice, there will be a minimum sensible value for the target thickness in order to perform the experiment in a reasonable time. In turn, this will affect the energies and angles measured for the particles produced in the reaction. As discussed above, the particles of interest are often of rather low energy. The energy that is measured may depend quite significantly on where the reaction takes place - at the front or at the back surface of the target, or somewhere in between. Also, for the lowest energy particles, the direction may be affected by multiple low-angle scattering of the charged particle as it leaves the target material.

In an experiment to identify and study the unknown excited states of an exotic nucleus, the kinematical formulae used to produce a plot such as Figure \ref{fig:11} will be inverted so that the measured energy and angle of a particle are used to calculate the excitation energy of the final state. Any process that modifies the measured energy and angle from the actual reaction values will lead to a limitation on the achieved resolution for excitation energy, even if the best possible computed corrections are applied. All of these factors were included in a detailed analysis of the resolution that could be expected from transfer reactions, under different experimental conditions \cite{WinfieldNIM}. The two basic categories of experiment were as follows:
\begin{enumerate}[I]
\item {\it rely on detecting the beam-like ejectile in a magnetic spectrometer}
\item {\it rely on detecting the light, target-like ejectile in a silicon detector}\\ ~\\with a third option arising which is\\
\item {\it detect decay gamma-rays in addition to the charged particles.}
\end{enumerate}

A magnetic spectrometer or a recoil separator is essential in the first case, in order to separate the reaction products from the direct beam and to measure the ejectile properties with sufficient accuracy. Operated at zero degrees, it will need to be instrumented to allow the accurate measurement of the scattering angles for the very forward-focussed beam particles. The degree of forward focussing, and hence the requirements placed upon the resolution of the angular measurements, become more and more demanding as the mass of the projectile increases. For heavier beams, it becomes impractical for existing detectors. Furthermore, any spread in the beam energy translates directly to a spread in the measured excitation energies, and any nucleon transfer reactions on contaminant material in the hydrogen targets (usually plastic) will contribute to the observed yield.

If the second method is employed, then we know from the discussion of the kinematics that the particles of interest are spread over a significant range of angles. In order to detect particles over this range, and with good resolution in both energy and angle, the most obvious choice is an array of semiconductor detectors, and silicon is by far the most versatile material at present. This method is less sensitive than the first, to a spread in the beam energy, but is limited as discussed above by the effects of the target thickness on the measured energies and angles. In practice, it is hard to imagine resolutions of better than 100-200 keV or so, for excitation energy, if the experiment demands targets of 0.5 mg/cm$^2$ or more \cite{WinfieldNIM}. (This assumes $(CD_2)_n$ deuterated polythene targets,
\footnote{For convenience, the $( \ldots )_n$ part of $(CD_2)_n$ will subsequently be omitted, and similarly for the non-deuterated $(CH_2)_n$ targets.}
and with a thickness determined by beam intensities that may be as low as $10^4$ pps). In some experiments, thinner targets could be used and hence better resolution achieved, if the beam intensity were to allow it. In any case, to achieve the best resolution, the detector array for the light particles should achieve good measurement of the particle angles. In the case of a silicon strip detector array, this requires a high degree of segmentation, or in some cases a resistive strip readout is possible.

A variant of the second method, which avoids the need for an extensive and highly segmented silicon array, is to use a magnetic solenoid aligned with the beam axis to collect and focus the charged particles onto a more modest array of silicon detectors located around the axis of the solenoid. This is the {\em HELIOS} concept, named after the first device of this type to be implemented \cite{HELIOSone}. The elegant feature of HELIOS is that it removes the kinematic compression of energies. Considering the kinematics of a typical (d,p) reaction as shown for example in Figure \ref{fig:11}, the lines for a difference of 1 MeV in excitation energy are separated in terms of proton energy by less than 1 MeV at a particular laboratory angle. In HELIOS, when the protons are focussed back to the axis of the solenoid, they are dispersed in distance according to a linear dependence on excitation energy. For a detector located at a a particular distance along the axis, it measures particles emitted at different angles for different excitation energies. The net result is to disperse excited states in energy in such a way that any limitation due to the intrinsic resolution of the silicon detector becomes significantly less important. However, if the limitation lies in the target thickness and the ensuing deleterious effects on the energy and angle of the particles leaving the target, then there is little benefit to be obtained from simply using a different method of measurement. Ultimately, if the experiment demands a relatively thick target, the resolution will be as estimated in ref. \cite{WinfieldNIM}. The helical detector concept is discussed again in section \ref{subsec:choosing}.

It may be that the limits to resolution imposed by a reasonable target thickness are not acceptable for a good measurement to be performed. This is likely to happen in the case of heavier nuclei where levels are more closely spaced than the light nuclei, or it can occur in any odd-odd nucleus for any mass. In this situation, which can be expected to be common, the third solution - measuring decay gamma-rays - becomes attractive.

The higher energy resolution that can be achieved with gamma-ray detection then gives a much better energy resolution for excited states. This of course works only for bound states in the final nucleus. In addition to the improved energy resolution, another feature of possibly comparable importance is that the gamma-ray decay pathway for a particular final state may help to identify the state more precisely. From the particle transfer measurement, it is only possible to infer the $\ell$-transfer, which leaves an uncertainty according to whether the spin is $(\ell + 1/2)$ or $(\ell - 1/2)$ since the transferred nucleon has spin 1/2. The gamma-ray decay branches may resolve this ambiguity. It should be noted that there is an experimental challenge in detecting the gamma-rays with a high enough efficiency and with the ability to apply a sufficiently good Doppler correction. For the typical beam energies discussed above, the final gamma-ray emitting nuclei will have velocities of the order of $0.10c$ in the laboratory reference frame, always aligned almost exactly along the beam direction ($c$ is the speed of light). In order to correct for the substantial Doppler shift that this implies, the gamma-ray detectors will also need to measure the angle of emission for the gamma-ray, relative to the incident beam. Doppler shift corrections are discussed further in section \ref{subsec:tiara}.

%-------------------------------------------------------------------------------------------------	
\section{Examples of light ion transfer experiments with radioactive beams}
\label{sec:4:lightion}

Having described the various approaches to designing an experiment in the previous section, these possibilities are now illustrated by means of specific examples. Mostly, the examples are early experiments which helped in developing the techniques and/or (for convenience) experiments by the author with collaborators.

\subsection{Using a spectrometer to detect the beam-like fragment}
\label{subsec:spectrometer}

\begin{figure}[h]
\sidecaption
\includegraphics[width=.5\textwidth]{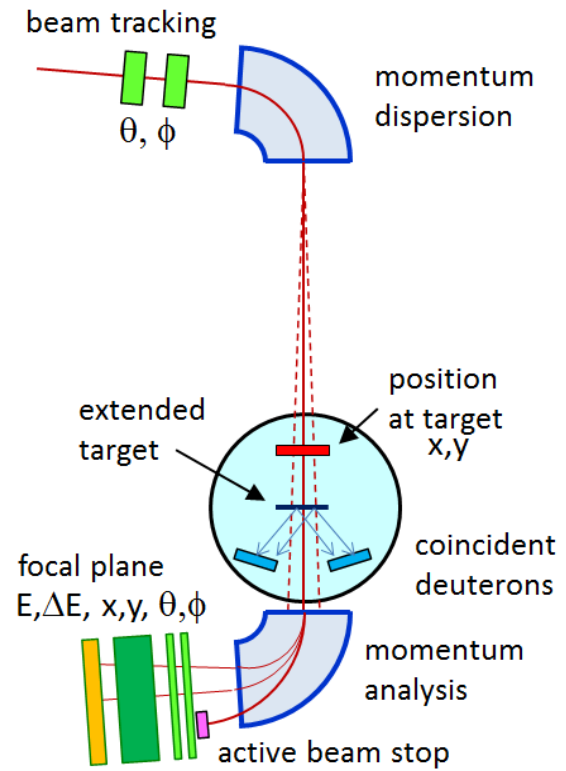}
%If the width of the Figure is less than 7.8 cm use the \texttt{sidecapion} command to flush the caption on the left side of the page. If the Figure is positioned at the top of the page, align the sidecaption with the top of the Figure -- to achieve this you simply need to use the optional argument \texttt{[t]} with the \texttt{sidecaption} command}
\caption{In this (p,d) study using a secondary $^{11}$Be beam \protect\cite{Fortier,WinfieldBe11}, the beam had a large energy spread, so a dispersion matched spectrometer was used. This, together with the limited spatial focussing of the beam required beam tracking detectors at the target and in the beam line. Coincident deuteron detection allowed background from the carbon in the $CH_2$ target to be removed in the analysis, but the $^{10}$Be measurement in the spectrometer gave all of the critical energy and angle information. The active beam stop comprised a plastic scintillator that allowed the intensity of the beam to be monitored.}
\label{fig:12}       % Give a unique label
\end{figure}

An example of an experiment in which the beam-like particle is measured, and used to extract all of the experimental information, is provided by the early experiment performed by the Orsay and Surrey groups at GANIL \cite{Fortier,WinfieldBe11} and illustrated in Figure \ref{fig:12}. The aim was to study the (p,d) reaction with $^{11}$Be in order to study the parentage of the $^{11}$Be halo ground state. Because the projectile is relatively light, then it is a reasonable approach to measure the beam-like particle (method I of section \ref{subsec:resolution}). A magnetic spectrometer was used, for two reasons. Firstly, the beam was produced by secondary fragmentation and therefore has significant spreads in both energy and angle. In order to resolve final states in $^{10}$Be that were separated by less energy than the spread in the beam, a dispersion matched spectrometer was required. This experiment used the {\em spectrom\`etre \`a perte d'energie du GANIL}, SPEG \cite{SPEG}. Secondly, in order to measure the scattering angle of the $^{10}$Be it was necessary to separate the $^{10}$Be from the beam and track its trajectory to a precision that required a spectrometer. In order to recover the scattering angle, it was also necessary to track the incident beam particles, which required detectors placed before the first (dispersive) dipole element of SPEG. The beam intensity was $3 \times 10^4$ particles per second (pps) and the mean beam energy was 35.3 A.MeV. The measured $^{10}$Be particles, at the focal plane, were dominated by the yield from carbon in the polythene $CH_2$ target. Reactions on just the protons in the target were isolated by recording the deuterons from the reaction in an array of ten large area silicon detectors. This experiment was successful in measuring the parentage of the $^{11}$Be ground state, which has a neutron halo, in terms of the $s$-wave and $d$-wave components (the latter with an excited $^{10}$Be core). In addition to the innovative experimental techniques, the experiment also highlighted some important complexities in the theory and made innovations in the theoretical interpretation. Specifically, it was necessary to go beyond the normal simplification of modelling the transferred nucleon in a potential well that is due to the core. It was necessary to use a dynamic picture of $^{11}$Be in terms of a particle-vibration coupling model, in order to calculate the overlap functions in the transfer amplitude directly from the nuclear structure model.

\subsection{Using a silicon array to detect the light (target-like) ejectile}
\label{subsec:silicon}

The first high resolution example of this kind of experiment, aimed at measuring spectroscopic quantities using a radioactive beam, was an experiment employing a previously-prepared source of radioactive $^{56}$Ni in order to measure the reaction (d,p) in inverse kinematics \cite{Rehm}. Useful and astrophysically relevant results were obtained. The experiment used silicon strip detectors arranged in the backward hemisphere with a solid target of $CD_2$ deuterated polythene
and a recoil separator device - in this case, the fragment mass analyser (FMA) at Argonne \cite{FMA}. The beam was produced in the normal way for a tandem accelerator using a source of radioactive nickel material, and had a typical intensity on target of $2.5 \times 10^4$ pps at an energy of 4.46 A.MeV. An additional challenge was the isobaric impurity of $^{56}$Co which was a factor of seven more intense than the $^{56}$Ni and was separated using differential stopping foils within the FMA.

A particular silicon array that was developed specifically for experiments with radioactive beams is MUST \cite{MUST}, which uses large area highly segmented silicon strip detectors with CsI detectors in a telescope configuration. MUST led the way in developing electronics that could cope with the many channels required for highly segmented detectors. Excellent particle identification is achieved. MUST has been used to study a range of reactions including inelastic scattering of a range of nuclei, and with regard to transfer it was very often targeted at experiments to study the structure of very light and even unbound exotic nuclei, for example $^{7,8}$He \cite{Valerie2,Valerie}. Another major silicon array is HiRA which was developed initially for experiments using radioactive beams produced by fragmentation at MSU \cite{HiRA}. The MUST array was combined with SPEG spectrometer in a study of neutron-rich argon isotopes with a pure reaccelerated beam of $2 \times 10^4$ pps of $^{46}$Ar at 10.7 A.MeV from SPIRAL at GANIL, incident on a $CD_2$ target of 0.4 mg/cm$^2$\cite{olivier-argon}. Good resolution in excitation energy was achieved, in part by exploiting the special optics of the SPEG beamline. The detection of argon ions in SPEG was useful in helping to identify and eliminate background from carbon in the target, and also allowed the identification of bound and unbound states in $^{47}$Ar according to whether $^{47}$Ar or $^{46}$Ar was detected in SPEG, although the spectrometer acceptance was limited and prevented a full coincidence experiment. Another interesting experiment that used a silicon array by itself was the study of (d,p) using a beam of $^{132}$Sn at 4.8 A.MeV from the Oak Ridge radioactive beam facility \cite{Kate}. As seen from the calculated cross sections in Figures \ref{fig:7} and \ref{fig:8}, this was not really the ideal energy for such a heavy beam, but it was the maximum possible. The resolution achieved for excitation energy was limited, for this heavy beam, not by the silicon array but by the target thickness of 0.16 mg/cm$^2$. As also suggested by the cross section plots in Figure \ref{fig:8}, the silicon detectors were optimised by mounting them in a range of angles around $90^\circ$ in the laboratory.

\begin{figure}[h]
\sidecaption
\includegraphics[width=.7\textwidth]{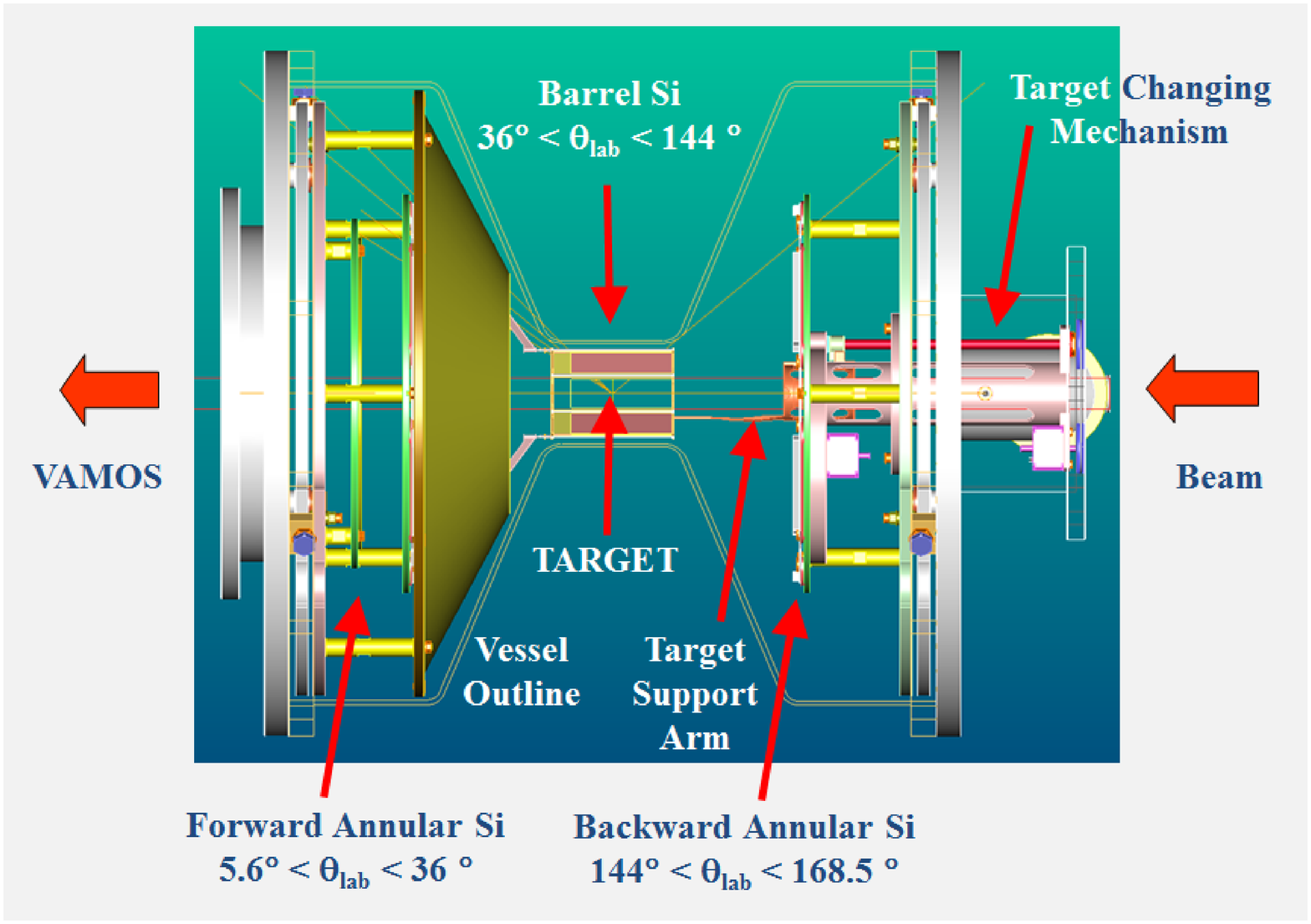}
%If the width of the Figure is less than 7.8 cm use the \texttt{sidecapion} command to flush the caption on the left side of the page. If the Figure is positioned at the top of the page, align the sidecaption with the top of the Figure -- to achieve this you simply need to use the optional argument \texttt{[t]} with the \texttt{sidecaption} command}
\caption{The TIARA array was designed specifically to measure nucleon transfer reactions in inverse kinematics with radioactive beams. It has an octagonal barrel of position-sensitive silicon detectors, with annular silicon arrays at forward and backward angles. In total, approximately 90\% of $4\pi$ is exposed to active silicon. The vacuum vessel is designed so that EXOGAM gamma-ray detectors can be placed very close to the target, achieving a gamma-ray peak efficiency of order 15\% at 1 MeV. A robotic target changing mechanism allows different targets to be placed at the centre of the barrel.}
\label{fig:13}       % Give a unique label
\end{figure}

The TIARA array \cite{Labiche} was the first purpose-built array to combine silicon charged particle detectors with gamma-ray detectors for transfer work and was first employed with a radioactive $^{24}$Ne beam at the SPIRAL facility at the GANIL laboratory \cite{Ne24}. Initial tests and benchmarking were performed with a stable beam and a reaction that was previously studied in normal kinematics \cite{Labiche}. TIARA was designed, taking into account the experience gained from using a high intensity radioactive beam of nearly $10^9$ pps of $^{19}$Ne in the TaLL experiment at Louvain-la-Neuve \cite{TaLL1,TaLL2,TaLL3}. This led to a design in which radioactive beam particles that are scattered at significant angles by the reaction target will be carried away from the immediate vicinity of the target, and hence away from the field of view of the gamma-ray array \cite{TaLL3}.

TIARA is shown schematically in Figure \ref{fig:13}. It is designed to be operated with four HPGe clover gamma-ray detectors from the EXOGAM array \cite{EXOGAM,TaLL3} mounted at $90^\circ$ and at a distance of only 50-55~mm from the centre of the target. The space available in the forward hemisphere was also severely restricted due to the design requirement of coupling to the VAMOS spectrometer \cite{VAMOS}. The spectrometer allows reaction products to be measured with high precision and to be identified according to $Z$ and $A$. The exceptionally large angular acceptance of VAMOS (up to $10^\circ$) also allows the efficient detection of recoils from the decay of unbound states via neutron emission. Examples of the gamma-ray and spectrometer performance are given in sections \ref{subsec:tiara}, \ref{subsec:zero} and \ref{subsec:unbound}.

\begin{figure}[h]
%\sidecaption
\centering
\includegraphics[width=.85\textwidth]{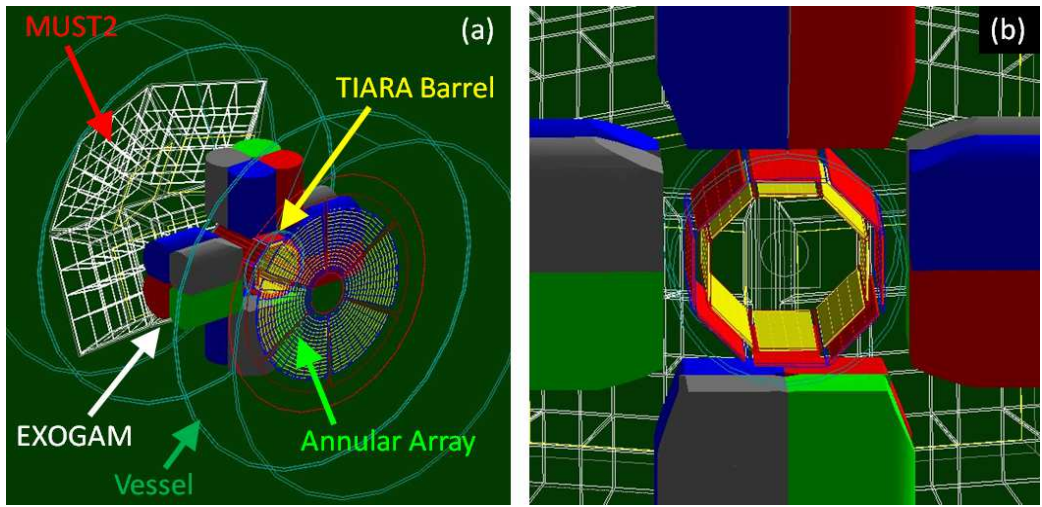}
%If the width of the Figure is less than 7.8 cm use the \texttt{sidecapion} command to flush the caption on the left side of the page. If the Figure is positioned at the top of the page, align the sidecaption with the top of the Figure -- to achieve this you simply need to use the optional argument \texttt{[t]} with the \texttt{sidecaption} command}
\caption{The TIARA setup as modelled in {\em g\'eant4} \protect\cite{geant}:~ (a) overview, including MUST2 \protect\cite{MUST2} and the EXOGAM clover HPGe gamma-ray detectors \protect\cite{EXOGAM}. The four leaves of each of the 4 are shown; (b) the central silicon array comprises two concentric octagonal barrels and the clover front faces are 54 mm from the beam axis. The view is looking with the beam from just in front of the annular array. Beyond the barrel, the detectors of MUST2 are glimpsed. The circular target is mounted at the centre of the barrel.}
\label{fig:14}       % Give a unique label
\end{figure}

Figure \ref{fig:14} shows in detail the geometry of the central barrel in TIARA relative to the segmented HPGe clover detectors of EXOGAM. The front faces of the clovers are mounted 54~mm from the centre of the target in this configuration with two layers of silicon in the barrel. The inner layer of silicon is position sensitive parallel to the beam direction, which is the most important direction in defining the scattering angle of any detected particle. Each of the 8 inner detectors has four resistive strips and is $400~\mu$m thick. The second layer of silicon is 1 mm thick but non-resistive. The 4 strips per detector align behind the strips on the inner barrel. The primary purpose of the second layer of the barrel is to indicate when particles punch through the inner layer. The target is placed at the geometric centre of the barrel. The targets are typically 0.5 mg/cm$^2$ self-supporting foils of $CD_2$ mounted on thin holders with holes of diameter 40~mm, where the large hole diameter is chosen so as to minimize the shadowing of the barrel by the target frame.

\begin{figure}[h]
\sidecaption
\includegraphics[width=.7\textwidth]{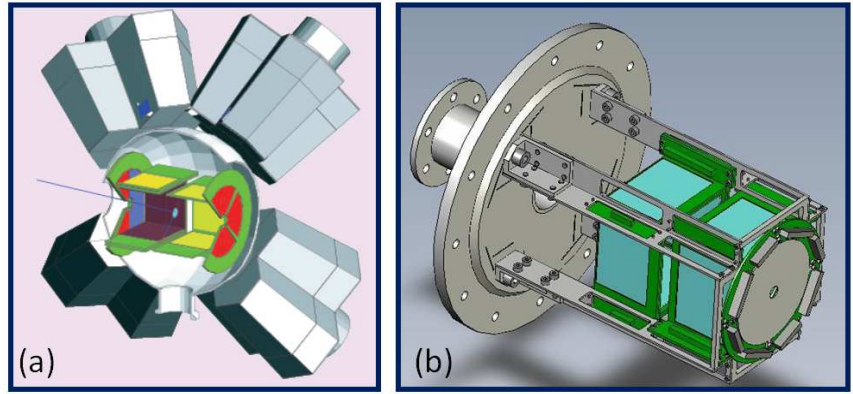}
%If the width of the Figure is less than 7.8 cm use the \texttt{sidecapion} command to flush the caption on the left side of the page. If the Figure is positioned at the top of the page, align the sidecaption with the top of the Figure -- to achieve this you simply need to use the optional argument \texttt{[t]} with the \texttt{sidecaption} command}
\caption{Two post-TIARA silicon arrays developed for use completely inside a large gamma-ray array: (a) T-REX \protect\cite{T-Rex}, which is operated inside the MINIBALL array of HPGe cluster detectors at ISOLDE \protect\cite{MINIBALL}, and (b) SHARC \protect\cite{SHARC} which is operated inside the TIGRESS array of segmented HPGe clover detectors \protect\cite{TIGRESS}. Both include silicon boxes situated forward and backward from the target.}
\label{fig:15}       % Give a unique label
\end{figure}

Subsequent developments of the TIARA approach are represented by T-REX \cite{T-Rex} and SHARC \cite{SHARC}, which are shown in Figure \ref{fig:15}. Another key development, with a barrel design similar to TIARA, is ORRUBA  \cite{Pain} (and its non-resistive strip version super-ORRUBA) which was developed at Oak Ridge. The most obvious feature of these arrays, relative to TIARA, is that they are designed to fit inside a more conventional gamma-ray array. To some extent, this is equivalent to accepting a limitation on the beam intensity that can be used - certainly at an intensity of $10^9$ pps as envisaged in the TIARA design, an enormous amount of radioactivity would be deposited inside the gamma-ray array by the elastic scattering of beam particles from a typical $CD_2$ target. However, at beam intensities of up to a about $10^8$ pps, the radioactivity deposited inside the array will be tolerable and there will be a real benefit in having the silicon array inside a more extensive array of gamma-ray detectors. The advantages lie in the energy resolution achievable with improved Doppler correction, and in simply having a wider range of gamma-ray angles included in the measurements. A wide range of gamma-ray angles may open up additional physics possibilities in the interpretation of the data. The planned deployment (GODDESS) of ORRUBA inside Gammasphere \cite{Gammasphere} with around 100 gamma-ray detectors is perhaps the pinnacle of this approach. The two arrays T-REX and SHARC, coincidentally, have extremely similar geometries. The choice of rectangular boxes allows the silicon detector designs to be relatively simple and hence economical, and the ends of the array are completed with compact annular detectors of a pre-existing design. T-REX (as in the case of ORRUBA, and the original TIARA) includes resistive strips, which helps to keep the number of electronics channels manageable using conventional electronics. However, the price that is paid for using resistive strips is quite high, in terms of performance. Firstly, such detectors typically have higher energy thresholds than non-resistive strips, because they have an electronic noise contribution related to the resistance of the strip \cite{resistive1,resistive2}. Secondly the position resolution that is achieved is dependent on the energy deposited, being proportional to $1/E$ \cite{owen-awcock}. SHARC is the first dedicated compact transfer array to utilise double-sided (non-resistive) silicon strip detectors completely, resulting in superior energy thresholds and a consistency in position resolution. This choice of detector was made possible by the availability of up to 1000 channels of high resolution electronics using the TIGRESS digital data acquisition system \cite{TIGRESS}.

\subsection{Choosing the right experimental approach to match the experimental requirements}
\label{subsec:choosing}
	
As will be apparent from the examples already discussed, a variety of experimental approaches are chosen by different experimenters, for transfer experiments. Largely, these are driven by specific experimental requirements, of which two of the most important are: beam intensity limitations, and the required resolution in excitation energy. One of the most versatile and complete approaches is the combination of a compact, highly segmented silicon array with an efficient gamma-ray detection (as adopted, for example, by TIARA) and hence the results from that approach are presented in some detail, in this document. In this section, we briefly review alternative choices made by experimenters.

\begin{figure}[h]
\sidecaption
\includegraphics[width=.7\textwidth]{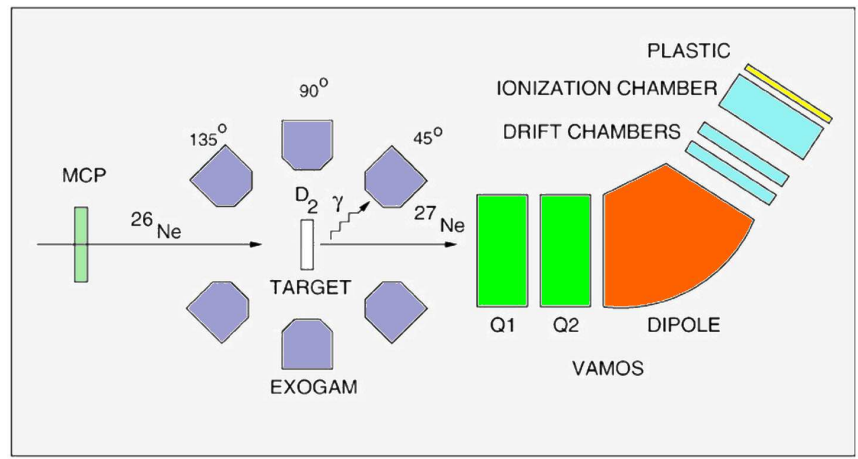}
%If the width of the Figure is less than 7.8 cm use the \texttt{sidecapion} command to flush the caption on the left side of the page. If the Figure is positioned at the top of the page, align the sidecaption with the top of the Figure -- to achieve this you simply need to use the optional argument \texttt{[t]} with the \texttt{sidecaption} command}
\caption{This is a variant on the technique of extracting spectroscopic information from the beam-like particle, rather than the light target-like particle. The aim was to use a thicker target to compensate for a low beam intensity, and the background from target contaminants such as carbon was minimized by using a solid deuterium target. Gamma-ray detection allowed precise excitation energy measurements. See text for definition of other terms. }
\label{fig:16}       % Give a unique label
\end{figure}

In the case of an experiment at SPIRAL at GANIL, aimed at studying $^{27}$Ne via the (d,p) reaction \cite{Ne27-obertelli}, the experimental limitation at the time was the available beam intensity. The solution adopted (see Figure \ref{fig:16}) was to employ a much thicker target, but this implied that the protons would have too low an energy to exit and be detected. Therefore the experiment was focussed on using the heavier beam-like particle, as in the $^{11}$Be experiment discussed in section \ref{subsec:spectrometer}. The final nucleus had a reasonably complex structure, and hence gamma-ray detection was considered vital and would possibly offer additional information on spin, since the proton differential cross sections could not be observed. The EXOGAM array of segmented Ge gamma-ray detectors was employed \cite{EXOGAM}. The required target thickness, in order to achieve sufficient gamma-ray detection, was then achieved by using a solid cryogenic pure $D_2$ target of 17 mg/cm$^2$. In terms of an equivalent $CD_2$ thickness of deuterons, the energy loss in the cryogenic target is reduced by a factor of three, so this is equivalent in energy terms to 6 mg/cm$^2$ of $CD_2$ but has three times the number of target nuclei. In addition, the absence of carbon in the target removes the problem of background reactions that was mentioned in section \ref{subsec:spectrometer}. A microchannel plate detector (MCP) before the target assisted in particle identification using the VAMOS spectrometer \cite{VAMOS}. Inside VAMOS, the particles were focussed by two quadrupole elements ($Q_1, Q_2$) through a dipole magnet and then detectors in the focal plane region recorded the particles' positions, angles and energies. An example of the particle identification that can be achieved in VAMOS is included in section \ref{subsec:zero}.

\begin{figure}[h]
\sidecaption
\includegraphics[width=.6\textwidth]{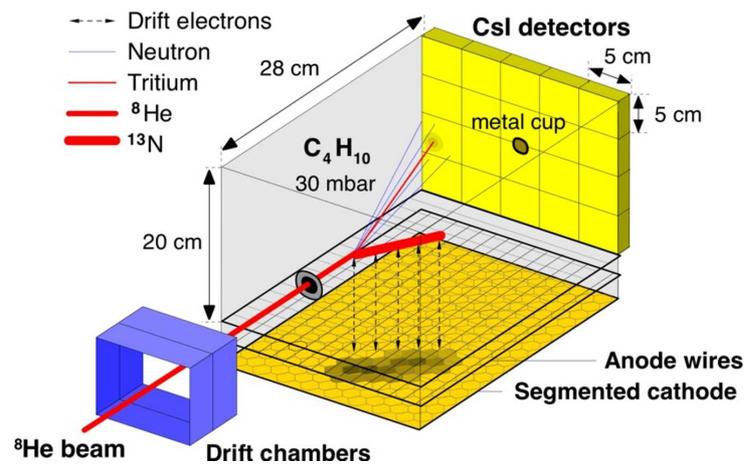}
%If the width of the Figure is less than 7.8 cm use the \texttt{sidecapion} command to flush the caption on the left side of the page. If the Figure is positioned at the top of the page, align the sidecaption with the top of the Figure -- to achieve this you simply need to use the optional argument \texttt{[t]} with the \texttt{sidecaption} command}
\caption{The MAYA detector \protect\cite{MAYA} is an {\it active target} in the sense that the gas that fills MAYA acts both as the target for the nuclear reactions and also as the fill gas of a time projection chamber. Ionisation paths in the gas are drifted to readout planes, and using the drift time it is possible to reconstruct every individual nuclear reaction in three dimensions (and with particle identification). The diagram shows a reaction on the $^{12}$C in the C$_4$H$_{10}$ gas, but reactions on the hydrogen, or other fill gases, can also be studied.}
\label{fig:17}       % Give a unique label
\end{figure}

Most experimental methods discussed here are limited in resolution by the energy loss effects in thick targets. However, this problem is largely removed if it is possible to determine the precise point of interaction within the target. By turning the target into an active detector, designs such as MAYA \cite{MAYA} (shown in Figure \ref{fig:17}) achieve this objective and hence can be used with the lowest beam intensities. In fact, for higher beam intensities it is usually necessary in this type of detector to place an electrostatic screen around the path of the beam itself. The classic model for this type of detector is IKAR, which was produced for high energy beams and operates with multiple atmospheres of H$_2$ gas \cite{IKAR}. In MAYA, the reaction can occur at any point through the gas. The ionisation by all the particles in the gas is drifted in an electric field to a readout plane where the position and amount of ionisation are recorded, along with the time of arrival (i.e. the drift time). This allows a full reconstruction in three dimensions of all charged particle trajectories, subject to various limitations in spatial and energy resolution. The measurement of the ionisation along the whole path of the particles in the gas allows the particle types to be identified. In order for proper drifting of the charge and proper readout, the choice of gas pressure is subject to some restrictions, and hence some particles might easily penetrate beyond the confines of the gas volume. The MAYA detector includes a forward wall of CsI detectors, to deal with these penetrating particles.

\begin{figure}[h]
\sidecaption
\includegraphics[width=.6\textwidth]{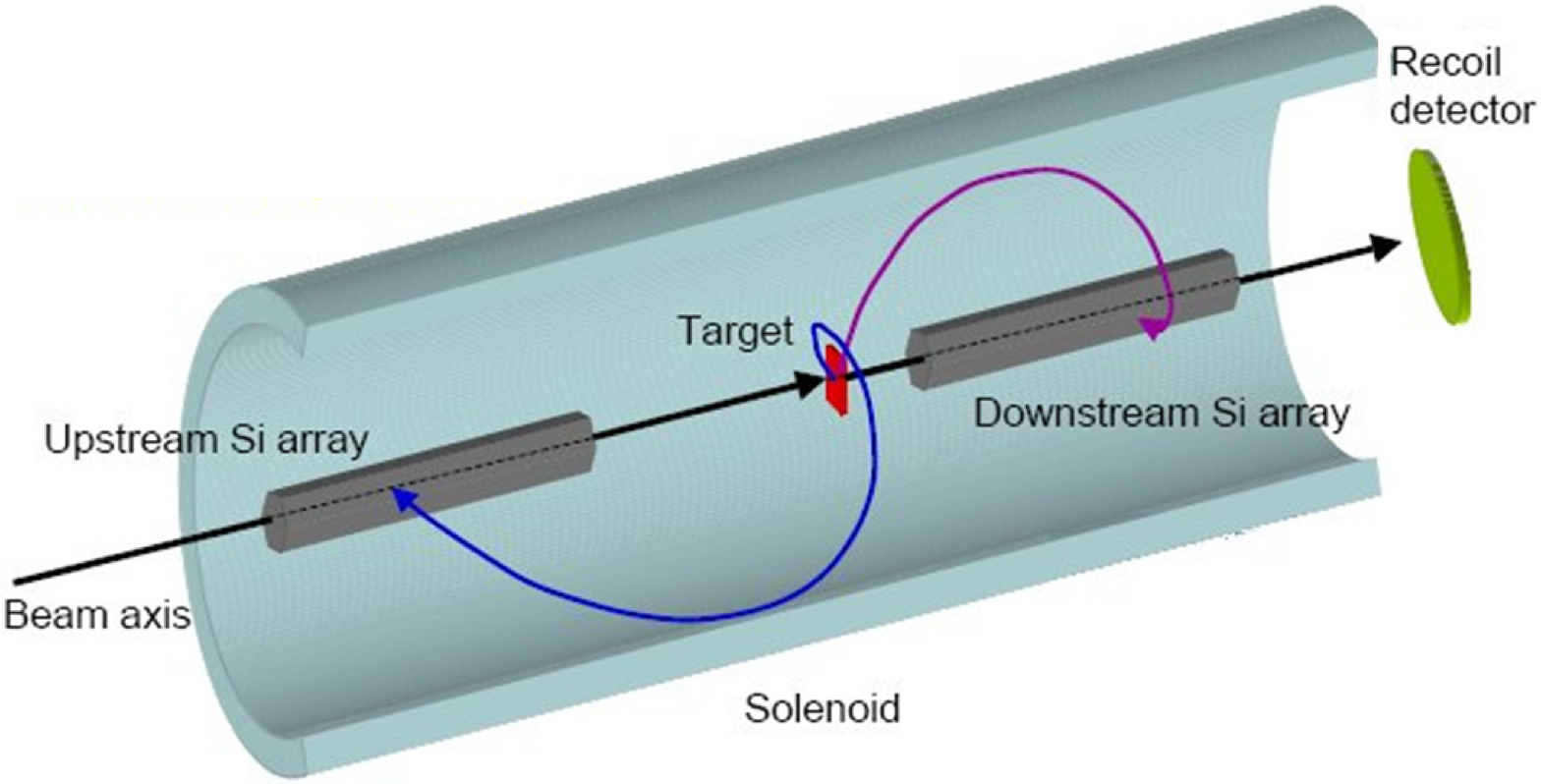}
%If the width of the Figure is less than 7.8 cm use the \texttt{sidecapion} command to flush the caption on the left side of the page. If the Figure is positioned at the top of the page, align the sidecaption with the top of the Figure -- to achieve this you simply need to use the optional argument \texttt{[t]} with the \texttt{sidecaption} command}
\caption{The HELIOS device \protect\cite{HELIOSone,HELIOStwo} collects particles magnetically at all angles and focusses them to compact detectors along the axis. The angular information is reconstructed from the measured energy and the distance from the target to its point of return to the axis, and is generally more accurate than can be obtained by direct angle measurements. The way in which the spectrometer operates has the effect of reducing the limitations arising from the detector energy resolutions themselves.}
\label{fig:18}       % Give a unique label
\end{figure}

A novel approach to achieving $4\pi$ detection efficiency is the HELIOS concept that has been developed by the Argonne group and collaborators \cite{HELIOSone}. Particles emerging at almost all angles from the target are focussed in a large-volume solenoidal field and are brought back to a position-sensitive silicon array aligned along the solenoid axis. This is shown schematically in Figure \ref{fig:18}, which is adapted from ref. \cite{HELIOSone}. The targets are typically $CD_2$ foils, but a gas cell target has also been constructed to allow the study of $^{3,4}$He-induced reactions. The ideal design parameters for the solenoid are remarkably similar to those for medical MRI scanners and indeed the original HELIOS is a decommissioned MRI device \cite{HELIOStwo}. The energy limitations arise not only from the field strength and radius, but also the length along the axis. It is shown in ref. \cite{HELIOSone} that the limitations are much more significant for a typical 0.5 m long device (or a 1 m long device with the target at the centre) than they are for a 1.5 m long device (Figure 8 of ref. \cite{HELIOSone}). The detection limits as a function of angle, for a device between the quoted lengths, are well matched to the kinematics of (d,p) in inverse kinematics. As shown in ref. \cite{HELIOSone}, the Q-value (excitation energy) is calculated directly from the measured anergy and distance along the axis for each detected particle (eq. (5) \cite{HELIOSone}). So also is the centre of mass angle (eq. (7) \cite{HELIOSone}). At an intermediate point in the calculation, the measured time of flight is used to measure the charge to mass ratio $A/q$ for the particle (eq. (2) \cite{HELIOSone}) but once the particle identification is made, the exact value is substituted in further calculations. Thus, apart from the measured energy and position, the calculations rely only upon the precise value/stability and the homogeneity of the magnetic field. The particle identification (apart from one $A/q$ ambiguity between deuterons and $^4$He$^{++}$) is a significant bonus, although it does have some implications for the time structure of pulsed beams. As mentioned in section \ref{subsec:resolution}, any impact on the excitation energy resolution arising from the detector energy resolution is significantly reduced in the HELIOS method, because particles are compared at the same $z$ (distance along the axis) rather than at the same $\theta _{{\rm lab}}$ (angle of emission in the laboratory frame). This turns out to have the effect of removing the {\em kinematic compression} observed in Figure \ref{fig:11}, wherein (particularly at backward angles in the laboratory) the kinematic lines are closer together in proton energy than in excitation energy.

An example of the use of HELIOS with an online produced radioactive beam is the study of $^{16}$C via the (d,p) reaction in inverse kinematics with a thin $CD_2$ target of 0.11 mg/cm$^2$ and a beam of $10^6$ pps of $^{15}$C \cite{c16-helios}. Interestingly, the $^{15}$C secondary beam was itself produced using the (d,p) reaction in inverse kinematics with a $^{14}$C primary beam. Good resolution was achieved, but one key doublet of states at 3.986/4.088 MeV in $^{16}$C could not be resolved. Each of these states gamma-decays to the 1.766 MeV level, and the 100 keV difference in the energies of these 2.2 MeV gamma-rays would be easily resolvable with a modern Ge gamma-ray array. It is a considerable challenge to combine the HELIOS technique with state-of-the-art gamma-ray detection. One very appealing future direction of development would be to combine the MAYA and HELIOS concepts, so that particles could be completely tracked in three dimensions but with the focussing and collection advantages of the magnetic field.

\subsection{Using (d,p) with gamma-rays, to study bound states}
\label{subsec:tiara}

\begin{figure}[h]
\sidecaption
\includegraphics[width=0.7\textwidth]{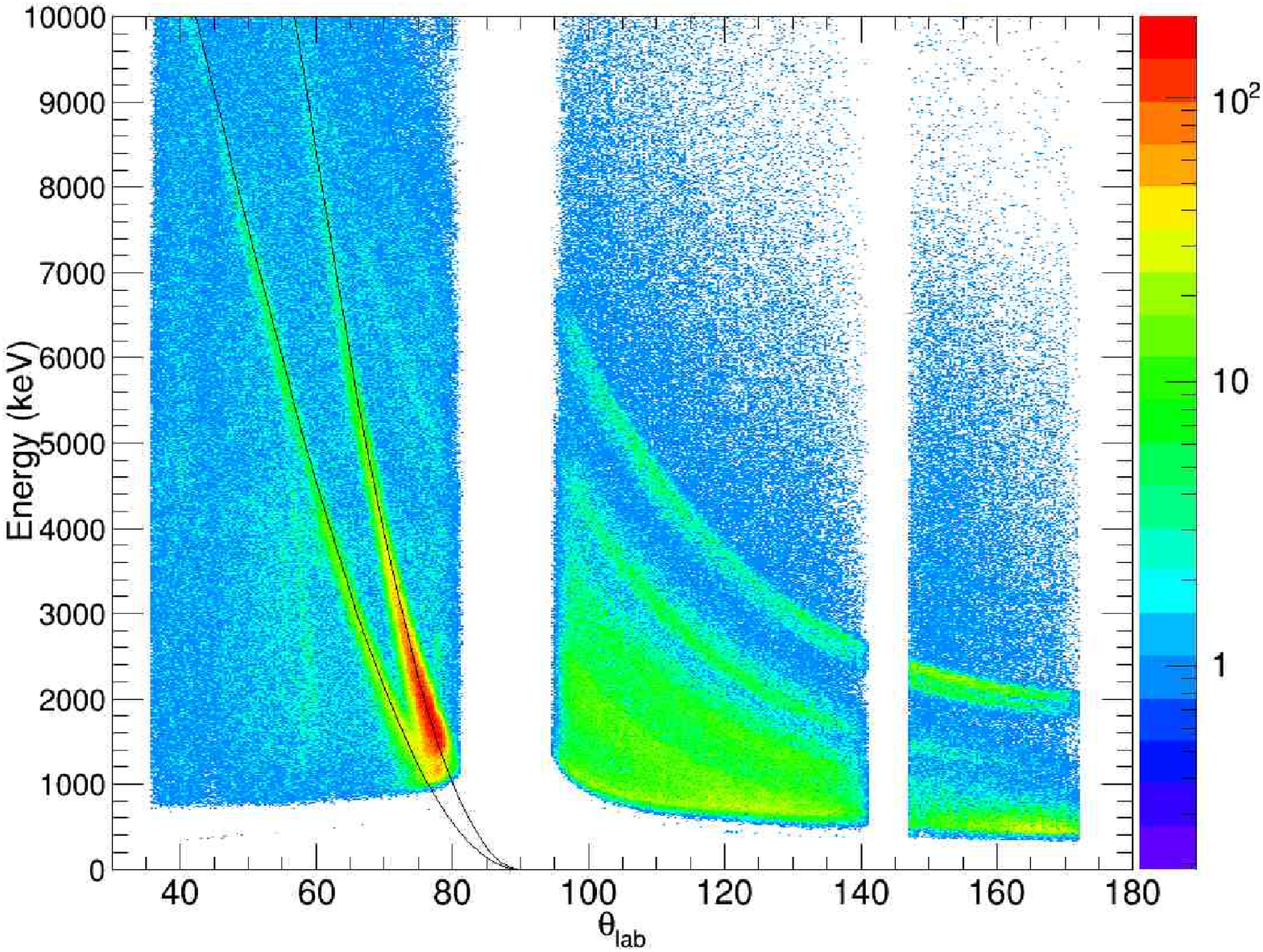}
%If the width of the Figure is less than 7.8 cm use the \texttt{sidecapion} command to flush the caption on the left side of the page. If the Figure is positioned at the top of the page, align the sidecaption with the top of the Figure -- to achieve this you simply need to use the optional argument \texttt{[t]} with the \texttt{sidecaption} command}
\caption{Raw data for a typical experiment \protect\cite{Wilson} using a silicon array to detect the light particles, to be compared with Figure \protect\ref{fig:11}. Kinematic lines are overlaid over the deuterons (higher energies) and protons from elastic scattering. At larger angles, the loci are clearly seen for protons from (d,p) reactions. The apparent angular dependence of the lower energy threshold is due to corrections that are applied to compensate for energy losses in the target. }
\label{fig:19}       % Give a unique label
\end{figure}

Typical data for the energies of the measured particles, as a function of their deduced laboratory angle, are shown in Figure \ref{fig:19} for an experiment using a silicon array with a large angular coverage\cite{Wilson}. This experiment was performed with a beam of $3 \times 10^7$ pps of $^{25}$Na at 5 A.MeV, using the SHARC array \cite{SHARC} at TRIUMF. Provided that calibrations have been performed in advance, this type of spectrum can be created online, during data acquisition. Once the kinematic lines are seen, the first hurdle is crossed, and the experiment is seen to be working correctly. Then, the discussion can turn to the specifics of the physics to be measured and the statistics that are required. The most intense lines will typically be those due to elastic scattering. In the Figure, the data show lines that are recognisable as coming from the elastic scattering of both deuterons and protons in the 0.5 mg/cm$^2$ $CD_2$ target. It is typical that any deuterated target will have some fraction of non-deuterated molecules. The intensity falls away, generally, as the energy increases and the centre of mass scattering angle also increases. The angular distribution may show oscillations, but the fall in intensity is the general tendency. In this particular experiment, there is a gap in the data near $90^\circ$ due to a physical gap in the array, related to the target mounting and changing mechanism. A further gap exists in the backward angle region due to the silicon detector support structure. In the region backward of $90^\circ$, the kinematic lines arising from (d,p) reactions are evident. In this angular range, there are no other deuteron-induced reactions (apart form (d,p)) that can contribute to the charged particle yield. From that perspective, no particle identification is needed for the backward angles. In fact, because of the low energies, no $\Delta E-E$ identification technique would be appropriate, but time-of-flight or silicon pulse-shape techniques would be feasible. The reason that particle identification could indeed be useful is that not all reactions will be induced on the deuterons in the target. Assuming a $CD_2$ target as in the experiment shown here, the reactions induced on carbon nuclei can produce charged particles at any angle. Typically, the compound nuclear reactions that arise from the carbon will produce both protons and alpha-particles (and possibly other species) by evaporation from the excited compound nucleus. Standard codes exist, to estimate the evaporation channels that will be important for a particular beam and energy combination (e.g. LISE++ \cite{LISE}, which includes the fusion-evaporation code PACE4 \cite{PACE}). These evaporated particles will not have a specific angle-energy relationship because several particles will be evaporated. Also, alpha-particles can deposit much more energy than protons in a given thickness of silicon because of their shorter range. Thus, the kinematic lines from (d,p) and elastic scattering will in general appear on a smooth background arising from evaporated charged particles from compound nuclear reactions. This is evident to some extent in Figure \ref{fig:19}, even though some experimental techniques have been applied so as to reduce the compound nuclear contribution (see below).

\begin{figure}[h]
\sidecaption
\includegraphics[width=1.0\textwidth]{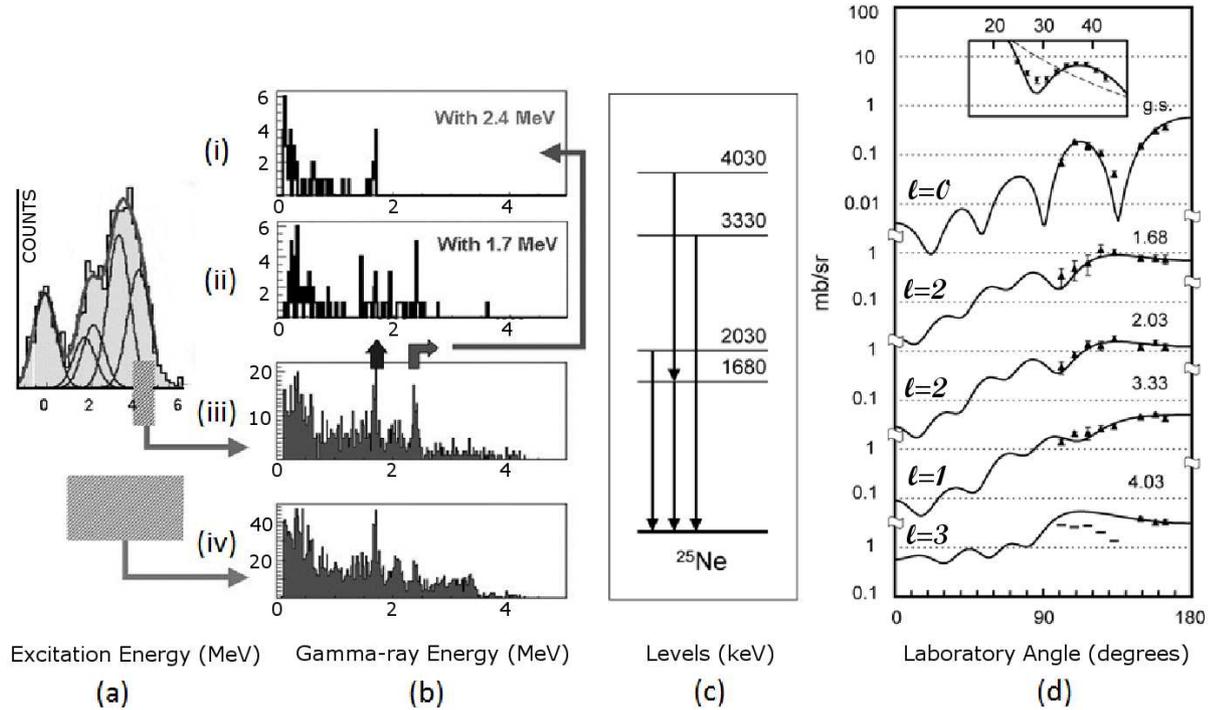}
%If the width of the Figure is less than 7.8 cm use the \texttt{sidecapion} command to flush the caption on the left side of the page. If the Figure is positioned at the top of the page, align the sidecaption with the top of the Figure -- to achieve this you simply need to use the optional argument \texttt{[t]} with the \texttt{sidecaption} command}
\caption{Results from a (d,p) study of $^{25}$Ne using a beam of $^{24}$Ne at 10 A.MeV \protect\cite{Ne24}: (a) example excitation energy spectrum reconstructed from the measured proton energies and angles and showing gating regions used to extract coincident gamma-ray spectra, (b) gamma-ray energy spectra (iii, iv) from $p-\gamma$ coincidences for highlighted regions of excitation energy in (a), spectra (i,ii) from $p-\gamma-\gamma$ data with the events and $\gamma$-ray gates indicated in (iii), (c) summary of the level and decay scheme deduced from this experiment, (d) differential cross sections for the indicated $\ell$ transfers to states in $^{25}$Ne. Elastic scattering data are inset (see text). }
\label{fig:20}       % Give a unique label
\end{figure}

Figure \ref{fig:20} summarises a range of experimental results from a (d,p$\gamma$) study using a radioactive beam of $2 \times 10^4$ pps of $^{24}$Ne at 10 A.MeV \cite{Ne24}. The energy and angle information as shown in the previous Figure can be combined to calculate the excitation energy in the final nucleus, assuming that the reaction was (d,p) initiated by the beam. Angular regions where other reactions dominate can be removed in the analysis. Figure \ref{fig:20}(a) shows an excitation energy spectrum for $^{25}$Ne calculated from the kinematic formulae, for one particular angle bin. The fit to the various excited state peaks in this spectrum was informed and constrained by the observed gamma-ray energies. The gamma-ray energy spectrum observed with specific limitations on the excitation energy are shown in part (b) of the Figure, where parts (iii) and (iv) correspond to the indicated excitation energy limits in $^{25}$Ne. For the events included in Figure \ref{fig:20}b(iii), the results of gating on particular gamma-ray peaks are shown in parts (i) and (ii). The $p-\gamma -\gamma$ triple coincidence statistics in these two spectra are sufficient (just) to deduce that the two observed gamma-ray transitions are in coincidence. (Actually, the experiment in ref. \cite{Ne24} also measured the heavy ($^{25}$Ne) particle after the reaction, so the data in Figure \ref{fig:20}b(i-ii) actually represent quadruple coincidence data). Taking into account the excitation energies at which the nucleus is fed by the (d,p) reaction, and the observed gamma-ray cascade, the level scheme in Figure \ref{fig:20}(c) was inferred. The angular distributions shown in Figure \ref{fig:20}(d) were used to deduce the transferred angular momentum carried by the neutron, according to the best-fit shape. The calculations that are shown were performed using the ADWA method. Different angular momenta were deduced for the various states. For example, the ground state has a clear $\ell =0$ distribution. The scaling of the theory to the experimental data gave the measured spectroscopic factors. In the case of the 4.03 MeV state, it was only possible to set a lower limit on the cross section at certain angles. This was related to the energy thresholds of the silicon detectors used for the proton detection. As shown in the kinematics diagrams in Figure \ref{fig:11}, and illustrated in the data of Figure \ref{fig:19}, the observed particles from (d,p) are lower in energy for states with higher excitation energy and hence the higher states are subject to this type of threshold effect. Raising the beam energy will give access to higher excitation energies. The observed lower limits on the cross section for the 4.03 MeV state were nevertheless sufficient to rule out the alternative angular momentum assignments and $\ell =3$ could be assigned. Finally, an inset in Figure \ref{fig:20}(d) shows the differential cross section for deuteron elastic scattering, measured as a function of the centre-of-mass scattering angle. This was derived from the rapidly rising locus of data points observed in the data, similar to that for the elastics shown in Figure \ref{fig:19}. This will be discussed further, in section \ref{subsec:elastic}.

The gamma-ray energy spectra of Figure \ref{fig:20} include a correction, applied event-by-event, for a very significant Doppler shift. At the recommended beam energies of 5-10 A.MeV, the projectiles have a velocity of approximately $0.10c$. Actually, the velocity is sufficient for the Doppler shift at $90^\circ$ due to the second-order terms to be easily measured. Hence, the full relativistically correct formula should be used, to apply Doppler corrections to the measured gamma-ray energies so that they accurately reflect the emission energies in the rest frame of the nucleus. The Doppler-corrected energy $E_{{\rm corr}}$ is given by
$$ E_{{\rm corr}} = \gamma ~(1 - \beta \cos \theta_{{\rm lab}} )~ E_{{\rm lab}} $$
where $\gamma = 1/\sqrt{1-\beta^2}$ and $\beta = \upsilon /c$ where $\upsilon$ is the velocity of the emitting nucleus. The angle $\theta_{{\rm lab}}$ is measured for the gamma-ray detector relative to the direction of motion of the nucleus. In practice, and taking into account the accuracy with which the gamma-ray angle can be determined, it is usually sufficient to assume that the emitting nucleus is travelling along the beam direction in these inverse kinematics experiments (although it is also easy to calculate it's angle from the measured light-particle angle). It will be relevant later, to note that another relativistic effect related to gamma-rays is significant at these beam energies. The angle of emission relative to the beam direction, as measured in the frame of the emitting nucleus, is different from the angle measured in the laboratory frame of reference. This consequence of relativistic abberation means that the gamma-rays emitted by a moving nucleus are concentrated conically towards its direction of motion, which is known as relativistic beaming or as the relativistic headlight effect. For isotropic centre of mass emission at $\beta = 0.1$, the fraction of gamma-rays emitted forward of $90^\circ$ in the laboratory will be about 55\%. The yield of gamma-rays observed at $10^\circ$ in the laboratory will be larger than the yield at $170^\circ$ by a factor of $1.22/0.82 = 1.49$. The relativistic aberration formula is given by
$$ \cos \theta_{{\rm lab}} = \frac{\cos \theta_{{\rm c.m.}} - \beta }{1 - \beta \cos \theta_{{\rm c.m.}}} $$
where $\theta_{{\rm c.m.}}$ is measured in the rest frame of the nucleus and other terms are as defined above.

\begin{figure}[h]
\sidecaption
\includegraphics[width=0.7\textwidth]{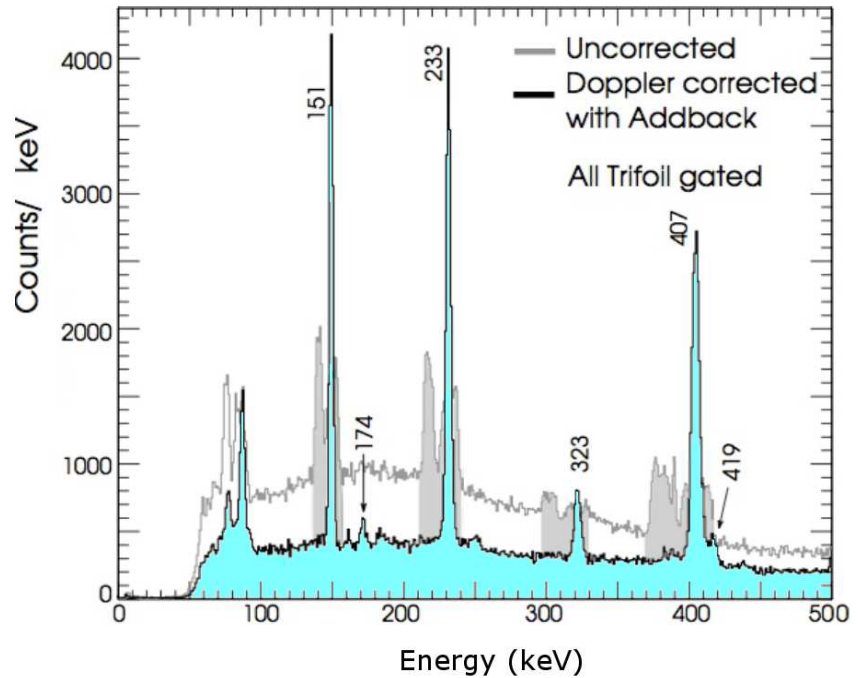}
%If the width of the Figure is less than 7.8 cm use the \texttt{sidecapion} command to flush the caption on the left side of the page. If the Figure is positioned at the top of the page, align the sidecaption with the top of the Figure -- to achieve this you simply need to use the optional argument \texttt{[t]} with the \texttt{sidecaption} command}
\caption{Results of the Doppler shift correction procedure applied to $^{26}$Na gamma-rays produced in the reaction of 5 A.MeV $^{25}$Na with deuterons \protect\cite{Wilson}. The upper spectrum (outlined and partly shaded in light grey) is uncorrected, with the shaded parts indicating the spread of counts contributing to four of the strongest peaks in the lower spectrum. The lower spectrum (darker shading) is corrected for the Doppler shift.
In addition to the Doppler correction, an add-back procedure has been applied to account for Compton scattering (see text). This lowers the continuum background. All of these data are "Trifoil gated" to remove or minimize events of a compound nuclear origin, as explained in section \protect\ref{subsec:zero}). }
\label{fig:21}       % Give a unique label
\end{figure}

The relativistic Doppler shift correction was already performed for the gamma-rays in Figure \ref{fig:20} and is shown in more detail for a different experiment, in Figure \ref{fig:21}. In the case of Figure \ref{fig:20}, the gamma-ray angle could be determined only according to which leaf (crystal) of the clover detector recorded the initial interaction. The resolution at 1 MeV was 65 keV  FWHM (full width at half maximum) after correction \cite{Ne24}, limited by the high value of $\beta = 0.1$, the close proximity of the detectors to the target (50 mm) and the lack of any further gamma-ray angle information. This is reduced by a third to just under 45 keV (FWHM) at 1 MeV in the TIARA configuration if the clover segmentation information is used \cite{Labiche}. In the experiments \cite{Wilson} with SHARC, using TIGRESS, the distance to the front face of the gamma-ray detectors was nearly three times larger than TIARA, at 145 mm. The gamma-ray clover detectors were centred at either $90^\circ$ or $135^\circ$ and each leaf of the clover was four-fold segmented electronically. An add-back procedure was applied, to account for Compton scattering between different leaves of the same clover. This involved adding the energies together and then adopting the segment with the highest energy as indicating the angle of the initial interaction (a criterion that is justified by simulations \cite{Labiche}). For a (d,p$\gamma$) gamma-ray at 1806 keV, the observed resolution after Doppler correction was 23 keV (FWHM) or 18 keV (FWHM) for detectors at $90^\circ$ and $135^\circ$ respectively (reflecting the Doppler broadening, as opposed to shift, that contributes at $90^\circ$). Scaling this to the previously quoted energy of 1 MeV gives a resolution of 10-12 keV (FWHM). This resolution is a factor of 10-50 better than the resolution in excitation energy obtained from using the measured energy and angle of the proton. Thus, the resolution in excitation energies for states populated in (d,p) reactions can be improved by a similar factor.

\subsection{The use of a zero-degree detector in (d,p) and related experiments}
\label{subsec:zero}
	
\begin{figure}[h]
\sidecaption
\includegraphics[width=0.7\textwidth]{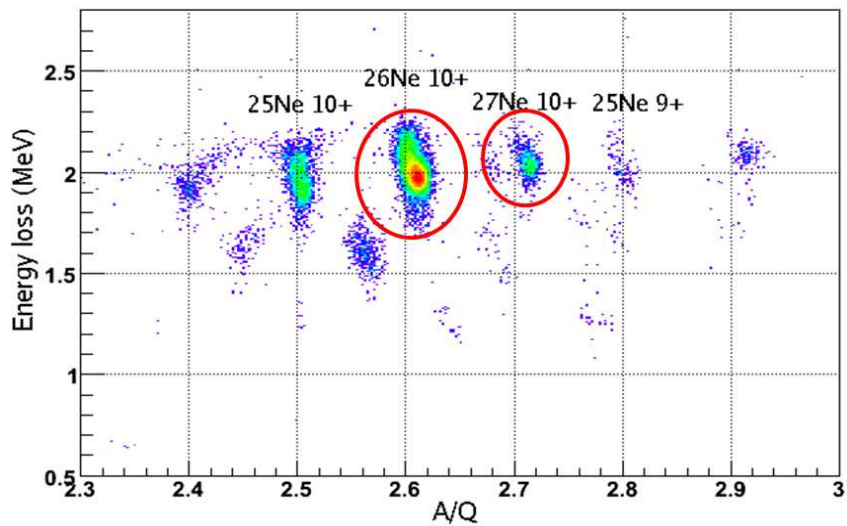}
%If the width of the Figure is less than 7.8 cm use the \texttt{sidecapion} command to flush the caption on the left side of the page. If the Figure is positioned at the top of the page, align the sidecaption with the top of the Figure -- to achieve this you simply need to use the optional argument \texttt{[t]} with the \texttt{sidecaption} command}
\caption{Data from a study of $^{26}$Ne at 10 A.MeV bombarding a $CD_2$ target \protect\cite{ne27,Simon}. Particles were detected in the wide-acceptance spectrometer VAMOS centred at zero degrees and were identified using the parameters measured at the focal plane. This determined the reaction channel and effectively eliminated any contribution from carbon in the target. }
\label{fig:22}       % Give a unique label
\end{figure}

The ability to detect the beam-like particle, as well as the light particle, from transfer reactions in inverse kinematics is a big advantage for several reasons. It was therefore a fundamental design constraint, for TIARA \cite{CAARI,Labiche}, that it should be coupled to the magnetic spectrometer VAMOS. The advantages are partly evident from inspection of Figure \ref{fig:22}. The different particle types observed at angles around zero degrees, following the bombardment of a $CD_2$ target with a $^{26}$Ne beam, are clearly identified. The beam in this case was 2500 pps at 10 A.MeV and the target thickness was 1.20 mg/cm$^2$. Two further features make this zero degree detection even more useful. Firstly, the silicon array will record the coincident particles only for the reactions induced on the hydrogen in the target; the recoil carbon nuclei for this constrained kinematics will essentially all stop in the target. Secondly, the spectrometer gives not only the particle identification but also the angle of emission for the heavy particle, which can be exploited, for example as in section \ref{subsec:unbound}. In the example shown here, the reaction products could be simultaneously collected and identified for (d,p) to bound states of $^{27}$Ne, (d,p) to unbound $^{27}$Ne that decays back to $^{26}$Ne, and (d,t) to bound states of $^{25}$Ne.

\begin{figure}[h]
\sidecaption
\includegraphics[width=0.7\textwidth]{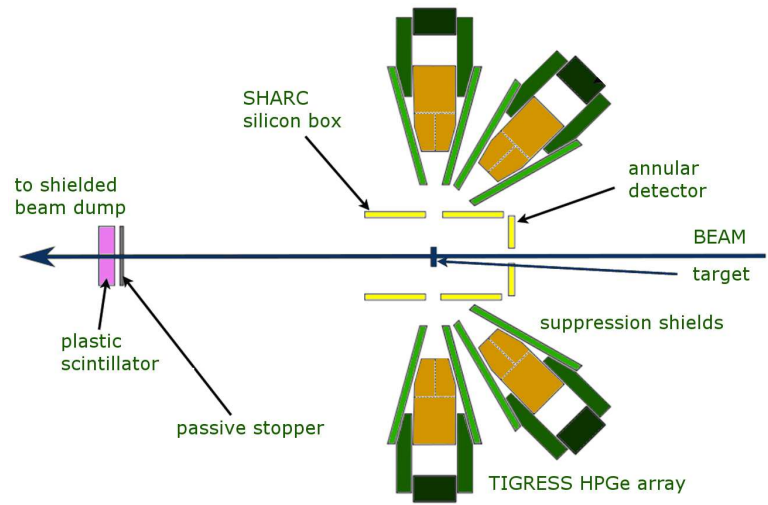}
%If the width of the Figure is less than 7.8 cm use the \texttt{sidecapion} command to flush the caption on the left side of the page. If the Figure is positioned at the top of the page, align the sidecaption with the top of the Figure -- to achieve this you simply need to use the optional argument \texttt{[t]} with the \texttt{sidecaption} command}
\caption{Schematic of the experimental setup for experiments combining the SHARC Si array with the TIGRESS gamma-ray array \cite{Wilson}. A plastic scintillator detector was introduced at zero degrees, 400 mm beyond the target, to help in identifying and eliminating events arising from reactions on the carbon component of the $CD_2$ target. The performance of this {\em trifoil} detector \protect\cite{GLWrutherford} is discussed in the text. }
\label{fig:23}       % Give a unique label
\end{figure}

In experiments currently performed at TRIUMF, there is no access to a spectrometer such as VAMOS, and hence a less elaborate solution was implemented, and is described here. Note that, in the longer term, the purpose-built fragment mass separator EMMA \cite{EMMA} will become available at TRIUMF. In the meantime, a detector developed at LPC Caen and called the {\it trifoil} was adapted \cite{GLWrutherford} from its original purpose, which was to provide a timing signal for secondary beams produced via projectile fragmentation at intermediate energies. The experimental layout for the first experiment \protect\cite{Wilson} using the trifoil in this fashion is shown in Figure \ref{fig:23}. In this implementation, the plastic scintillator in the trifoil will record signals arising from unreacted beam particles or transfer and similar reactions in the target, i.e. where the beam-like particle is not slowed down. If the reaction in the $CD_2$ target was induced by the carbon, then it could be either a transfer reaction (if peripheral) or a compound nuclear reaction. In the former case, no particle would be observable in the silicon array SHARC. In the second case, the evaporated charged particles could be observed, but also the product at zero degrees would be slower moving and would have a higher $Z$ than for a transfer reaction induced by the hydrogen in the target. The compound nuclear products are then stopped by a passive layer of aluminium, whilst still leaving the direct reaction products with sufficient energy to be recorded in the trifoil and then pass through to a remote beam dump. The present trifoil detector is big enough to span the cone of recoil beamlike particles corresponding to protons from (d,p) collected over a wide range of angles. Compound nuclear events are completely prevented from producing a valid trifoil signal, by means of the passive stopper, but depending on the beam rate there may be random coincidences with other beam particles arriving in the same bunch of the pulsed beam. (Ideally, the detector would be insensitive to unreacted beam particles, and this was achieved to some extent.)

\begin{figure}[h]
\sidecaption
\includegraphics[width=1.0\textwidth]{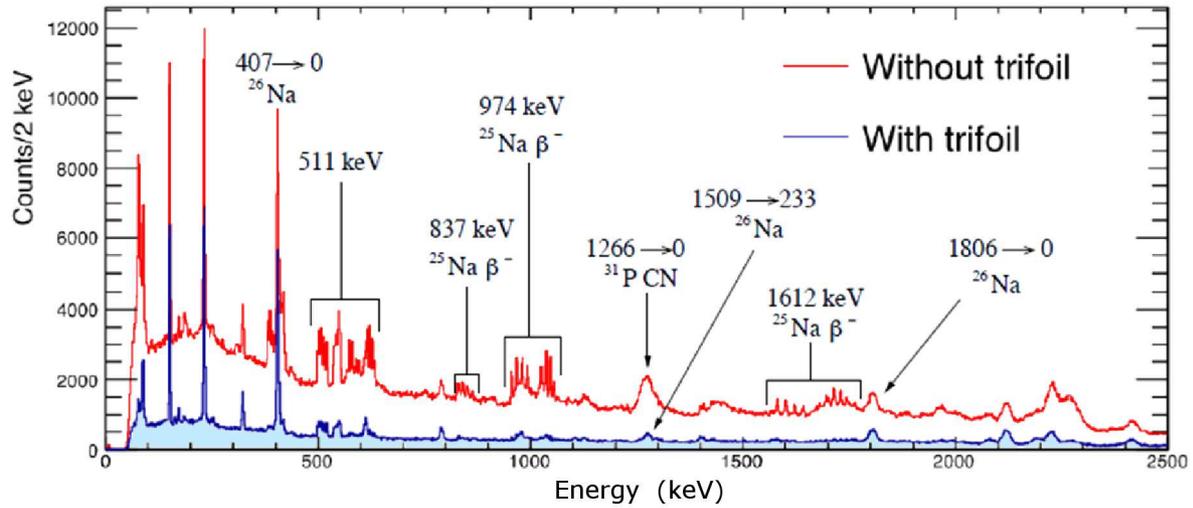}
%If the width of the Figure is less than 7.8 cm use the \texttt{sidecapion} command to flush the caption on the left side of the page. If the Figure is positioned at the top of the page, align the sidecaption with the top of the Figure -- to achieve this you simply need to use the optional argument \texttt{[t]} with the \texttt{sidecaption} command}
\caption{Gamma-ray energy spectrum acquired for a beam of $^{25}$Na at 5 A.MeV incident on a $CD_2$ target, using the full TIGRESS array (see text). The requirement of a trifoil signal eliminates a large fraction of the smooth background, and largely removes the peaks due to scattered radioactivity and compound nuclear reactions. The radioactivity peaks are dispersed by the Doppler correction. }
\label{fig:24}       % Give a unique label
\end{figure}

The effect of the zero degree trifoil detector in reducing the background in the gamma-ray energy spectra is illustrated in Figure \ref{fig:24}. This spectrum was acquired for a beam of $^{25}$Na at 5 A.MeV incident on a $CD_2$ target with an average intensity of $3 \times 10^7$ pps. The spectrum includes data from the full TIGRESS array, comprising 8 detectors with 4 placed at $90^\circ$ and 4 at $135^\circ$ in this experiment \cite{Wilson}. The spectrum is Doppler corrected as described above, and hence the gamma-rays produced by a source at rest (such as the 511 keV annihilation gamma-ray and those originating from the radioactive decay of scattered and then stopped $^{26}$Na projectiles) have been transformed into multiple peaks depending on their angle of detection relative to the target. Escape suppression has also been applied, using the signals from the scintillator shields for each clover detector. The first thing to note is that the smooth background, arising from unsuppressed Compton scattering events due to higher energy gamma-rays, is massively reduced by applying the trifoil requirement. This is quantified below. Secondly, with regard to the peaks, it can be seen for example that the 1806 keV peak arising from the (d,p) product $^{26}$Na is retained in the trifoil-gated spectrum with high efficiency whereas the 1266 keV peak arising from the compound nuclear product $^{31}$P is mostly eliminated. In fact, the elimination of the $^{31}$P peak reveals an underlying $^{26}$Na peak at 1276 keV.

\begin{figure}[h]
\sidecaption
\includegraphics[width=1.0\textwidth]{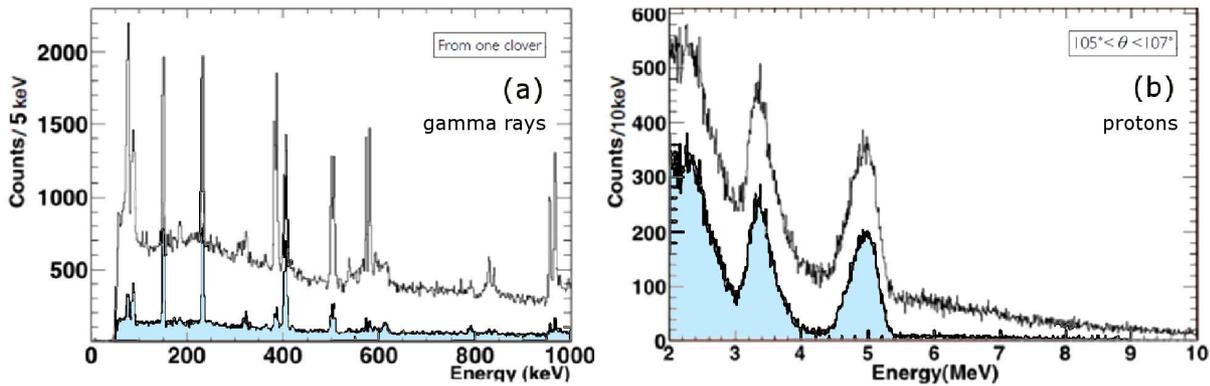}
%If the width of the Figure is less than 7.8 cm use the \texttt{sidecapion} command to flush the caption on the left side of the page. If the Figure is positioned at the top of the page, align the sidecaption with the top of the Figure -- to achieve this you simply need to use the optional argument \texttt{[t]} with the \texttt{sidecaption} command}
\caption{Energy spectra accumulated for a beam of $^{25}$Na at 5 A.MeV incident on a $CD_2$ target, showing the rejection of background using the trifoil detector as discussed in the text: (a) expanded view of the low energy {\it gamma-ray} spectrum, for a single clover crystal at $82^\circ$ to the beam direction, (b) example {\it proton} energy spectrum for measured proton angles between $105^\circ$ and $107^\circ$ in the laboratory frame, i.e. a vertical slice in Figure \protect\ref{fig:19}. }
\label{fig:25}       % Give a unique label
\end{figure}

In order to quantify the improvement in peak:background ratio that was achieved by using the trifoil, spectra such as those in Figure \ref{fig:25} were produced. The gamma-ray energy spectrum in Figure \ref{fig:25}(a) is for a single clover at a single laboratory angle. The data were analysed in this way, in order to be sure to separate as much as possible the gamma-rays arising from transfer and compound nuclear reactions. The optimal value of the velocity $\beta$ for the Doppler correction is of course different for these two different categories of reaction, so the correction procedure produces relative movement in energy between counts from transfer and compound reactions depending upon the angle of the gamma-ray detection. The proton energy data in Figure \ref{fig:25}(b) are for a thin slice in a spectrum of energy versus angle such as that shown in Figure \ref{fig:19}. Already, in Figure \ref{fig:19}, the trifoil requirement was applied and this reduced a smooth background arising from compound nuclear events. The extent of this background reduction can be measured using Figure \ref{fig:25}(b). In this particular experiment, the average efficiency for successfully tagging a genuine proton or a genuine gamma-ray (i.e. one arising from a transfer or other direct reaction) was about 80\%. The shortfall relative to 100\% was due to the intrinsic efficiency properties of this particular trifoil detector. The average probability for incorrectly tagging a charged particle or gamma-ray of compound nuclear origin was about 15\%, or for a gamma-ray from radioactive decay it was about 10\%. The origin of this unwanted probability lay in the high beam intensity and the chance of recording an unreacted beam particle in the same nanosecond sized beam bunch as a compound reaction. Taken overall, the peak:background ratio in each of the proton energy spectrum and the gamma-ray energy spectrum was improved by nearly an order of magnitude. The two reductions of the background are not independent. For a particular gamma-ray peak, an enhancement in the peak:background ratio of typically a factor of 40 was observed, and there is scope for improvement upon this as noted above.

\subsection{Simultaneous measurements of elastic scattering distributions}
\label{subsec:elastic}

In the experiments with TIARA \cite{Ne24,ne27,bea} and SHARC \cite{Wilson}, the absolute normalisation was provided by a simultaneous measurement of the elastic scattering cross section. An example of the data obtained for the cross section, plotted as a function of the centre of mass scattering angle, is shown as the inset in Figure \ref{fig:20}. This technique works well, so long as the elastic scattering can be measured sufficiently close to $90^\circ$ in the laboratory that it includes the small values of the centre of mass angle where the elastic cross section can be calculated reliably. The method relies upon being able to evaluate the cross section theoretically using an optical model calculation. At small centre of mass angles, the deviation from Rutherford scattering will be small and the cross section will be reliable. Assuming that the measurements can be made, there are significant advantages in using this technique. The three main advantages concern (a) the beam integration, (b) the target thickness and (c) the dead time in the data acquisition system. The beam integration would normally require the direct counting of every incident beam particle, with a detector of a known and consistent efficiency. The target thickness would normally be required to be known precisely. However, the measurement of the yield for elastic scattering allows the product of these two quantities ((a) and (b)) to be measured, including any necessary correction for the dead time (c) of the acquisition. In the experiments described, the trigger for the acquisition was for a particle to be detected in the silicon array. The elastic scattering and (d,p) reaction events were then subject to the same dead time constraints. It is still necessary to have a reasonable measurement of the target thickness, so that corrections can be applied for the energy lost by the incident beam and by charged particles as they leave the target.

\subsection{Extending (d,p) studies to unbound states}
\label{subsec:unbound}

The extension of (d,p) studies to include transfer to states in the continuum of the final nucleus is relatively straightforward experimentally compared to the theoretical interpretation. In fact, this issue highlights situations in the development of the reaction theory that have remained unresolved, or partially unresolved, from the days when (d,p) reactions in normal kinematics were a major topic of research.

An experimental example that is relatively simple to treat, both experimentally and theoretically, is provided by a study of the lowest $7/2^-$ state in $^{27}$Ne, populated via (d,p) with a $^{26}$Ne beam \cite{ne27}. This state is observed as an unbound resonance at an excitation energy of 1.74 MeV in $^{27}$Ne, compared to the neutron separation energy of 1.43 MeV. For reasons of both the relatively small energy above threshold and the relatively large neutron angular momentum of $\ell =3$, this unbound state is quite narrow. In fact, the experiment implies the natural width to be 3-4 keV (but in the data it is observed with a peak width of 950 keV due primarily to target thickness effects). In the case of a relatively narrow resonance, meaning a resonance with a natural width that is small compared to its energy above threshold, it is possible to carry out a theoretical analysis with relatively small modifications to the theory. One method is to make the approximation that the state is bound, say by 10 keV, in order to calculate the form factor (i.e. overlap integral) for the neutron in the transfer; this can satisfactorily describe the wave function in the region of radii where the transfer takes place. An improved approach is to use a resonance form factor, following the method of Vincent and Fortune \cite{VincentFortune}. In this theory, the magnitude of the differential cross section scales in proportion to the width of the resonance. If a barrier penetrability calculation is used, to estimate the width for a pure single particle state, then the cross section can again be interpreted in terms of a spectroscopic factor. The Vincent and Fortune method has been implemented \cite{Cooper} in the Comfort extension \cite{Comfort} of the widely used DWBA code DWUCK4 \cite{Kunz}. For these narrow, almost bound resonances, the structure of the differential cross section retains its characteristic shape, determined by the transferred angular momentum.

It has long been known \cite{Akram} that the oscillatory features of the differential cross sections, which allow the transferred angular momentum to be inferred from experimental data, are less prominent or even absent when the final state is unbound and broad in energy. The method of Vincent and Fortune also ceases to be applicable, for these broad resonances. Because of the lack of structure, it becomes difficult to interpret the experimental data so as to determine the spins of final states. An experimental example is provided by the study of unbound states in $^{21}$O via the (d,p) reaction with a beam of $^{20}$O ions \cite{bea}. The analysis in ref. \cite{bea} included calculations using the CDCC model mentioned in section \ref{subsec:adwa}, wherein the continuum in $^{21}$O was considered to be divided into discrete energy bins with particular properties.

\begin{figure}[h]
\sidecaption
\includegraphics[width=0.7\textwidth]{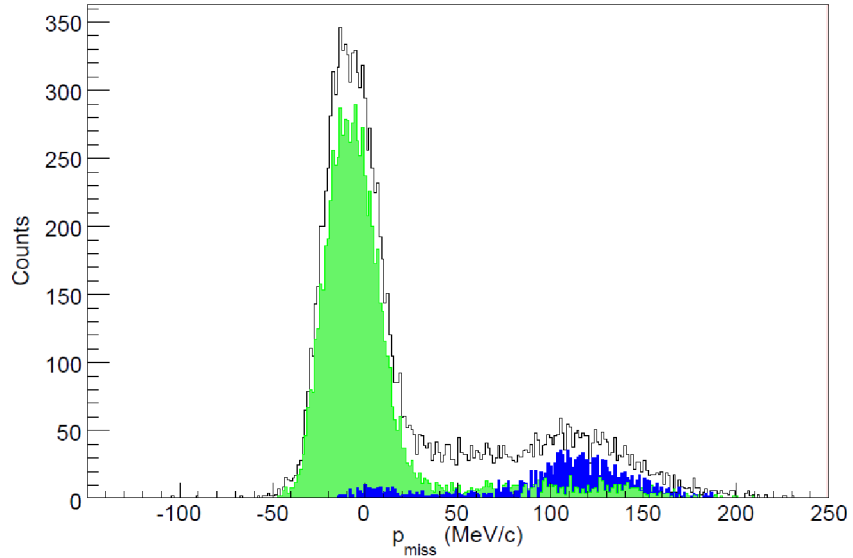}
%If the width of the Figure is less than 7.8 cm use the \texttt{sidecapion} command to flush the caption on the left side of the page. If the Figure is positioned at the top of the page, align the sidecaption with the top of the Figure -- to achieve this you simply need to use the optional argument \texttt{[t]} with the \texttt{sidecaption} command}
\caption{Reconstructed magnitude for the momentum of any missing particle when $^{26}$Ne and a light charged particle (assumed to be a proton) are detected from the reaction of a $^{26}$Ne beam on a $CD_2$ target \protect\cite{Simon}. The upper histogram is for all data where a $^{26}$Ne was positively identified. The green shaded area giving mainly a peak near zero is for data selected to highlight elastic scattering, which in fact is d($^{26}$Ne,$^{26}$Ne)d. The dark blue shaded area with fewer counts is selected to highlight the reaction d($^{26}$Ne,$^{27}$Ne$^*$ $\rightarrow$ $^{26}$Ne + n )p where the neutron was undetected.  }
\label{fig:25a}       % Give a unique label
\end{figure}

It may be possible to recover some sensitivity to the transferred angular momentum by observing the sequential decay of the resonance states. The observed angular distribution should reflect the angular momentum of the decay of the resonance, with a dependence on the magnetic substate populations for the resonant state in the transfer reaction. An attempt to exploit this effect was made in the study of d($^{26}$Ne,$^{27}$Ne)p mentioned above \cite{ne27,Simon}. The $^{26}$Ne products were identified in a magnetic spectrometer as shown in Figure \ref{fig:22}. By a process of ray tracing \cite{VAMOS} it was also possible to reconstruct the magnitude and direction ($\theta,\phi$) of the $^{26}$Ne momentum. Combining this with the momentum of the incident beam and the light particle detected in TIARA, it was possible to reconstruct the missing momentum \cite{Simon}. It was assumed that the light particle was a proton, arising from (d,p). The primary aim of this particular analysis was to be able to separate the events arising from (d,p) from those arising from (d,d) or (p,p) in the part of the TIARA array forward of $90^\circ$. In this sense, it was very successful, as shown by the separation of the main elastic peak from the sequential decay peak in Figure \ref{fig:25a}. A threshold of 40 MeV/c effectively discriminates between these two reaction channels. Unfortunately, the resolution in terms of the reconstructed angle (rather than the magnitude) of the unobserved neutron momentum was inadequate to take this further. No useful angular correlation could be extracted, for the sequential $^{27}$Ne$^*$ $\rightarrow $ $^{26}$Ne + $n$ decays.
	
\subsection{Simultaneous measurement of other reactions such as (d,t)}
\label{subsec:dt}
	
Radioactive beams are so difficult to produce that an experiment should make the best possible use of the beam delivered to the target. The compact silicon arrays such as TIARA were designed to cover the whole range of laboratory angles with particle detectors that would assist in this aim. The detectors in the forward hemisphere can record the particles from reactions such as (d,t) or (d,$^3$He), at the same time as those just forward of $90^\circ$ record elastic scattering and those in (predominantly) the backward hemisphere record the (d,p) reaction products. Indeed, the experiment using TIARA to study $^{21}$O via (d,p) with a beam of $^{20}$O \cite{bea} was also designed to measure the (d,t) reaction to $^{19}$O at the same time. The (d,t) measurements \cite{Alexis,Ramus} employed the telecopes of MUST2 \cite{MUST2} which were mounted at the angles forward of the TIARA barrel (cf. Figure \ref{fig:14}). The gamma-ray coincidence measurements with EXOGAM allowed new spin assignments as well as the spectroscopic factor measurements for $^{19}$O states \cite{Alexis,Ramus}. Any studies with (d,t) are immediately useful for comparing to the sorts of knockout studies described in section \ref{subsec:knockout}. The work of ref. \cite{bea} was able to take the additional step of combining the spectroscopic factors measured for (d,p) and (d,t) from $^{20}$O. In an analysis based on sum rules and the formalism of \cite{Baranger} and \cite{Signoracci}, it was possible to derive experimental numbers for the single particle energies for this nucleus. The values were in good agreement \cite{bea} with the effective single particle energies of the USDA and USDB shell model interactions for the $sd$-shell obtained in ref. \cite{BrownRichter}. The previously discussed experiment using a beam of $^{26}$Ne \cite{ne27,Simon} used the same TIARA + MUST2 experimental setup as the $^{20}$O experiment. The data for the (d,t) reaction from $^{26}$Ne are still under analysis \cite{JST} but an interesting feature here is that the (p,d) reaction was also able to be measured at the same time. The separation experimentally of the d and t products of the (p,d) and (d,t) was possible in MUST2 with a suitable combination of time-of-flight identification and kinematical separation.

\subsection{Taking into account gamma-ray angular correlations in (d,p)}
\label{subsec:gamma}

It is well known that gamma-ray angular correlations will be observed for gamma-rays de-exciting states that are populated in nuclear reactions. These correlations have been widely exploited to reveal information about transition multipolarities and mixing, and hence to deduce spin assignments. For a nucleus produced in a reaction, and having some spin $J$, the angular distribution of gamma-rays measured relative to some $z$-axis (such as the beam direction) will depend on the population distribution for the magnetic substates $m_j = -J$ up to $+J$. If $J=0$, the gamma-rays will necessarily be isotropic. However, for other $J$-values the population of substates will be determined by the reaction mechanism and other details of the reaction. Thus, in (d,p) reactions for example, the gamma-ray angular distribution can depend on details such as the angle of detection of the proton. Certain simplifications can be made. For example, if $J=1/2$ then, for an unpolarised incident beam, and for protons detected symmetrically around zero degrees (with respect to the beam) the gamma-rays will necessarily be isotropic. Historically, experiments performed with stable beams and targets were designed to restrict the detection parameters in such a way as to simplify the angular momentum algebra, so as to remove any need to understand the magnetic substate populations, and hence the reaction mechanism, in detail. One of the most widely used classifications for angular correlation experiments are the Methods I and II of Litherland and Ferguson \cite{Litherland}. These methods are discussed in some detail in various texts, for example ref. \cite{England}. A simple and relevant example of the application of Method II is a study of the $^{26}$Mg(d,p$\gamma$)$^{27}$Mg reaction, in which the spins of the first three states in $^{27}$Mg were deduced from the measured gamma-ray angular correlations \cite{Eswaran}. Method I of Litherland and Ferguson involves measuring a $\gamma$-$\gamma$ angular correlation relative to a particular fixed angle for the first gamma-ray. The quantisation axis is defined by the direction of the incident beam. Method II, the more relevant here, is to measure a particle-$\gamma$ angular correlation where the outgoing particle from the reaction is measured at either $0^\circ$ or $180^\circ$. This limits the orbital magnetic quantum numbers of the projectile and ejectile to be $m_\ell =0$ and the consequences of this eliminate the need for any detailed knowledge of the reaction in order to know the substate populations for the final nucleus.

In the present work, we consider a more general situation where we retain one major simplification, namely the cylindrical symmetry of the particle detection, around the beam axis. The discussion is based around the previously discussed experiment using a  $^{25}$Na beam to study (d,p) reactions populating states in $^{26}$Na \cite{Wilson}. The SHARC experimental setup (cf. Figure \ref{fig:23}) gives essentially cylindrically symmetrical detection of the protons. The simplification that is produced by this symmetry in the angular description of the angular correlation is dramatic and is described in sections III.E and III.F of the article by Rose and Brink \cite{RoseBrink}. Rose and Brink define an {\it alignment condition} which means that $w(-M_1)=w(M_1)$ for all values of the magnetic substate quantum number $M_1$ of the emitting nucleus with spin $J_1$. Here, $w$ is the weight (i.e. population probability) for a given magnetic substate and is subject to the normalisation
$$ \sum _{M_1} w(M_1) = 1~.  $$
As described in their method 2 of section III.F, entitled {\it the alignment is achieved by a particle-particle reaction}, the alignment condition will be achieved if the outgoing particle is detected with cylindrical symmetry (assuming that the beam and target particles are unpolarised).  Method II of Litherland and Ferguson is simply a very restricted instance of this stipulation. The results used here to describe angular correlations are taken from Rose and Brink's article \cite{RoseBrink}, and they have also been summarised and discussed in the book by Gill \cite{Gill}.
	
Suppose we have an experiment where the outgoing particle (for example, the proton in a (d,p) reaction) is detected in a cylindrically symmetric fashion at some particular angle with respect to the beam direction. Let the spin of the excited state be $J_1$. Suppose also, for simplicity, that the gamma-ray transition by which the excited state decays is a pure transition of a particular mulitpole $L$ (the more general cases of mixed multipolarity transitions with a mixing ratio $\delta$ are discussed in refs. \cite{RoseBrink, Gill}). If a gamma-ray detector with a fixed solid angle were then to be moved sequentially to various angles $\theta$ with respect to the beam direction, then the angular distribution observed for the gamma-rays would be given by equation (3.38) of ref. \cite{RoseBrink},
$$ W_{{\rm exp}} (\theta ) = \sum _K a_K P_K (\cos \theta ) $$
where it can be shown that $K$ runs from 0 to $2L$ and is even, the $P_K$ are the Legendre polynomials and the $a_K$ can be calculated (as described below) provided that we know the magnetic substate populations of the initial state $J_1$ and the spin of the final state $J_2$. Outside of the summation, there will also be an additional factor, usually denoted $A_0$, to normalise $W$ to the data. The definition of $W(\theta )$ is chosen so that isotropic emission corresponds to $W(\theta )=1$. Note that this implies that the constant term in the expansion is always $a_0 = 1$. The number of gamma-rays in total that are emitted at an angle $\theta$ into an angular range $d\theta $ is given by $W(\theta ) \times 2\pi \sin \theta d\theta $. In the case of a transition with pure multipolarity (i.e. with a mixing parameter of $\delta =0$) equation (3.47) of ref. \cite{RoseBrink} states that the theoretical form for the angular distribution is given by
$$ W(\theta ) = \sum _K B_K(J_1) \times R_K(LLJ_1 J_2) \times P_K (\cos \theta) $$
where the $R_K$ are independent of the reaction mechanism and basically contain coefficients to describe the angular momentum coupling. The expression for $R_K$ is given by equation (3.36) of ref. \cite{RoseBrink},
$$ R_K (LL^\prime J_1 J_2) = (-)^{1+J_1 - J_2 + L^\prime  - L - K} \times \sqrt{(2J_1 +1) (2L+1) (2L^\prime +1)} \times (LL^\prime 1-1 \mid K0)
\times W(J_1 J_1 L L^\prime ; K J_2 )  $$
where the final two terms are the Clebsch-Gordon coefficient and the Racah W-coefficient describing the indicated angular momentum couplings. These coefficients may be obtained from tables or recursion formulae or from a suitable computer code such as ref. \cite{PDS}. In the present case, for a pure multipolarity, we have $L^\prime = L$. It is the $B_K$ coefficients that contain the information from the reaction mechanism, via the magnetic substate population parameters, $w(M_1)$. The expression for $B_K$ is given by equation (3.62) of ref. \cite{RoseBrink},
$$ B_K (J_1 ) = \sum _{M_1 = 0 {\rm ~or~} 1/2}^{M_1 = J_1 } w(M_1) \times \rho _K (J_1 M_1)  $$
where the statistical tensor coefficients $\rho _K$ are given by
$$ \rho _K (J_1 M_1) = (2-\delta _{{M_1},0} ) \times (-)^{J_1 - M_1} \times \sqrt{2J_1 + 1} \times (J_1 J_1 M_1 -M_1 \mid K0)  $$
and the final term is again a Clebsch-Gordon coefficient. For most normally-arising cases, the values of $\rho _K$ and $R_K$ are tabulated in the appendix of ref. \cite{RoseBrink}. The above description has followed exclusively the formulation of Rose and Brink \cite{RoseBrink}. Other authors have also presented formulae to describe these angular correlations, but it should be remembered that the different authors often adopt different phase conventions, etc., and hence the tables of symbols appropriate to one description can not be assumed to be appropriate for a different description: one particular formulation must be used consistently. Also, in ref. \cite{RoseBrink} the formalism is extended  to the case where a gamma-ray cascade occurs, and the second (or subsequent) gamma-ray is the one that is observed. In this case, as given by equation (3.46) of ref. \cite{RoseBrink}, the $R_K$ coefficient in the expression for $W(\theta )$ is replaced by a product of coefficients $U_K R_K$ where $U_K$ depends on $J_1$ and $J_2$ for the initial gamma-ray transition and $R_K$ depends on $J_2$ and $J_3$ for the second gamma-ray transition. The extension to a longer gamma-ray cascade is straightforward.

Thus, if the spins of the states are known, it is possible to calculate the $a_K$ coefficients, $a_2$, $a_4, \ldots$, of the Legendre polynomials in the gamma-ray angular distribution $W(\theta )$ provided that the magnetic substate weights $w(M_1)$ are known - at least, for pure multipolarity transitions. These expressions all rely on the particle detection being cylindrically symmetric at some angle (or range of angles) with respect to the beam direction. This ensures that $w(-M_1)=w(M_1)$ for all $M_1$.

\begin{figure}[h]
\sidecaption
\includegraphics[width=1.0\textwidth]{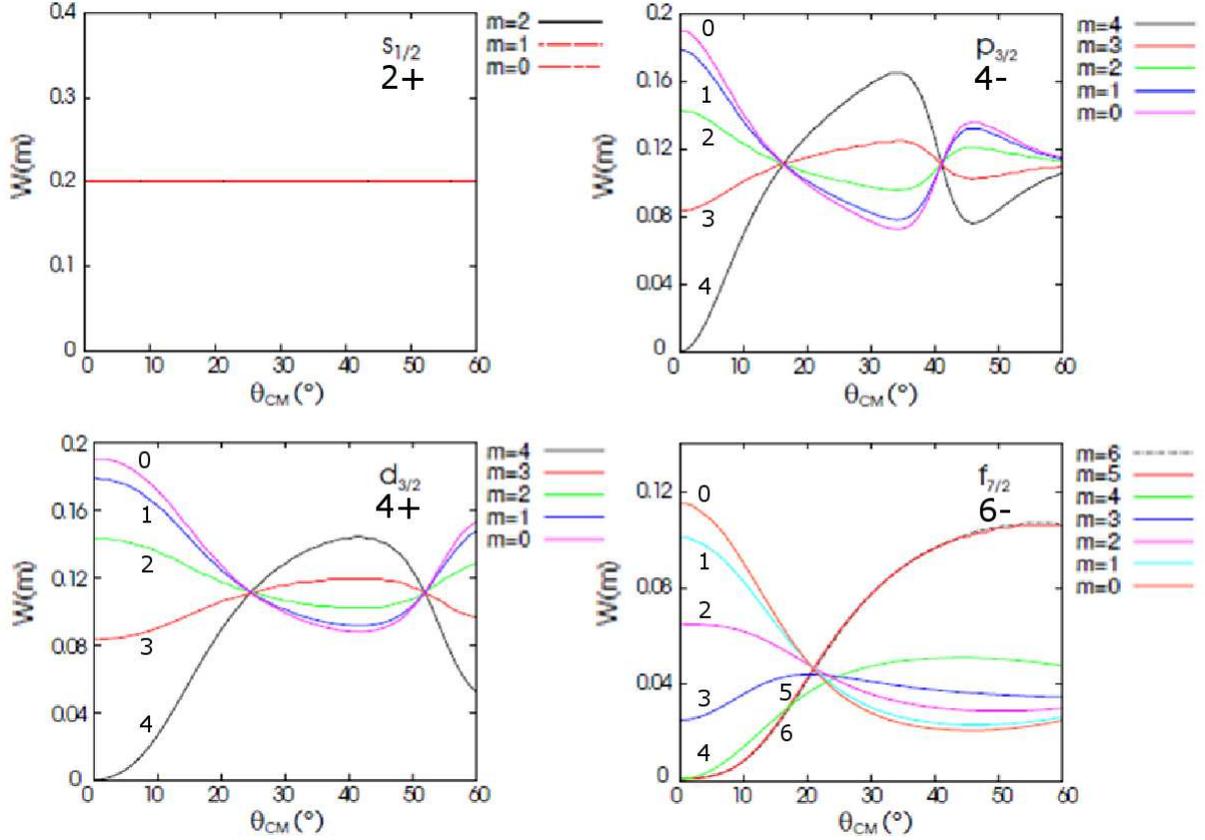}
%If the width of the Figure is less than 7.8 cm use the \texttt{sidecapion} command to flush the caption on the left side of the page. If the Figure is positioned at the top of the page, align the sidecaption with the top of the Figure -- to achieve this you simply need to use the optional argument \texttt{[t]} with the \texttt{sidecaption} command}
\caption{Calculations of magnetic substate population parameters as a function of centre of mass angle, performed using the ADWA model with the code TWOFNR \cite{TWOFNR}. The calculations all suppose a final state at 2.2 MeV excitation, formed in the (d,p) reaction with $^{25}$Na to make $^{26}$Na. The orbital into which the neutron is transferred is indicated, along with the assumed final state spin. It can be seen that, in general, the populations vary dramatically. In the experiment, centre of mass angles out to approximately $30^\circ$ were studied. }
\label{fig:26}       % Give a unique label
\end{figure}

The values of the population parameters $w(M_1)$ depend on the reaction mechanism and, in general, on the angle of the particle detection. An ADWA calculation for a (d,p) reaction can be used to calculate the population parameters $w(M_1)$ and their evolution with the detection angle of the proton. Examples of this are shown in Figure \ref{fig:26}, for the (d,p) study discussed above, using a beam of $^{25}$Na at 5 A.MeV \cite{Wilson}. The different panels correspond to different assumptions about the final orbital for the transferred neutron, and also for the final spin in $^{26}$Na. The different panels are for $\ell$ transfers of $\ell = 0,1,2$ and 3. The different lines are for different values of $M_1$ from 0 to $J_1$. The main point to note is that in general the populations change dramatically, for different angles of observation. The obvious counter example is the panel for $s_{1/2}$ transfer. The symmetry imposed by $s$-wave transfer forces all five substates, from $M_1 = -2$ to $+2$ to have equal weights of 0.2 at every observation angle and the gamma-ray emission will always be isotropic in this case.

\begin{figure}[h]
\sidecaption
\includegraphics[width=0.65\textwidth]{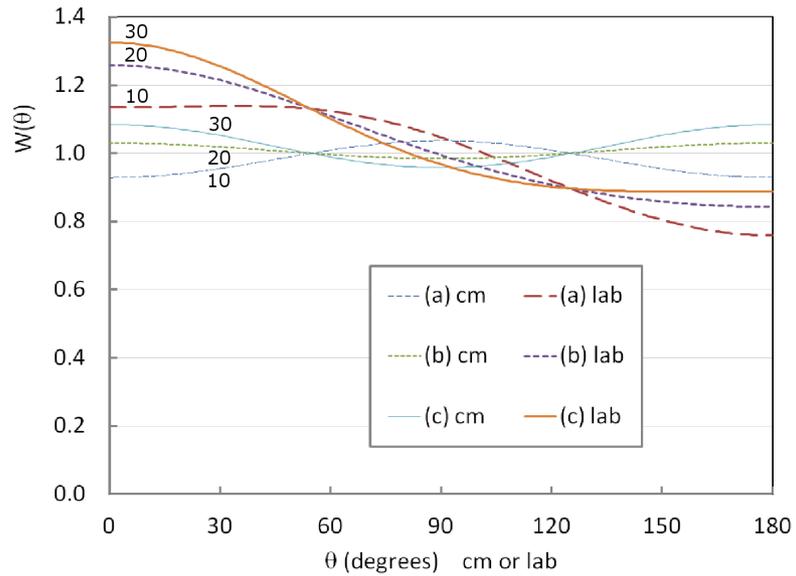}
%If the width of the Figure is less than 7.8 cm use the \texttt{sidecapion} command to flush the caption on the left side of the page. If the Figure is positioned at the top of the page, align the sidecaption with the top of the Figure -- to achieve this you simply need to use the optional argument \texttt{[t]} with the \texttt{sidecaption} command}
\caption{Gamma-ray angular distributions for different detection angles $\theta _{{\rm cm}}$(proton) for the proton from (d,p). Calculated for $^{25}$Na incident on deuterons at 5 A.MeV, with $1p_{3/2}$ transfer populating a hypothetical $4^-$ state at 2.2 MeV excitation energy. For the three symmetric curves, the horizontal axis shows the gamma-ray angle in the centre of mass frame of the emitting nucleus. For the other three curves, the horizontal angle is the gamma-ray angle measured in the laboratory, with respect to the beam direction. The proton centre of mass angles are (a) $10^\circ$, (b) $20^\circ$, (c) $30^\circ$.}
\label{fig:27}       % Give a unique label
\end{figure}

In Figure \ref{fig:27} the gamma-ray angular distributions determined by the substate populations are plotted, for the upper right hand case in Figure \protect\ref{fig:26}, namely $1p_{3/2}$ transfer populating a hypothetical $4^-$ state at 2.2 MeV excitation energy in $^{26}$Na. The gamma-ray decay is assumed to be a pure dipole decay to the $3^+$ ground state. Since the multipolarity of this decay is $L=1$, the maximum value of $K$ for the $a_K$ coefficients is 2. In the centre of mass frame (rest frame) of the emitting nucleus, the gamma-ray angular distribution with respect to the beam axis is given by a constant term plus a term proportional to $a_2 P_2 (\cos \theta )$, and the value of $a_2$ depends on the detection angle of the proton. It is assumed that, for a given proton angle $\theta$(proton) with respect to the beam direction, the protons are detected with cylindrical symmetry at all polar angles, $\phi$. For the centre of mass gamma-ray angular distributions, the functions are necessarily symmetric around $90^\circ$. The three curves intersecting the axis higher up at $\theta =0$ are plotted with the horizontal axis representing the gamma-ray angle as measured in the laboratory frame. There is a focussing of the gamma-rays towards zero degrees, due to the relativistic headlight effect as discussed in section \ref{subsec:tiara}.

\begin{figure}[h]
%\sidecaption
\centering
\includegraphics[width=0.9\textwidth]{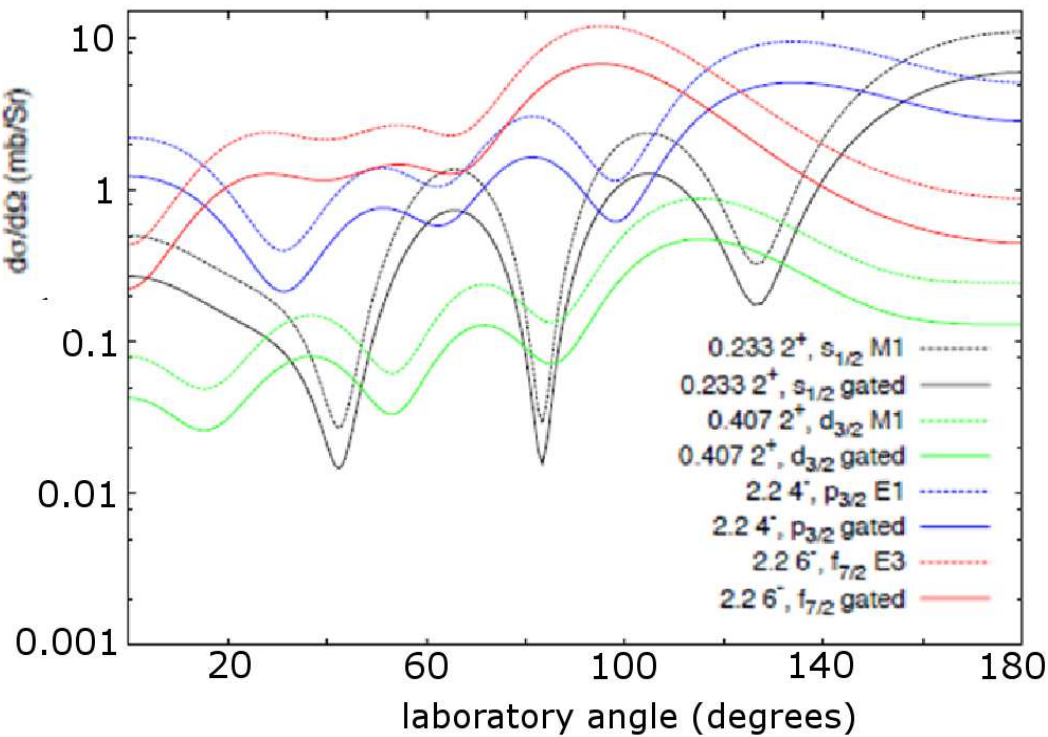}
%If the width of the Figure is less than 7.8 cm use the \texttt{sidecapion} command to flush the caption on the left side of the page. If the Figure is positioned at the top of the page, align the sidecaption with the top of the Figure -- to achieve this you simply need to use the optional argument \texttt{[t]} with the \texttt{sidecaption} command}
\caption{Differential cross sections in the laboratory frame, calculated for the (d,p) reaction leading to four different states in $^{26}$Na for an experiment at 5 A.MeV, in inverse kinematics. Pairs of almost parallel curves are shown for (a) $1s_{1/2}$ transfer to a $2^+$ state at 0.233 MeV, (b) $1p_{3/2}$ transfer to a hypothetical $4^-$ state at 2.2 MeV, (c) $0d_{3/2}$ transfer to a $2^+$ state at 0.407 MeV, (d) $0f_{7/2}$ transfer to a hypothetical $6^-$ state at 2.2 MeV in $^{26}$Na. In each case, the upper curve is the ADWA calculation and the lower curve is the calculated curve for a gamma-ray coincidence requirement (see text).  }
\label{fig:28}       % Give a unique label
\end{figure}

In Figure \ref{fig:28}, the differential cross sections in the laboratory frame are shown, for the population of states in $^{26}$Na via the (d,p) reaction in inverse kinematics. The curves for $\ell = 0, 1, 2$ and 3 show the expected movement of the main peak progressively further away from $180^\circ$ as $\ell$ increases. The parallel curves with the lower cross sections are actually the computed curves, assuming a gamma-ray coincidence requirement. The angular distributions for a gamma-decay to the ground state were computed using TWOFNR and the ADWA model, for each proton laboratory angle. The gamma-ray angular distributions were then integrated over the appropriate range of angles, corresponding to the laboratory angles spanned by the TIGRESS detectors in the experiment \cite{Wilson}. The relativistic aberration effect was also taken into account. The important point here is that the curves, whilst not perfectly parallel, are very little modified in shape from the ungated curves, i.e. those that have no coincidence requirement. This means that the experimental data can simply be corrected for the measured efficiency of the gamma-ray array and then compared with the unmodified ADWA calculations. This simplification was achieved in this experiment by the large angular range spanned by the gamma-ray array, which meant that the various changes in the angular distributions of the gamma-rays had little net effect after integration. The slight distortions that do occur are negligible (in this case) compared to the statistical errors in the data points and to the inevitable discrepancies that typically occur, between the theoretical and experimental shapes of the differential cross sections. The results from this experiment \cite{Wilson} are currently being prepared for publication.

\clearpage
%-------------------------------------------------------------------------------------------------
\subsection{Summary}
\label{subsec:summary}

Section \ref{sec:4:lightion} was headed {\em examples of light ion transfer experiments with radioactive beams} and in this section a range of different experimental approaches have been reviewed. With a relatively light projectile such as $^{11}$Be it was possible to make all of the detailed spectroscopic measurements using the beam-like particle. For the alternative approach using a silicon array for the light (target-like) particle, the TIARA array and subsequent developments such as T-REX and SHARC were described. Gamma-ray detection was shown to be useful, or in many cases essential, in order to resolve different excited states and to identify them on the basis of their gamma-ray decay pathways. Hence, the related issues of Doppler correction and angular correlations were discussed. The use of a detector centred at zero degrees for the beam-like reaction products was shown to be a great advantage. Whilst a large-acceptance spectrometer such as VAMOS gives superior performance including full particle identification, it was shown that even a simple detector such as the {\it trifoil} can substantially assist in the reduction of background. The background arises from compound nuclear reactions induced by the beam on contaminant materials in the target, such as carbon. A common target choice is to use normal $(CH_2)_n$ or deuterated $(CD_2)_n$ polythene self-supporting foils. The option of using a helical orbit (solenoidal) spectrometer instead of a conventional silicon array, for the light particle detection, was described. An example of the use of a cryogenic target of deuterium was included: in the example described, the target was thick and largely absorbed the low energy target-like particles, but it is worth noting that there is research aimed at producing much thinner cryogenic targets that could be used with light particle detection. Finally, an important different approach was described, wherein the target thickness is essentially removed as a limitation because the target becomes the detector itself. This is sometimes called an {\it active target}. With a time projection chamber (TPC) such as MAYA, the fill-gas of the detector includes within its molecules the target nuclei, and the measurements make it possible to reconstruct the full kinematics of the nuclear reaction in three dimensions. This makes an active target the ideal choice for very low intensity beams, where a thick target is indispensable. With more development to improve the resolution and dynamic range, this type of detector could eventually have the widest applicability of all experimental approaches.
	
\section{Heavy ion transfer reactions}
\label{sec:5:heavyion}
	
For the transfer of a nucleon between two heavy ions, there is an important selectivity in favour of certain final states which allows the spins of the final states to be deduced. This is known as $j_>/j_<$ selectivity because it can tell us whether the final orbital for the transferred nucleon has $j = \ell + 1/2$ or $j = \ell - 1/2$. The origin of the effect is two-fold \cite{bond-comment}. Firstly, a heavy ion at the appropriate energies will have a small de Broglie wavelength because of its large mass, and hence its path can be reasonably described as a classical trajectory. Secondly, the transfer must take place in a peripheral encounter between projectile and target because a smaller impact parameter will result in a strongly absorbed compound nuclear process and a larger impact parameter will keep the nuclei from interacting except through the large repulsive coulomb interaction. Therefore, we can consider classical trajectories for peripheral transfer and take into account quantum mechanical factors in a semiclassical fashion. Of course, a full quantum mechanical treatment using the normal reaction theories is possible. The advantage of the semiclassical model is that it allows the origins of the particular selectivity in heavy ion transfer to be understood more readily.

\subsection{Selectivity according to j$_>$ and j$_<$ in a semi-classical model}
\label{subsec:bond}

\begin{figure}[h]
\sidecaption
\includegraphics[width=.35\textwidth]{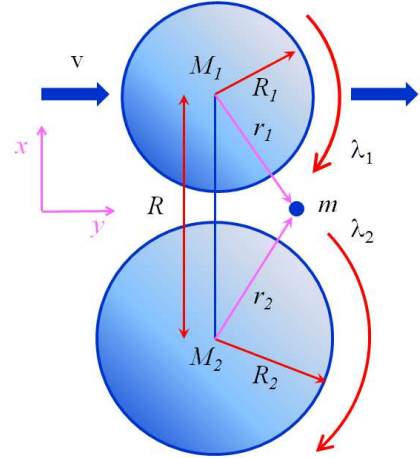}
%If the width of the Figure is less than 7.8 cm use the \texttt{sidecapion} command to flush the caption on the left side of the page. If the Figure is positioned at the top of the page, align the sidecaption with the top of the Figure -- to achieve this you simply need to use the optional argument \texttt{[t]} with the \texttt{sidecaption} command}
\caption{Sketch of the transfer of a mass $m$ from the projectile $M_1$ to the target $M_2$ in a heavy ion collision, showing the variables used to derive the Brink matching conditions \protect\cite{Brink} (see text). }
\label{fig:29}       % Give a unique label
\end{figure}

The semiclassical model for nucleon transfer between heavy ions has been described by Brink \cite{Brink} and is represented in Figure \ref{fig:29}. At the moment of transfer, the mass $m$ has some linear momentum in the beam direction due to the beam velocity $\upsilon$ and also due to the rotational motion of $m$ around $M_1$. Just after  the transfer, it is orbiting $M_2$ which is at rest, and all of the linear momentum is due to the orbital motion. The initial and final linear momenta should be approximately equal by conservation of momentum. Quantum mechanically, they need not be exactly equal because of the uncertainty in momentum introduced by the spatial uncertainty in the precise point of transfer as measured in the beam direction (which can be estimated). A similar condition can be formulated for the angular momentum of the transferred mass $m$. Before the transfer, this has contributions from the relative motion between the two colliding heavy ions and from the internal orbital angular momentum of the transferred nucleon. These are the only parts that change: the former due to the adjustments in mass and possibly charge, and the second due to the change of orbital. Once again, the initial and final values should be almost equal.

The two kinematical conditions given by Brink \cite{Brink} are:
$$ \Delta k = k_0 - \lambda_1 / R_1 - \lambda_2 / R_2 \approx 0 $$
$$ \Delta L = \lambda_2 - \lambda_1 + \frac{1}{2} k_0 (R_1 - R_2 ) + Q_{{\rm eff}} R / \hbar \upsilon \approx 0  $$
where the orbital angular momentum and projection on the $z$-axis for the transferred particle are given by $(\ell, \lambda)$ with subscripts 1 and 2 for before and after the transfer, respectively. The quantity $Q_{{\rm eff}}$ is equal to the reaction Q-value in the case of neutron transfer, but otherwise has an adjustment due to changes in Coulomb repulsion: $Q_{{\rm eff}} = Q - \Delta ( Z_1 Z_2 e^2 /R)$. The beam direction is $y$ and the $z$ direction is chosen perpendicular to the reaction plane. A further pair of conditions arise from the requirement that the transfer should take place in the reaction plane, where the two nuclei meet, and hence the spherical harmonic functions $Y_{\ell m}$ should not be zero in that plane:
$$ \ell_1 + \lambda_1 = {\rm even} $$
$$ \ell_2 + \lambda_2 = {\rm even}. $$

The two kinematical conditions arising from linear momentum and angular momentum conservation will each, separately, imply a particular {\it well matched} angular momentum value, for a given reaction, bombarding energy and final state energy (Q-value). Alternatively, for a given $\ell$-transfer they will each imply a particular excitation energy at which the matching is optimal. If the values implied by the two equations are equal, then the reaction to produce a state of the given spin and excitation energy will have a large cross section (if such a state exists, with the correct structure in the final nucleus). If the two values are not equal, then the cross section will be reduced by an amount that depends on the degree of mismatch.

\subsection{Examples of selectivity observed in experiments}
\label{subsec:hi-selectivity}

\begin{figure}[h]
\sidecaption
\includegraphics[width=.4\textwidth]{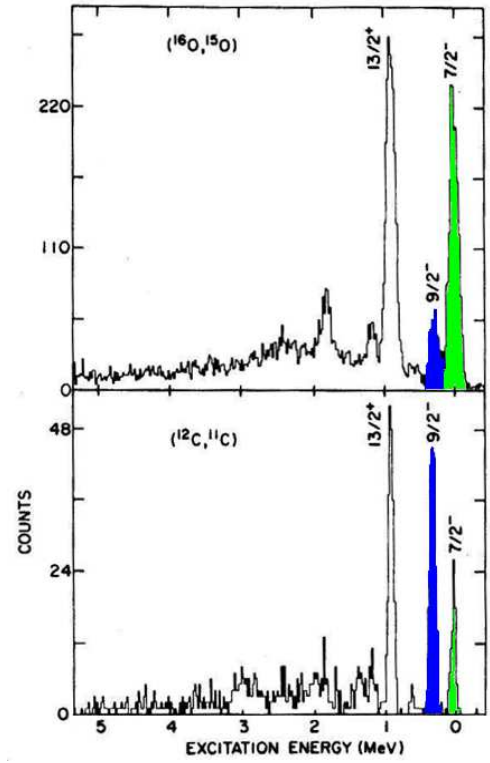}
%If the width of the Figure is less than 7.8 cm use the \texttt{sidecapion} command to flush the caption on the left side of the page. If the Figure is positioned at the top of the page, align the sidecaption with the top of the Figure -- to achieve this you simply need to use the optional argument \texttt{[t]} with the \texttt{sidecaption} command}
\caption{Illustration of the $j_> /j_<$ selectivity exhibited in heavy ion transfer when the Q-value is large and negative. The data are for the reactions ($^{16}$O,$^{15}$O) and ($^{12}$C,$^{11}$C) on a $^{148}$Sm target with the same beam velocity, defined by a beam energy of 7.5 A.MeV. The selectivity is reversed due to the parent orbitals of the transferred neutron being $0p_{1/2}$ ($j_<$) and $0p_{3/2}$ ($j_>$) respectively. Therefore the upper panel favours $j_>$ states and the lower panel favours $j_<$ states. The two highlighted peaks correspond to populating the $0h_{9/2}$ ($j_<$) and $1f_{7/2}$ ($j_>$) orbitals. The biggest peak (unshaded) corresponds to the $0i_{13/2}$ ($j_>$) orbital. Figure adapted from ref. \cite{bond-comment}}
\label{fig:30}       % Give a unique label
\end{figure}

A detailed inspection of the Brink matching conditions for $\Delta k$  and $\Delta L$, given above, implies that a reaction with a large negative Q-value will favour final states with high spin, or more specifically a large value of $\lambda _2$ in the notation of Figure \ref{fig:29}. This arises because the conservation of linear momentum favour a high value of $\lambda _2 + \lambda_1$ and the conservation of angular momentum implies a large value for $\lambda_2 - \lambda_1$. This selectivity, which occurs for heavy ion transfer with a negative Q-value, is discussed in detail by Bond \cite{Bond} with a derivation in terms of DWBA formalism.  As further noted by Bond \cite{bond-comment} the large negative Q-value will imply that the projectile has reduced kinetic energy after the collision and hence is slowed down, which implies a significant transfer of angular momentum. Being heavy ions, the angular momentum of relative motion is large, and hence a relatively small reduction corresponds to transfer into a relatively high spin orbital. In Figure \ref{fig:30} for the ($^{16}$O,$^{15}$O) reaction, which has a large negative Q-value, the neutron is transferred from the $0p_{1/2}$ orbital. For the best matching, there will be a maximum $\ell$-transfer which implies that the nucleon will change $\lambda$, i.e. the projection of the angular momentum in the direction perpendicular to the reaction plane, as much as possible. For example, from a $0p_{1/2}$ orbital (with orbital angular momentum $\ell =1$) and an initial projection of $m_\ell = -1$ (which implies also that $m_s = +1/2$) the transfer will favour $m_\ell = +\ell$ for a high-$\ell$ orbital in the final nucleus. It is reasonable to assume that there is no interaction in the transfer to change the direction (projection) of the intrinsic spin of the nucleon. Therefore the relative directions of orbital and spin angular momentum for the nucleon become swapped in the transfer process. The preferred transfer in this case is from $\ell - 1/2$ (denoted as $j_<$) to $\ell + 1/2$ (denoted as $j_>$). In general, if the Q-value is negative, the transfer from an orbital with $j_<$ ($j_>$) will favour the population of orbitals with $j_>$ ($j_<$) in order to achieve the largest change in $\lambda$ for the transferred nucleon. Therefore, in Figure \ref{fig:30}, the reaction ($^{12}$C,$^{11}$C) shows the opposite selectivity to ($^{16}$O,$^{15}$O). In the upper panel we see a favouring of the ($j_>$) $7/2^-$ state corresponding to the $1f_{7/2}$ orbital, and a relative suppression of the ($j_<$) $9/2^-$ state corresponding to the $0h_{9/2}$ orbital. This selectivity is reversed in the lower panel, and we also see that the ($j_>$) $13/2^+$ state (corresponding to the $0i_{13/2}$ orbital) follows the ($j_>$) $7/2^-$ in becoming weaker relative to the favoured ($j_<$) $9/2^-$ state.

\begin{figure}[h]
\sidecaption
\includegraphics[width=.45\textwidth]{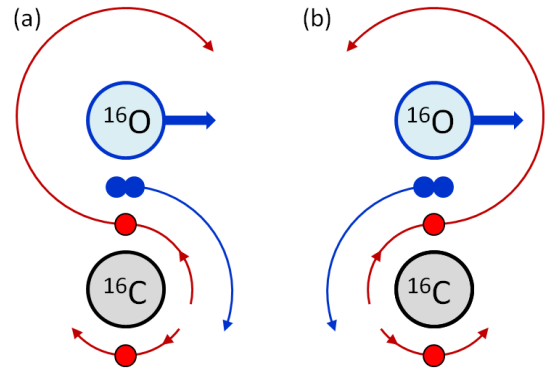}
%If the width of the Figure is less than 7.8 cm use the \texttt{sidecapion} command to flush the caption on the left side of the page. If the Figure is positioned at the top of the page, align the sidecaption with the top of the Figure -- to achieve this you simply need to use the optional argument \texttt{[t]} with the \texttt{sidecaption} command}
\caption{The semiclassical model of Brink \protect\cite{Brink,AnyasWeiss} can be extended to two-step transfer reactions, such as this ($^{18}$O,$^{17}$F) reaction on a target of $^{18}$O (Figure adapted from ref. \cite{n19o21}). The reaction is modelled as a dineutron transfer from the $^{18}$O projectile and the pickup of a proton from the $^{18}$O target: (a) the strongly favoured senses for the two transfers, (b) the less favoured transfer directions.  }
\label{fig:31}       % Give a unique label
\end{figure}

The discussion for single nucleon transfer can be simply extended to include cluster transfer \cite{Brink}. A further step is to describe reactions in which nucleons are transferred in both directions, to and from the projectile, or in two independent transfers in the same direction. In the work of ref. \cite{n19o21}, the ideas developed by Brink \cite{Brink} and described by Anyas-Weiss {\it et al.} \cite{AnyasWeiss} are extended to describe the reactions ($^{18}$O,$^{17}$F) and ($^{18}$O,$^{15}$O) where one of the two steps is the transfer of a dineutron cluster. The trajectories of the transferred particles between the two heavy ions are represented in Figure \ref{fig:31} for the favoured (well matched) and unfavoured trajectories. The proton is required in each case to make a transition from $j_<$ to $j_>$ in a stretched trajectory as shown, so as to form the $5/2^+$ ground state in $^{17}$O, which was observed in the experiment \cite{n19o21}. Figure \ref{fig:31}(a) shows that the favoured final states in $^{19}$N will have a total spin where 1/2 from the $0p_{1/2}$ proton is added collinearly with the orbital angular momentum transferred by the dineutron cluster. This type of selectivity was observed in the experiment and was used to interpret the states populated in $^{19}$N and $^{21}$O. In the case of the $^{21}$O there has been independent verification of the interpretation via the previously-mentioned study of the (d,p) reaction with a beam of $^{20}$O using TIARA \cite{bea}.

%-------------------------------------------------------------------------------------------------	
\section{Perspectives}
\label{sec:6:perspectives}

\begin{figure}[h]
\sidecaption
\includegraphics[width=.7\textwidth]{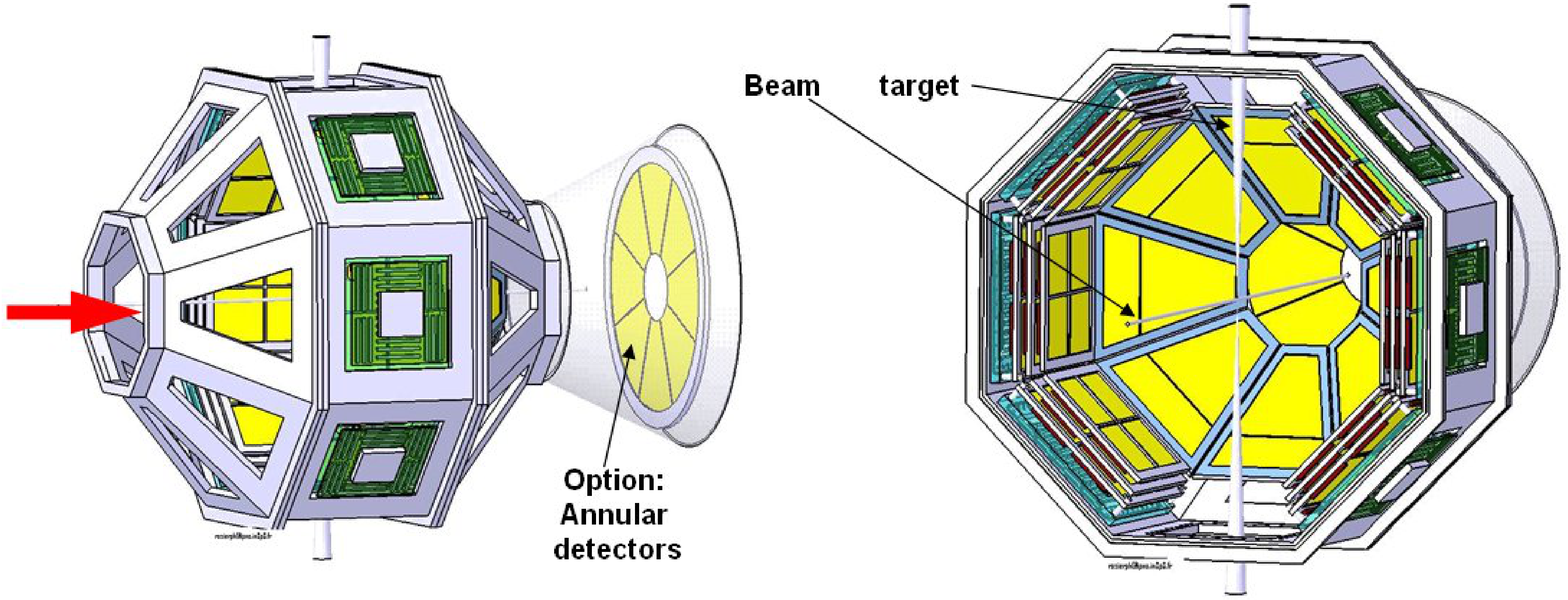}
%If the width of the Figure is less than 7.8 cm use the \texttt{sidecapion} command to flush the caption on the left side of the page. If the Figure is positioned at the top of the page, align the sidecaption with the top of the Figure -- to achieve this you simply need to use the optional argument \texttt{[t]} with the \texttt{sidecaption} command}
\caption{Preliminary design for a new array GASPARD \cite{GASPARD} which would represent a new generation of device for the approach using a compact particle array with coincident gamma-ray detection. Multilayer highly segmented particle detectors with enhanced particle identification properties, plus the ability to use cryogenic targets, are amongst its advantages. }
\label{fig:32}       % Give a unique label
\end{figure}

It is always dangerous to speculate about the future directions for the development of instrumentation or experimental techniques. The experimental devices described here are all likely to deliver a range of new results in nucleon transfer, as new facilities and more beams at the appropriate energies become available. It is, however, perhaps worth taking note of some of the new developments that might be expected. These developments will in part be enabled by an increased capability to deal with large numbers of electronics channels, due to innovations in electronics design. One development is to take the simple idea of a highly efficient silicon array (i.e. with a large geometrical coverage) mounted inside a highly efficient gamma-ray array (as adopted by TIARA, SHARC, T-REX, ORRUBA, $\ldots$) and improve it. This is the aim of GASPARD \cite{GASPARD} which is an international initiative based originally around the new SPIRAL2 Phase 2 facility but also able to be deployed potentially at HIE-ISOLDE. A preliminary design is shown in Figure \ref{fig:32}. The particle detection is based on one to three layers of silicon, depending on angle. The segmentation of the silicon is sub-millimetre, but with the detectors still able to supply particle identification information based on the pulse shape. The array is sufficiently compact to fit inside newly developed gamma-ray arrays such as AGATA or PARIS. The geometry is chosen to allow innovative target design, and in particular to have operation with the thin solid hydrogen target CHyMENE, currently under development at Saclay. Another current development is the AT-TPC detector at MSU \cite{ATTPC} which aims to combine the advantages of the active target MAYA and the helical spectrometer HELIOS. As noted previously, a key advantage of an active target is that it can, in principle, remove the limitations on energy resolution (or, indeed the limitations on even being able to the detect reaction products) that arise from target thickness. An alternative approach to minimising the target thickness effect is to use an extremely thin target but to compensate by passing the beam through it many times, say $10^6$ times. Under certain circumstances, a beam of energy 5-10 A.MeV as suitable for transfer could be maintained and recirculated in a storage ring for this many revolutions. A thin gas jet target would allow transfer reactions to be studied in inverse kinematics. The ring could be periodically refilled and the beam cooled, in a procedure that was synchronised with the time structure of the beam production. This is one of the ideas behind the proposed operation of the TSR storage ring with reaccelerated ISOL beams at ISOLDE \cite{TSR}.

In summary, there are some very powerful experimental devices already available and able to exploit the existing and newly developed radioactive beams. In addition, there are challenging and exciting developments underway, that will create even better experimental possibilities to exploit the beams from the next generation of facilities. Because of their unique selectivity, and because the states that are populated have a simple structure that should be especially amenable to a theoretical description and interpretation, transfer reactions will always be at the forefront of studies using radioactive beams to extend our knowledge of nuclear structure.

%\begin{figure}
%\begin{center}\includegraphics[width=0.6\textwidth]{wfres_o17_spar.eps}\end{center}
%\caption{\label{wfres} Radial part of the  $d_{3/2}$ single-particle resonance wavefunction in $^{17}$O at $E_r=0.95$~MeV compared with a slightly bound %wavefunction ($E=-0.1$~MeV) and a bin wavefunction, centered at the nominal energy of the resonance and with a width of 0.5 MeV.}
%\end{figure}

%---------------------------------------  ACKNOWLEDGEMENTS ------------------------------------
\begin{acknowledgement}
Thank you to all of my colleagues who have assisted with putting together, performing and analysing the results from the experiments described in this work: these include N.A. Orr, M. Labiche, R.C. Lemmon, C.N. Timis, B. Fernandez-Dominguez, J.S. Thomas, S.M. Brown, G.L. Wilson, C. Aa. Diget, the VAMOS group at GANIL and the TIGRESS group at TRIUMF, the nuclear theory group at Surrey, plus all of the other colleagues from the TIARA, MUST2, TIGRESS and Charissa collaborations: {\it we together weathered many a storm} \cite{Bob}.
\end{acknowledgement}

%\input{referenc}
%%%%%%%%%%%%%%%%%%%%%%%% referenc.tex %%%%%%%%%%%%%%%%%%%%%%%%%%%%%%
% sample references
% %
% Use this file as a template for your own input.
%
%%%%%%%%%%%%%%%%%%%%%%%% Springer-Verlag %%%%%%%%%%%%%%%%%%%%%%%%%%
%
% BibTeX users please use
%\bibliographystyle{elsart-num}
% \bibliography{catford-euroschool-transfer}
%

\end{document}